\documentclass[12pt,a4paper]{article}%
\usepackage{amsmath,amssymb,amsfonts}
\usepackage{graphicx,epsfig}
\usepackage{color}
\usepackage{amsmath}
\usepackage{amsfonts}
\usepackage{amssymb}
\usepackage{float}
\usepackage{cancel}
\usepackage{caption2}
\usepackage{graphics}
\usepackage{hyperref}
\usepackage{graphicx}%
\numberwithin{equation}{section} 
\setcaptionmargin{1cm}

\def\beq{\begin{equation}}
\def\eeq{\end{equation}}
\def\bea{\begin{eqnarray}}
\def\eea{\end{eqnarray}}
\newcommand{\newc}{\newcommand}
\newc{\gsim}{\lower.7ex\hbox{$\;\stackrel{\textstyle>}{\sim}\;$}}
\newc{\lsim}{\lower.7ex\hbox{$\;\stackrel{\textstyle<}{\sim}\;$}}
\newc{\ie}{{\it i.e.}}
\newc{\etal}{{\it et al.}}
\newc{\mev}{\hbox{\rm\,MeV}}
\newc{\gev}{\hbox{\rm\,GeV}}
\newc{\tev}{\hbox{\rm\,TeV}}
\begin{document}

\begin{titlepage}
\begin{flushright}
\end{flushright}
\vskip 1.0cm
\begin{center}
{\Large \bf Bounding scalar operator dimensions in 4D CFT} \vskip 1.0cm
{\large Riccardo Rattazzi$^a$,\ Vyacheslav S.~Rychkov$^b$,\ Erik Tonni$^{c}$,\ Alessandro Vichi$^a$} \\[1cm]
{\it $^a$ Institut de Th\'eorie des Ph\'enom\`enes Physiques, EPFL,  CH--1015 Lausanne, Switzerland}\\[5mm]
{\it $^b$ Scuola Normale Superiore and INFN, Piazza dei Cavalieri 7, 56126 Pisa, Italy} \\[5mm]
{\it $^c$ Dipartmento di Fisica, Universit\`a di Pisa and INFN,
sezione di Pisa,\\
Largo Bruno Pontecorvo 3, 56127 Pisa, Italy}\\[5mm]
\vskip 1.0cm \abstract{ In an arbitrary unitary 4D CFT we
consider a scalar operator $\phi$, and the operator $\phi^2$ defined
as the lowest dimension scalar which appears in the OPE $\phi\times
\phi$ with a nonzero coefficient. Using general considerations of
OPE, conformal block decomposition, and crossing symmetry, we derive
a theory-independent inequality $[\phi^2]\leq f([\phi])$ for the
dimensions of these two operators. The function $f(d)$ entering this
bound is computed numerically. For $d\rightarrow1$ we have
$f(d)=2+O(\sqrt{d-1}$), which shows that the free theory limit is
approached continuously. We perform some checks of our bound. We
find that the bound is satisfied by all weakly coupled 4D conformal
fixed points that we are able to construct. The Wilson-Fischer fixed
points violate the bound by a constant $O(1)$ factor, which must be
due to the subtleties of extrapolating to $4-\varepsilon$
dimensions. We use our method to derive an analogous bound in 2D,
and check that the Minimal Models satisfy the bound, with the Ising
model nearly-saturating it. Derivation of an analogous bound in 3D
is currently not feasible because the explicit conformal blocks are
not known in odd dimensions. We also discuss the main
phenomenological motivation for studying this set of questions:
constructing models of dynamical ElectroWeak Symmetry Breaking
without flavor problems.}
\end{center}
\end{titlepage}

\newpage

\tableofcontents

\newpage

\section{The problem and the result}

Operator dimensions in unitary Conformal Field Theories (CFT) are subject to
important constraints known as unitarity bounds. In the simplest case of a
scalar primary operator $\phi$, the unitarity bound states
that\footnote{Unless explicitly noted otherwise, all statements of this paper
refer to $D=4$ spacetime dimensions.}
\begin{align}
&  d\equiv\lbrack\phi]\geq1,\label{bound}\\
d  &  =1\Longleftrightarrow\phi\text{ is free.} \label{limit}%
\end{align}
This classic result invites the following question: What happens if
$d=1+\varepsilon$? In particular, is there any sense in which the CFT (or at
least its subsector not decoupled from $\phi$) should be close to the free
scalar theory if $d$ is close to 1? For instance, do all operator dimensions
in this subsector approach their free scalar theory values in the limit
$d\rightarrow1$? The standard proof of the unitarity bound \cite{mack} does
not shed light on this question.

\begin{figure}
[ptb]
\begin{center}
\includegraphics[
height=2.041in,
width=3.0848in
]%
{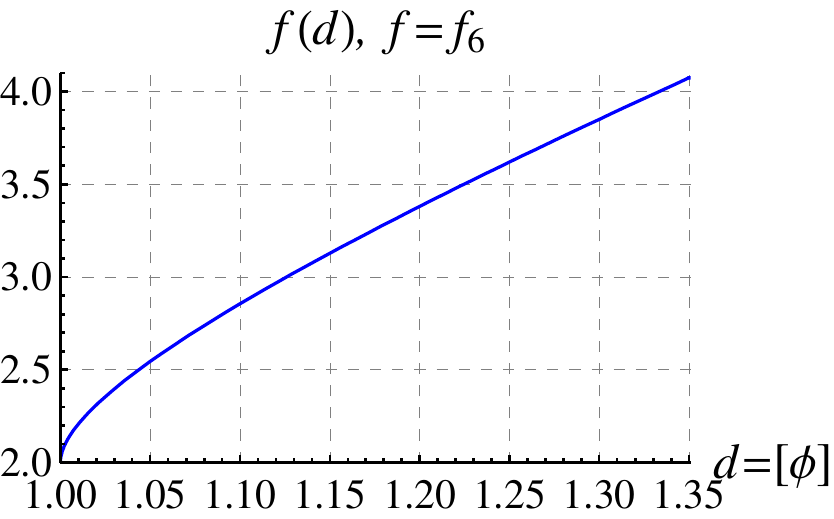}%
\caption{The best current bound (\ref{main}), obtained by the method described
in Section~\ref{results}. The subscript in $f_{6}$ refers to the order of
derivatives used to compute this bound.}%
\label{fig:bound-intro}%
\end{center}
\end{figure}

In this paper we will show that such continuity indeed holds for the operator
`$\phi^{2}$'$,$ by which we mean the lowest dimension \textit{scalar} primary
which appears in the OPE of $\phi$ with itself:%
\begin{equation}
\phi(x)\phi(0)\sim(x^{2})^{-d}(1+C|x|^{\Delta_{\min}}\phi^{2}(0)+\ldots
)\,,\quad C\neq0\,. \label{OPE0}%
\end{equation}
In free theory $\Delta_{\min}\equiv\lbrack\phi^{2}]=2,$ and we will show that
$\Delta_{\min}\rightarrow2$ in any CFT as $d\rightarrow1.$ More precisely, we
will show that in any 4D CFT
\begin{equation}
\Delta_{\min}\leq f(d), \label{main}%
\end{equation}
where $f(d)$ is a certain continuous function such that $f(1)=2$. We will
evaluate this function numerically; it is plotted in Fig.
\ref{fig:bound-intro} for $d$ near $1.$%

%TCIMACRO{\FRAME{ftbpFU}{3.0848in}{2.041in}{0pt}{\Qcb{The best current bound
%(\ref{main}), obtained by the method described in Section~\ref{results}. The
%subscript in $f_{6}$ refers to the order of derivatives used to compute this
%bound.}}{\Qlb{fig:bound-intro}}{bound-intro.pdf}%
%{\special{ language "Scientific Word";  type "GRAPHIC";  display "USEDEF";
%valid_file "F";  width 3.0848in;  height 2.041in;  depth 0pt;
%original-width 3.333in;  original-height 2.3004in;  cropleft "0";
%filename 'bound-intro.pdf';file-properties "XNPEU";}}}%
%BeginExpansion
%EndExpansion

We stress that bound (\ref{main}) applies to the OPE $\phi\times\phi$ of an
arbitrary scalar primary $\phi$. However, since the function $f(d)$ is
monotonically increasing, the bound is strongest for the scalar primary of
minimal dimension.

Our analysis will use only the most general properties of CFT, such as
unitarity, OPE, conformal block decomposition, and crossing symmetry. The
resulting bound (\ref{main}) is thus model independent. In particular, it
holds independently of the central charge, of the spectrum of whatever other
operators which may be present in the CFT, and of the coefficients with which
these operators may appear in the OPE (\ref{OPE0}). Our analysis is
non-perturbative and does not assume that the CFT may be continuously
connected to the free theory.

We do not know of any 4D CFT which comes close to saturating the bound of
Fig.\ \ref{fig:bound-intro}. We are not claiming that this bound is the best
possible, and in fact we do know that it can be somewhat improved using our
method and investing some more computer time. In spite of not being the best
possible, the curve of Fig.~\ref{fig:bound-intro} is a valid bound, and
represents a necessary condition which should be satisfied in any unitary CFT.\

The paper is organized as follows. In Section \ref{pheno} we explain the
phenomenological motivations behind this question, which are related to the
naturalness problem of the electroweak scale. In Section \ref{CFT-review} we
review the necessary CFT techniques. In Section \ref{sum-rule} we derive a sum
rule for the contributions of all primary fields (with arbitrary spins and
dimensions) appearing in the $\phi\times\phi$ OPE. In Section \ref{results} we
explain how the sum rule is used to derive the bound (\ref{main}). In Section
\ref{sec:comparison} we check our bound against operator dimensions in various
calculable CFTs in $D=4$ and $4-\varepsilon$. We also present and similarly
check an analogous bound in $D=2$. In Section \ref{sec:connection} we discuss
to what extent our result in its current form addresses the phenomenological
problem from Section \ref{pheno}. In Section \ref{sec:conclusions} we conclude
and indicate future research directions.

%The \textsc{Mathematica} codes used to derive our result are available at
%\cite{download}.

\section{Phenomenological motivation}

\label{pheno}

The phenomenological motivation of our study is given by one declination of
the hierarchy problem, which was lucidly discussed in a paper by Luty and Okui
\cite{luty} (see also \cite{Strassler:2003ht}). This section is to a
significant extent a review of the discussion in that paper. The bulk of the
paper is logically independent of this section, and the reader who is mainly
interested in the formal aspects of our result may skip to Section
\ref{CFT-review}.

The issue of mass hierarchies in field theory can be conveniently depicted
from a CFT viewpoint. Indeed the basic statement that a given field theory
contains two widely separated mass scales $\Lambda_{\text{IR}}\ll
\Lambda_{\text{UV}}$ already implies that the energy dependence of physical
quantities at $\Lambda_{\text{IR}}\ll E\ll\Lambda_{\text{UV}}$ is
\textit{small}, corresponding to approximate scale (and conformal) invariance.
In the case of perturbative field theories the CFT which approximates the
behaviour in the intermediate mass region is just a free one. For instance, in
the case of non-SUSY GUT's, $\Lambda_{\text{IR}}$ and $\Lambda_{\text{UV}}$
are respectively the Fermi and GUT scale, and the CFT which approximates
behaviour at intermediate scales is just the free Standard Model. From the CFT
viewpoint, the naturalness of the hierarchy $\Lambda_{\mathrm{IR}}\ll
\Lambda_{\mathrm{UV}}$, or equivalently its stability, depends on the
dimensionality of the scalar operators describing the perturbations of the CFT
Lagrangian around the fixed point. In the language of the RG group,
naturalness depends on the relevance of the deformations at the fixed point.
If the theory possesses a scalar operator $\mathcal{O}_{\Delta}$, with
dimension $\Delta<4$, one generically expects\footnote{See concrete examples
in Section \ref{sec:quant} below.} UV physics to generate a perturbation
\begin{equation}
\mathcal{L}_{pert}=c\Lambda_{\mathrm{UV}}^{4-\Delta}\mathcal{O}_{\Delta}\,,
\end{equation}
corresponding roughly to an IR scale
\begin{equation}
\Lambda_{\mathrm{IR}}=c^{\frac{1}{4-\Delta}}\Lambda_{\mathrm{UV}}\,.
\end{equation}
Absence of tuning corresponds to the expectation that $c$ be not much smaller
than $O(1)$. If $4-\Delta$ is $O(1)$ (\textit{strongly relevant} operator) a
hierarchy between $\Lambda_{\mathrm{IR}}$ and $\Lambda_{\mathrm{UV}}$ can be
maintained only by tuning $c$ to be \textit{hierarchically} smaller than one.
This corresponds to an unnatural hierarchy. On the other hand when $4-\Delta$
is close to zero (\textit{weakly relevant} operator) a mass hierarchy is
obtained as soon as both $4-\Delta$ and $c$ are just \textit{algebraically}
small\footnote{We stole this definition from ref. \cite{Strassler:2003ht}.}.
For instance for $4-\Delta=c=0.1$ the mass hierarchy spans 10 orders of
magnitude. Therefore for a weakly relevant operator a hierarchy is considered
natural. The hierarchy between the confinement and UV scale in Yang-Mills
theory is an example in this second class, albeit a limiting one\footnote{This
is because the corresponding deformation, the glueball field $G_{\mu\nu}%
^{A}G_{A}^{\mu\nu}$, is marginally relevant: its scaling dimension is
$4-ag^{2}$ and becomes exactly $4$ at the gaussian fixed point.}. The only
exception to the above classification of naturalness concerns the case in
which the strongly relevant operators transform under some global approximate
symmetry. In that case it is natural to assume that the corresponding $c$'s be
small, even hierarchically small. The stability of the hierarchy depends then
on the dimension $\Delta_{S}$ of the scalar singlet (under all global
symmetries) of lowest dimension. If $4-\Delta_{S}\ll1$ the hierarchy is natural.

According to the above discussion, in the SM the hierarchy between the weak
scale and any possible UV scale is unnatural because of the presence of a
scalar bilinear in the Higgs field $H^{\dagger}H$ which is a total singlet
with dimension $\sim2$. On the other hand in supersymmetric extensions of the
SM, such scalar bilinears exist but their coefficient can be naturally chosen
to be small. In a general supersymmetric model the weak scale is then
naturally generated either by a marginally relevant deformation (dynamical
supersymmetry breaking) or simply by adding strongly relevant supersymmetry
breaking deformations with small coefficients (soft supersymmetry breaking).
Technicolor models are instead similar to the case of YM theory: at the
gaussian fixed point there are no gauge invariant scalars of dimension $<4$.

As far as the hierarchy is concerned these extensions are clearly preferable
to the SM. However as far as flavor physics is concerned the SM has, over its
extensions, an advantage which is also a simple consequence of operator
dimensionality. In the SM the flavor violating operators of lowest
dimensionality, the Yukawa interactions, have dimension $=4$,
\begin{equation}
\mathcal{L}_{Y}=y_{ij}^{u}H\bar{q}_{L}u_{R}+y_{ij}^{d}H^{\dagger}\bar{q}%
_{L}d_{R}+y_{ij}^{e}H^{\dagger}\bar{L}_{L}e_{R}\quad(\text{SM})\,,
\end{equation}
and provide a very accurate description of flavor violating phenomenology. In
particular, the common Yukawa origin of masses and mixing angles leads to a
critically important suppression of Flavor Changing Neutral Currents (FCNC)
and CP violation. This suppression is often called Natural Flavor Conservation
or GIM mechanism \cite{Glashow:1970gm}. Once the hierarchy
%%%%%%%%%%%%%%%%%%%%%%%%%
$v\ll\Lambda_{\mathrm{UV}}$
%I REMOVED _F from v to avoid confusion with _F below, which denotes Flavor
is taken as a fact, no matter how unnatural, extra unwanted sources of flavor
violation are automatically suppressed. In particular the leading effects are
associated to 4-quark interactions, with dimension 6, and are thus suppressed
by $v^{2}/\Lambda_{\mathrm{UV}}^{2}$. The situation is not as good in
supersymmetry, where in addition to the Yukawa interactions flavor is violated
by operators of dimension 2 and 3 involving the sfermions. The comparison with
technicolor brings us to discuss the motivation for our paper. In technicolor
the Higgs field is a techni-fermion bilinear $H=\bar{T}T$ with dimension
$\sim3$. The SM fermions instead remain elementary, i.e. with dimension $3/2$.
The Yukawa interactions are therefore irrelevant operators of dimension 6,
\begin{equation}
\mathcal{L}_{Y}=\frac{y_{ij}}{\Lambda_{\text{F}}^{2}}H\bar{q}q\quad
\text{(TC),}%
\end{equation}
and are associated to some new dynamics \cite{ETC}, the flavor dynamics, at a
scale $\Lambda_{\text{F}}$, which plays the role of our $\Lambda_{\mathrm{UV}%
}$. Very much like in the SM, and as it is found in explicit models
\cite{ETC}, we also expect unwanted 4-quark interactions
\begin{equation}
\frac{c_{ijkl}}{\Lambda_{\mathrm{F}}^{2}}\bar{q}_{i}q_{j}\bar{q}_{k}q_{l}
\label{4q}%
\end{equation}
suppressed by the same flavor scale. Unlike in the SM, in technicolor the
Yukawa interactions are \textit{not} the single most relevant interaction
violating flavor. This leads to a tension. On one hand, in order to obtain the
right quark masses, $\Lambda_{\mathrm{F}}$ should be rather low. On the other
hand, the bound from FCNC requires $\Lambda_{\mathrm{F}}$ to be generically
larger. For instance the top Yukawa implies $\Lambda_{\mathrm{F}}%
\lesssim10\,\mathrm{{TeV}}$. On the other hand the bound from FCNC on
operators like Eq.~(\ref{4q}) is rather strong. Assuming $c_{ijkl}\sim1$,
flavor mixing in the neutral kaon system puts a generic bound ranging from
$\Lambda_{\mathrm{F}}>10^{3}\,\mathrm{{TeV}}$, assuming CP conservation and
left-left current structure, to $\Lambda_{\mathrm{F}}>10^{5}\,\mathrm{{TeV}}$,
with CP violation with left-right current. Of course assuming that $c_{ijkl}$
have a nontrivial structure controlled by flavor breaking selection rules one
could in principle obtain a realistic situation. It is however undeniable that
the way the SM disposes of extra unwanted sources of flavor violation is more
robust and thus preferable. The origin of the problem is the \textit{large}
dimension of the Higgs doublet field $H$. Models of walking technicolor (WTC)
\cite{Holdom} partially alleviate it. In WTC, above the weak scale the theory
is assumed to be near a non-trivial fixed point, where $H=\bar{T}T$ has a
sizeable negative anomalous dimension. WTC is an extremely clever idea, but
progress in its realization has been slowed down by the difficulty in dealing
with strongly coupled gauge theories in 4D. Most of our understanding of WTC
relies on gap equations, a truncation of the Schwinger-Dyson equations for the
$\bar{T}T$ self-energy. Although gap equations do not represent a fully
defendable approximation, they have produced some interesting results. In case
of asymptotically free gauge theory they lead to the result that $H=\bar{T}T$
can have dimension 2 at the quasi-fixed point, but not lower \cite{Georgi}. In
this case the Yukawa interactions would correspond to dimension 5 operators,
which are more relevant than the unwanted dimension 6 operators in
Eq.\ (\ref{4q}). However some tension still remains: the top Yukawa still
requires a Flavor scale below the bound from the Kaon system, so that the
absence of flavor violation, in our definition, is not robust. It is quite
possible that the bound $[H]\geq2$ obtained with the use of gap equation will
not be true in general. Of course the closer $[H]$ is to 1, the higher the
flavor scale we can tolerate to reproduce fermion masses, and the more
suppressed is the effect of Eq.\ (\ref{4q}). However if $[H]$ gets too close
to 1 we get back the SM and the hierarchy problem! More formally, a scalar
field of dimension exactly 1 in CFT is a free field and the dimension of its
composite $H^{\dagger}H$ is trivially determined to be 2, that is strongly
relevant. By continuity we therefore expect that the hierarchy problem strikes
back at some point as $[H]$ approaches 1. However the interesting remark made
by Luty and Okui \cite{luty} is that, after all, we do not really need $[H]$
extremely close to 1. For instance $[H]=1.3$ would already be good, in which
case the corresponding CFT is not weakly coupled and it could well be that
$[H^{\dagger}H]$ is significantly bigger than $2[H]$ and maybe even close to
4. The motivation of our present work is precisely to find, from prime
principles, what is the upper bound on $\Delta_{S}=[H^{\dagger}H]$ as $d=[H]$
approaches 1. In simple words this may be phrased as the question: how fast do
CFTs, or better a subsector of CFTs, become free as the dimension of a scalar
approaches 1?

In the following subsection we would like to make a more quantitative analysis
of the tension between flavor and electroweak hierarchy in a scenario where
the electroweak symmetry breaking sector sits near a fixed point between the
EW scale $\Lambda_{\mathrm{IR}}$ and some $\Lambda_{\mathrm{UV}}$ at which, or
below which, Flavor dynamics must take place.

\subsection{Quantitative analysis}

\label{sec:quant}

Let us normalize fields and couplings in the spirit of Naive Dimensional
Analysis (NDA). The Lagrangian will thus be written as
\begin{equation}
\mathcal{L}={\frac{1}{16\pi^{2}}}F(\phi,\lambda,M)\,,
\end{equation}
where fields, couplings and physical mass scales are indicated collectively.
With this normalization, Green's functions in the coordinate representation
have no factors of $\pi$, the couplings $\lambda$ are loop counting
parameters, and the mass scales $M$ correspond to physical masses (as opposed
to decay constants).

Our hypothesis is that below some UV scale $\Lambda_{\mathrm{UV}}$ the theory
splits into the elementary SM without Higgs and a strongly coupled CFT which
contains a (composite) Higgs scalar doublet operator $H$. These two sectors
are coupled to each other via weak gauging and the Yukawa interactions. As a
warmup exercise consider then the top Yukawa
\begin{equation}
{\frac{1}{16\pi^{2}}}\lambda_{t}H\bar{Q}_{L}t_{R}+\mathrm{h.c.}%
\end{equation}
and its lowest order correction to the CFT action
\begin{align}
\Delta\mathcal{L}  &  =\left(  {\frac{1}{16\pi^{2}}}\right)  ^{2}\lambda
_{t}^{2}\int d^{4}xd^{4}y\,H(x)^{\dagger}H(y)\,\bar{Q}_{L}t_{R}(x)\,\bar
{t}_{R}Q_{L}(y)\nonumber\\
&  \sim\left(  {\frac{1}{16\pi^{2}}}\right)  ^{2}\lambda_{t}^{2}\int
d^{4}xd^{4}y\,(H^{\dagger}H)(x)|x-y|^{\Delta_{S}-2d}|x-y|^{-6}\nonumber\\
&  \sim{\frac{1}{16\pi^{2}}}\int d^{4}x\,\lambda_{t}^{2}\Lambda_{\mathrm{UV}%
}^{2+2d-\Delta_{S}}(H^{\dagger}H)(x)\nonumber\\
&  \equiv{\frac{1}{16\pi^{2}}}\int d^{4}x\,[\bar{\lambda}_{t}(\Lambda
_{\mathrm{UV}})]^{2}\Lambda_{\mathrm{UV}}^{4-\Delta_{S}}(H^{\dagger}H)
\label{deformation}%
\end{align}
where we used the $H^{\dagger}\times H$ OPE\footnote{Notice that $H^{\dagger
}H$ is \textit{defined }as the scalar $SU(2)$ singlet operator of lowest
dimension in the $H^{\dagger}\times H$ OPE. Weak $SU(2)$ is assumed to be a
global symmetry of the CFT. The $SU(2)$ invariance of the fermion propagators
realizes the projection on the singlet.} and cut off the $d^{4}y$ integral at
a UV distance $\Lambda_{\mathrm{UV}}^{-1}$ (ex.: there exist new states with
mass $\Lambda_{\mathrm{UV}}$). The quantity $\bar{\lambda}_{t}(\Lambda
_{\mathrm{UV}})=\lambda_{t}\Lambda_{\mathrm{UV}}^{d-1}$ represents the
dimensionless running coupling evaluated at the scale $\Lambda_{\mathrm{UV}}$.
Given our NDA normalization, $\bar{\lambda}_{t}^{2}$ is the loop counting
parameter (no extra $\pi$'s). For $\Delta_{S}<4$ Eq.\ (\ref{deformation})
represents a relevant deformation of the CFT. Allowing for a fine tuning
$\epsilon_{t}$ between our naive estimate of the top loop and the true result,
the deformation lagrangian is
\begin{equation}
{\mathcal{L}_{\Delta}=}{\frac{1}{16\pi^{2}}}\bar{\lambda}_{t}(\Lambda
_{\mathrm{UV}})^{2}\epsilon_{t}\Lambda_{\mathrm{UV}}^{4-\Delta_{S}}H^{\dagger
}H\,, \label{topeffect}%
\end{equation}
corresponding to a physical infrared scale
\begin{equation}
\Lambda_{\mathrm{IR}}=[\bar{\lambda}_{t}(\Lambda_{\mathrm{UV}})^{2}%
\epsilon_{t}]^{\frac{1}{4-\Delta_{S}}}\Lambda_{\text{UV}}\equiv(c_{t}%
)^{\frac{1}{4-\Delta_{S}}}\Lambda_{\text{UV}}\,,
\end{equation}
where we made contact with our previous definition of the coefficient $c$. If
the above were the dominant contribution, then a hierarchy would arise for
$4-\Delta_{S}<1$ provided $\bar{\lambda}_{t}(\Lambda_{\mathrm{UV}})<1$ and/or
a mild tuning $\epsilon_{t}<1$ exists. However, unlike the normal situation
where the Higgs is weakly self-coupled and the top effects dominate, in our
scenario the Higgs is strongly self-coupled. Therefore we expect a leading
contribution to $\mathcal{L}_{\Delta}$ to already be present in the CFT
independently of the top:
\begin{equation}
\mathcal{L}_{\Delta}=\frac{c}{16\pi^{2}}\Lambda_{\mathrm{UV}}^{4-\Delta_{S}%
}H^{\dagger}H\,.
\end{equation}
The presence of such an effect basically accounts for the fact that a CFT with
a relevant deformation does not even flow to the fixed point unless the
deformation parameter is tuned. This is in line with our initial discussion.
So we shall work under the assumption that $c$ is somewhat less than 1.

We can describe the generation of the electroweak scale by writing the
effective potential for the composite operator vacuum expectation value
$\langle(H^{\dagger}H)\rangle=\mu^{\Delta_{S}}$. Compatibly with scale
invariance there will also be a term
\begin{equation}
V_{\text{CFT}}=\frac{a}{16\pi^{2}}\mu^{4}%
\end{equation}
where $a$ is a numerical coefficient that depends on the CFT and on the
direction of the VEV in operator space. The full effective potential has then
the form
\begin{equation}
V_{\text{eff}}={\frac{1}{16\pi^{2}}}\left[  -\Lambda_{\mathrm{IR}}%
^{4-\Delta_{S}}\mu^{\Delta_{S}}+\mu^{4}\right]  \,,
\end{equation}
which is stationary at $\mu\sim\Lambda_{\mathrm{IR}}$. Here we put for example
$a=1$ and chose a negative sign for the scale breaking contribution. Notice
that the vacuum dynamics picture that we just illustrated is analogous to the
Randall-Sundrum model \cite{rs} with Goldberger-Wise radius stabilization
\cite{gw} with the identification of $1/\mu$ with the position of the IR brane
in conformal coordinates. This fact should not be surprising at all given the
equivalence of that model to a deformed CFT \cite{rscft}.

Our analysis of the top sector is however useful to discuss the two basic
constraints on this scenario.

The first constraint is the request that $\Lambda_{\mathrm{UV}}$ be below the
scale where the top Yukawa becomes strong, at which point the SM becomes
strongly coupled to the CFT and our picture breaks down. The running top
coupling in its standard normalization is $y_{t}(E)=4\pi\bar{\lambda}(E)$.
Using the known experimental result $y_{t}(\Lambda_{\mathrm{IR}})\sim1$ we
thus have
\begin{equation}
y_{t}(\Lambda_{\mathrm{UV}})=y_{t}(\Lambda_{\mathrm{IR}})\left(  \frac
{\Lambda_{\mathrm{UV}}}{\Lambda_{\mathrm{IR}}}\right)  ^{d-1}\sim
(c)^{-\frac{d-1}{4-\Delta_{S}}}\,.
\end{equation}
Perturbativity corresponds to $y_{t}(\Lambda)\lesssim4\pi$, that is
\begin{equation}
\frac{d-1}{4-\Delta_{S}}\,\ln(1/c)\lesssim\ln(4\pi)\sim\ln10\,. \label{weak}%
\end{equation}
Clearly, this bound is better satisfied the closer $d$ is to 1.

A second constraint is presented by the request of robust decoupling of
unwanted flavor breaking effects. Assuming any generic interaction among SM
states is present at the scale $\Lambda_{\mathrm{UV}}$ with strength
comparable to $\bar{\lambda}_{t}^{2}(\Lambda)$, we can parameterize flavor
violation by
\begin{equation}
\mathcal{L}_{\text{fermion}}={\frac{1}{16\pi^{2}}}\left[  \bar{q}\!\!\not \!
\!Dq+\frac{\bar{\lambda}_{t}^{2}(\Lambda_{\mathrm{UV}})}{\Lambda_{\mathrm{UV}%
}^{2}}(\bar{q}q)^{2}\right]  \,, \label{4quarks}%
\end{equation}
which by going to canonical normalization becomes
\begin{equation}
\mathcal{L}_{\text{fermion}}=\bar{q}\!\!\not \!  \!Dq+\frac{1}{\Lambda
_{\mathrm{F}}^{2}}(\bar{q}q)^{2}%
\end{equation}
with
\begin{equation}
\Lambda_{\mathrm{F}}=\Lambda_{\mathrm{IR}}\left(  \frac{\Lambda_{\mathrm{UV}}%
}{\Lambda_{\mathrm{IR}}}\right)  ^{2-d}=\Lambda_{\mathrm{IR}}(c)^{-\frac
{2-d}{4-\Delta_{S}}}\,. \label{dmeno2}%
\end{equation}
By taking $\Lambda_{\mathrm{IR}}=1\,\mathrm{{TeV}}$, the bound from FCNC can
be parametrized as $\Lambda_{\mathrm{F}}>10^{F}\,\mathrm{{TeV}}$. Making the
conservative assumption that all quark families appear in Eq.\ (\ref{4quarks}%
), compatibility with the data requires $\Lambda_{\mathrm{F}}\geq10^{3}%
\div10^{4}$ TeV.\footnote{We consider the limit $\Lambda_{\text{F}}=10^{5}$
TeV really an overkill.} Thus robust suppression of FCNC corresponds to
\begin{equation}
F>3\div4.\text{\quad(robust)\thinspace.} \label{robust}%
\end{equation}
On the other hand if only the third family appears in Eq.\ (\ref{4quarks}),
the mixing effects involving the lighter generations are generally suppressed
by extra powers of the CKM angles. In that case the bound on $\Lambda
_{\mathrm{F}}$ is weaker, and $F>0.5$ is basically enough. In the latter case
the detailed structure of the Flavor theory matters. Notice that for
conventional walking technicolor models, for which $d\geq2$, we always have
$\Lambda_{\mathrm{F}}\leq\Lambda_{\mathrm{IR}}$ so that even the weaker bound
is somewhat problematic. We are however interested to see to what extent we
can neglect this issue by focussing on the robust bound (\ref{robust}). From
Eq.\ (\ref{dmeno2}) we must have%
\begin{equation}
\frac{2-d}{4-\Delta_{S}}\,\ln(1/c)\geq F\ln10\,. \label{flavor}%
\end{equation}
Eqs. (\ref{weak},\ref{flavor}) together imply
\begin{equation}
\frac{F}{2-d}<\frac{\ln(1/c)}{(4-\Delta_{S})\ln10}<\frac{1}{d-1}\qquad
\qquad\Longrightarrow\qquad\qquad d<1+\frac{1}{1+F}\,, \label{dbound}%
\end{equation}
which for the robust bound (\ref{robust}) requires $d<1.25$. At the same time
the amount of tuning needed to generate the hierarchy between $\Lambda
_{\mathrm{IR}}$ and the flavor scale $\Lambda_{\mathrm{UV}}$ is
\[
c=10^{(4-\Delta_{S})F/(d-2)}\,.
\]
A reasonable request $c>0.1$ then reads
\begin{equation}
4-\Delta_{S}<\frac{2-d}{F}\,. \label{deltadbound}%
\end{equation}
The ultimate goal of our study is thus to find a prime principle upper bound
on $\Delta_{S}$ as a function of $d$. This bound should provide an extra
important constraint which together with eqs.~({\ref{dbound},\ref{deltadbound}%
) may or may not be satisfied. Our main result (\ref{main}) is a step towards
this goal, although is not yet a complete solution. The point is that the
lowest dimension scalar in the }$\phi\times\phi$ OPE, whose dimension $\Delta$
appears in~(\ref{main}), is not necessarily a singlet. Nonetheless we think
our result already represents some interesting piece of information. We
postpone a {detailed discussion of this connection until Section
\ref{sec:connection}. }

\section{Necessary CFT techniques}

\label{CFT-review}

To make the paper self-contained, in this section we will review a few
standard CFT concepts and results, concentrating on those which are crucial
for understanding our result and its derivation. Our personal preferred list
of CFT literature includes \cite{mack-salam},\cite{ferrara},\cite{todorov}%
,\cite{fradkin},\cite{df},\cite{adscft}. We will mostly work in the Euclidean signature.

\subsection{Primary fields and unitarity bounds}

In perturbative field theories, classification of local operators is
straightforward: we have a certain number of fundamental fields, from which
the rest of the operators are obtained by applying derivatives and
multiplication. In CFTs, a similar role is played by the \textit{primary
fields}. These local operators $\mathcal{O}(x)$ are characterized by the fact
that they are annihilated by the Special Conformal Transformation generator
$K_{\mu}$ (at $x=0$). Thus a primary field $\mathcal{O}(x)$ transforms under
the little group$\equiv$the subgroup of conformal transformations leaving
$x=0$ invariant (this includes Lorentz transformations $M_{\mu\nu}$,
dilatations $D$, and Special Conformal Transformations $K_{\mu}$) as follows:%
\begin{align}
\lbrack M_{\mu\nu},\mathcal{O}(0)]  &  =\Sigma_{\mu\nu}\mathcal{O}%
(0)\,,\label{reps}\\
\lbrack D,\mathcal{O}(0)]  &  =i\Delta\,\mathcal{O}(0)\,,\nonumber\\
\lbrack K_{\mu},\mathcal{O}(0)]  &  =0\,.\nonumber
\end{align}
Here we assume that $\mathcal{O}$ has well-defined quantum numbers: the
scaling dimension $\Delta$, and the Lorentz\footnote{Or Euclidean rotation, if
one is working in the Euclidean.} spin $(j,\tilde{j})$ (the matrices
$\Sigma_{\mu\nu}$ are the corresponding generators).

Once all the primary operators are known, the rest of the field content is
obtained by applying derivatives; the fields obtained in this way are called
descendants. The multiplication operation used to generate composite operators
in perturbative field theories has a CFT analogue in the concept of the OPE,
which will be discussed in Section \ref{sec:OPE} below. To avoid any possible
confusion, we add that this picture applies equally well also to conformal
gauge theories, \textit{e.g.} to $\mathcal{N}=4$ Super Yang-Mills, provided
that only physical, gauge invariant fields are counted as operators of the theory.

Knowing (\ref{reps}), one can determine the transformation rules at any other
point $x$ using the conformal algebra commutation relations \cite{mack-salam}.
In principle, one could also imagine representations where $K_{\mu}$ acts as a
nilpotent matrix (type Ib in \cite{mack-salam}) rather than zero as in
(\ref{reps}). However, as proven in \cite{mack}, only representations of the
form (\ref{reps}) occur in unitary CFTs. Moreover, unitarity implies important
lower bounds on the operator dimensions. We are mostly interested in symmetric
traceless fields $\mathcal{O}_{(\mu)}$, $(\mu)\equiv\mu_{1}\ldots\mu_{l}$,
which correspond to $j=\tilde{j}$ tensors:%
\begin{equation}
\mathcal{O}_{\mu_{1}\ldots\mu_{l}}\equiv\sigma_{\alpha_{1}\dot{\alpha}_{1}%
}^{\mu_{1}}\ldots\sigma_{\alpha_{l}\dot{\alpha}_{l}}^{\mu_{l}}\phi^{\alpha
_{1}\ldots\alpha_{l}\dot{\alpha}_{1}\ldots\dot{\alpha}_{l}}\,.
\label{sim-traceless}%
\end{equation}
This is traceless in any pair of $\mu$ indices because $\sigma_{\alpha
\dot{\alpha}}^{\mu}\sigma_{\beta\dot{\beta}}^{\mu}\propto\varepsilon
_{\alpha\beta}\varepsilon_{\dot{\alpha}\dot{\beta}}.$ For such primaries
the\textit{ unitarity bound} reads \cite{mack}:%
\begin{align}
&  l=0\text{: }\Delta\geq1,\quad\quad\quad\Delta=1\text{ only for a free
scalar;}\nonumber\\
&  l\geq1\text{: }\Delta\geq l+2,\quad~\Delta=l+2\text{ only for a conserved
current.} \label{unitarity}%
\end{align}
Notice a relative jump of one unit when one passes from $l=0$ to $l\geq1$. In
particular, a conserved spin-1 current has $\Delta=3$, while the
energy-momentum tensor has $\Delta=4.$ The full list of unitarity bounds,
which includes also fields with $j\neq\tilde{j}$, can be found in \cite{mack}.
Recently \cite{intriligator}, some of these bounds were rederived in a very
physically transparent way, by weakly coupling a free scalar theory to the CFT
and studying the unitarity of the S-matrix generated by exchanges of CFT operators.

\subsection{Correlation functions}

\label{sec:corr}

As is well known, conformal symmetry fixes the coordinate dependence of 2- and
3-point functions of primary fields. For example, for scalar primaries we
have:%
\begin{align}
\langle\phi(x)\phi(y)\rangle &  =\frac{1}{|x-y|^{2\Delta_{\phi}}%
}\,,\label{2pt}\\
\langle\phi(x)\tilde{\phi}(y)\rangle &  =0\qquad(\phi\neq\tilde{\phi})\,.
\label{nondiag}%
\end{align}
As it is customary, we normalize $\phi$ to have a unit coefficient in the RHS
of (\ref{2pt}). Correlators of two fields with unequal dimensions vanish by
conformal symmetry. Even if several primaries of the same dimension exist, by
properly choosing the basis we can make sure that the nondiagonal correlators
(\ref{nondiag}) vanish. Notice that we are working with real fields,
corresponding to hermitean operators in the Minkowski space description of the theory.

The 3-point functions are also fixed by conformal symmetry:
\begin{align*}
\langle\phi_{1}(x_{1})\phi_{2}(x_{2})\phi_{3}(x_{3})\rangle &  =\frac
{\lambda_{123}}{|x_{12}|^{\Delta_{1}+\Delta_{2}-\Delta_{3}}|x_{23}%
|^{\Delta_{2}+\Delta_{3}-\Delta_{1}}|x_{13}|^{\Delta_{1}+\Delta_{3}-\Delta
_{2}}}\,,\\
x_{12}  &  \equiv x_{1}-x_{2}\text{ etc.}%
\end{align*}
The constants $\lambda_{123}$, which become unambiguously defined once we
normalize the fields via the 2-point functions, are an important
characteristic of CFT dynamics. These constants appear as the OPE coefficients
(see below), and if they are all known, any $n$-point function can be
reconstructed via the OPE. Thus in a sense finding these constants, together
with the spectrum of operator dimensions, is equivalent to solving, or
constructing, the theory.

Also the correlator of two scalars and a spin $l$ primary $\mathcal{O}_{(\mu
)}$ is fixed up to a constant \cite{fradkin}:%
\begin{align}
\langle\phi_{1}(x_{1})\phi_{2}(x_{2})\mathcal{O}_{(\mu)}(x_{3})\rangle &
=\frac{\lambda_{12\mathcal{O}}}{|x_{12}|^{\Delta_{1}+\Delta_{2}-\Delta
_{\mathcal{O}}+l}|x_{23}|^{\Delta_{2}+\Delta_{\mathcal{O}}-\Delta_{1}%
-l}|x_{13}|^{\Delta_{1}+\Delta_{\mathcal{O}}-\Delta_{2}-l}}Z_{\mu_{1}}\ldots
Z_{\mu_{l}}\,,\label{3ptl2}\\
Z_{\mu}  &  =\frac{x_{13}^{\mu}}{x_{13}^{2}}-\frac{x_{23}^{\mu}}{x_{23}^{2}%
}\,.\nonumber
\end{align}
The OPE coefficients $\lambda_{12\mathcal{O}}$ are real, once a real field
basis (\ref{2pt}) (and similarly for higher spin primaries) is chosen. This
reality condition follows from the reality of Minkowski-space correlators of
hermitean operators at spacelike separation, see Appendix \ref{reality} for a
more detailed discussion.

When it comes to 4-point functions, conformal symmetry is no longer sufficient
to fix the coordinate dependence completely. In the case of 4 scalar
operators, the most general conformally-symmetric expression is%
\begin{equation}
\langle\phi_{1}(x_{1})\phi_{2}(x_{2})\phi_{3}(x_{3})\phi_{4}(x_{4}%
)\rangle=\left(  \frac{|x_{24}|}{|x_{14}|}\right)  ^{\Delta_{1}-\Delta_{2}%
}\left(  \frac{|x_{14}|}{|x_{13}|}\right)  ^{\Delta_{3}-\Delta_{4}}%
\frac{g(u,v)}{|x_{12}|^{\Delta_{1}+\Delta_{2}}|x_{34}|^{\Delta_{3}+\Delta_{4}%
}}, \label{4pt}%
\end{equation}
where $g(u,v)$ is an arbitrary function of the conformally-invariant
cross-ratios:
\begin{equation}
u=\frac{x_{12}^{2}x_{34}^{2}}{x_{13}^{2}x_{24}^{2}},\quad v=\frac{x_{14}%
^{2}x_{23}^{2}}{x_{13}^{2}x_{24}^{2}}.
\end{equation}

\subsection{Operator Product Expansion}

\label{sec:OPE}

A very powerful property of CFT is the Operator Product Expansion (OPE), which
represents a product of two primary operators at finite separation as a sum of
local primaries:%
\begin{align}
\phi_{1}(x)\phi_{2}(0)  &  =\sum_{\mathcal{O}}\lambda_{12\mathcal{O}}%
[C_{(\mu)}(x)\mathcal{O}_{(\mu)}(0)+\ldots]\,,\label{OPE}\\
C_{(\mu)}(x)  &  =\frac{1}{|x|^{\Delta_{1}+\Delta_{2}-\Delta_{\mathcal{O}}}%
}\frac{x^{\mu_{1}}\cdots x^{\mu_{l}}}{|x|^{l}}\,.\nonumber
\end{align}
Here we wrote an OPE appropriate for a pair of scalars $\phi_{1}\times\phi
_{2}$.

The \ldots\ in (\ref{OPE}) stands for an infinite number of terms, less
singular in the $x\rightarrow0$ limit, involving the derivatives of the
primary $\mathcal{O}_{(\mu)}$ (i.e.~its descendants). The coordinate
dependence of the coefficients of these descendants is in fact completely
fixed by the conformal symmetry, so that $\lambda_{12\mathcal{O}}$ appears as
an overall coefficient for the full contribution of $\mathcal{O}_{(\mu)}$ and
its descendants. We can write schematically:%
\[
\phi_{1}\times\phi_{2}=\sum_{\mathcal{O}}\lambda_{12\mathcal{O}}%
\,\raisebox{-7pt}{\begin{picture}(20,22) \put(0,0){\line(1,1){10}}
\put(0,22){\line(1,-1){10}}
\put(10,10){\line(1,0){10}}\put(10,12){\line(1,0){10}}
\end{picture}}~\mathcal{O\,,}%
\]
denoting by $\mathcal{O}$ the contribution of the whole conformal family.

For example, for a scalar operator $\mathcal{O}$ appearing in the OPE
$\phi\times\phi$ the first few subleading terms are (\cite{fradkin}, p. 125,
$\Delta=\Delta_{\mathcal{O}}$)%
\begin{equation}
\phi(x)\phi(0)\sim\frac{\lambda_{12\mathcal{O}}}{|x|^{2\Delta_{\phi}-\Delta}%
}[1+\frac{1}{2}x_{\mu}\partial_{\mu}+\frac{1}{8}\frac{\Delta+2}{\Delta
+1}x_{\mu}x_{\nu}\partial_{\mu}\partial_{\nu}-\frac{1}{16}\frac{\Delta}%
{\Delta^{2}-1}x^{2}\partial^{2}+\ldots]\mathcal{O}(0) \label{few}%
\end{equation}

There are several ways to determine the precise form of the coefficients of
the descendants. One, direct, way \cite{ferrara} is to demand that the RHS of
(\ref{OPE}) transform under the conformal algebra in the same way as the known
transformation of the LHS. A second way \cite{do1} is to require that the full
OPE, with the descendant contributions included, correctly sum up to reproduce
the 3-point function (\ref{3ptl2}) not only in the limit $x_{1}\rightarrow
x_{2}$, but at finite separation as well. The last, seemingly the most
efficient way, is via the so called \textit{shadow field formalism}, which
introduces conjugate auxiliary fields of dimension $4-\Delta$ and uses them to
compute \textquotedblleft amputated" 3-point functions, which turn out to be
related to the OPE coefficient functions \cite{fradkin}.

Using the OPE, any $n$-point function $\left\langle \phi_{1}(x)\phi
_{2}(0)\ldots\right\rangle $ can be reduced to a sum of $(n-1)$-point
functions. Applying the OPE recursively, we can reduce any correlator to
3-point functions which are fixed by the symmetry. Of course, this procedure
can be carried out in full only if we already know which operators appear in
the OPE, and with which coefficients. Consistency of (\ref{OPE}) and
(\ref{3ptl2}) in the limit $x_{1}\rightarrow x_{2}$ requires that the same
constant $\lambda_{12\mathcal{O}}$ appear in both equations. Thus the sum in
(\ref{OPE}) is taken over all primaries $\mathcal{O}_{(\mu)}$ which have
non-zero correlators (\ref{3ptl2}). It is not difficult to show that the
correlator $\langle\phi_{1}\phi_{2}\mathcal{O}\rangle$ vanishes if
$\mathcal{O}$ has $j\neq\tilde{j}$,\footnote{Fields with $j\neq\tilde{j}$
correspond to antisymmetric tensors. The correlator $\left\langle \phi
_{1}(x)\phi_{2}(-x)\mathcal{O}_{j\neq\tilde{j}}(0)\right\rangle $ must vanish
for this particular spacetime configuration, since we cannot construct an
antisymmetric tensor out of $x_{\mu}$. Any other configuration can be reduced
to the previous one by a conformal transformation.} and thus such fields do
not appear in the OPE of two scalars (see e.g.\ \cite{mackOPE}, p.156).

We stress that in CFT, the OPE is \textit{not} an asymptotic expansion but is
a bona fide convergent power-series expansion\footnote{In general, it will
involve fractional powers depending on the dimensions of the entering
fields.}. The region of expected convergence can be understood using the
state-operator correspondence in the radial quantization of CFT (see
\cite{pol1}, Sections 2.8,2.9 for a lucid discussion in 2D). In this picture,
every state $|\Psi\rangle$ defined on a sphere of radius $r$ around the origin
can be expanded in a basis of states generated by local operator insertions at
the origin acting on the vacuum: $\mathcal{O}(0)|0\rangle$. For example,
consider the Euclidean 4-point function $\left\langle \phi_{1}(x)\phi
_{2}(0)\phi_{3}(x_{3})\phi_{4}(x_{4})\right\rangle $, and suppose that%
\begin{equation}
0<|x|<\min\left(  |x_{3}|,|x_{4}|\right)  \,, \label{eq:OPEconv}%
\end{equation}
so that there exists a sphere centered at the origin which contains $0,x$ but
not $x_{3},x_{4}$. Cutting the path integral along the sphere, we represent
the 4-point function as a Hilbert-space product%
\[
\left\langle \phi_{1}(x)\phi_{2}(0)\phi_{3}(x_{3})\phi_{4}(x_{4})\right\rangle
\text{\thinspace=}\left\langle \Psi_{S}|\phi_{3}(x_{3})\phi_{4}(x_{4}%
)|0\right\rangle \,,\qquad\langle\Psi_{S}|\equiv\langle0|\phi_{1}(x)\phi
_{2}(0)\,,
\]
where the radial quantization state $|\Psi_{S}\rangle$ lives on the sphere,
and can be expanded in the basis of local operator insertions at $x=0$. Thus
we expect the OPE to converge if (\ref{eq:OPEconv}) is satisfied. To quote
\cite{pol1}, the convergence of the OPE is just the usual convergence of a
complete set in quantum mechanics.
%This construction shows that the OPE applied within the 4-point function  on
%trespectintoWe can divide whatever Consider the state generated by the The
%state in the Euclidean domain. Let us apply the OPE (\ref{OPE}) inside the
%correlator. If we keep $x_{3}$,$x_{4}$ fixed and vary $x,$ the only
%singularities are at the coincident points $x=0,x_{3},x_{4}$. Thus we expect
%the OPE to converge in the punctured disk $0<|x|<\min\left(  |x_{3}%
%|,|x_{4}|\right)  $.\footnote{If we analytically continue $x$ to the Minkowski
%space, while keeping the other 3 points in the Euclidean $t=0$ section
%(similarly to how it's done in Appendix \ref{spacetime}), then the
%singularities lie on the light cones of $x=0,x_{3},x_{4}$.} Moreover, by
%analytic continuation we should be able to recover the 4-point function in the
%whole spacetime.
See also \cite{luscherOPE},\cite{mackOPE} for rigorous proofs of OPE
convergence in 2D and 4D CFT, based on the same basic idea.

The concept of OPE is also applicable in theories with broken scale
invariance, e.g.~in asymptotically free perturbative field theories, such as
QCD, which are well defined in the UV. These theories can be viewed as a CFT
with a relevant deformation associated to a scale $\Lambda_{\text{QCD}}$. The
question of OPE\ convergence in this case is more subtle. In perturbation
theory, the OPE provides an \textit{asymptotic expansion} of correlation
functions in the $x\rightarrow0$ limit \cite{WZ}, and is unlikely to be
convergent because of non-perturbative ambiguities associated with the
renormalons and the choice of the operator basis. It has however been
conjectured in \cite{MackDuality} that full non-perturbative correlators
should satisfy a convergent OPE also in theories with broken scale invariance.
This presumably includes QCD, but so far\ has been proved, by direct
inspection, only for free massive scalar (see \cite{schroer} and Ref.~[6] of
\cite{MackDuality}).

\subsection{Conformal blocks}

\label{sec:blocks}

As we mentioned in Section \ref{sec:corr}, conformal invariance implies that a
scalar 4-point function must have the form (\ref{4pt}), where $g(u,v)$ is an
arbitrary function of the cross-ratios. Further information about $g(u,v)$ can
be extracted using the OPE. Namely, if we apply the OPE to the LHS of
(\ref{4pt}) both in $12$ and in $34$ channel, we can represent the 4-point
function as a sum over primary operators which appear in both OPEs:%
\begin{align}
\langle\phi_{1}\phi_{2}\phi_{3}\phi_{4}\rangle &  =\sum_{\mathcal{O}}%
\lambda_{12\mathcal{O}}\lambda_{34\mathcal{O}}\,\text{\texttt{CB}%
}_{\mathcal{O}}\,,\label{CB}\\
\text{\texttt{CB}}_{\mathcal{O}}  &
=\raisebox{-7pt}{\begin{picture}(35,22) \put(0,0){\line(1,1){10}}
\put(0,22){\line(1,-1){10}} \put(10,10){\line(1,0){15}}
\put(10,12){\line(1,0){15}} \put(25,12){\line(1,1){10}}
\put(25,10){\line(1,-1){10}}
\end{picture}}\,.
\end{align}
The nondiagonal terms do not contribute to this equation because the 2-point
functions of nonidentical primaries $\mathcal{O}\neq\mathcal{O}^{\prime}$
vanish, and so do 2-point functions of any two operators belonging to
different conformal families. The functions \texttt{CB}$_{\mathcal{O}}$, which
receive contributions from 2-point functions of the operator $\mathcal{O}$ and
its descendants, are called \textit{conformal blocks}. Conformal invariance of
the OPE implies that the conformal blocks transform under the conformal group
in the same way as $\langle\phi_{1}\phi_{2}\phi_{3}\phi_{4}\rangle.$ Thus they
can be written in the form of the RHS of (\ref{4pt}), with an appropriate
function $g_{\mathcal{O}}(u,v).$ In terms of these functions, (\ref{CB}) can
be rewritten as
\begin{equation}
g(u,v)=\sum_{\mathcal{O}}\lambda_{12\mathcal{O}}\lambda_{34\mathcal{O}%
}\,g_{\mathcal{O}}(u,v). \label{eq:g=sum}%
\end{equation}

In general, functions $g_{\mathcal{O}}(u,v)$ depend on the spin $l$ and
dimension $\Delta$ of the operator $\mathcal{O}$, as well as on the dimensions
$\Delta_{i}=[\phi_{i}].$ Various power-series representations for these
functions were known since the 70's, but it seems that simple closed-form
expressions were obtained only recently by Dolan and Osborn \cite{do1}%
,\cite{do2}. In what follows we will heavily use their result, in the
particular case when all $\Delta_{i}$ are equal. In this case it takes the
form independent of $\Delta_{i}$:$\footnote{This happens because for equal
$\Delta_{i}$ the coefficients of conformal descendants in the OPE are
determined only by $\Delta$ and $l$, see e.g.~Eq.~(\ref{few}).}$%
\begin{align}
g_{\mathcal{O}}(u,v)  &  \equiv g_{\Delta,l}(u,v)=\frac{(-)^{l}}{2^{l}}%
\frac{z\bar{z}}{z-\bar{z}}\left[  \,k_{\Delta+l}(z)k_{\Delta-l-2}(\bar
{z})-(z\leftrightarrow\bar{z})\right]  ,\nonumber\\
k_{\beta}(x)  &  \equiv x^{\beta/2}{}_{2}F_{1}\left(  \beta/2,\beta
/2,\beta;x\right)  , \label{DO}%
\end{align}
where the variables $z,\bar{z}$ are related to $u,v$ via%
\begin{equation}
u=z\bar{z},\quad v=(1-z)(1-\bar{z}), \label{uvzzbar}%
\end{equation}
or, equivalently\footnote{Notice that the RHS of (\ref{DO}) is invariant under
$z\leftrightarrow\bar{z}$.},%
\[
z,\bar{z}=\frac{1}{2}\left(  u-v+1\pm\sqrt{(u-v+1)^{2}-4u}\right)  .
\]

\begin{figure}
[h]
\begin{center}
\includegraphics[
height=2.1024in,
width=2.7968in
]%
{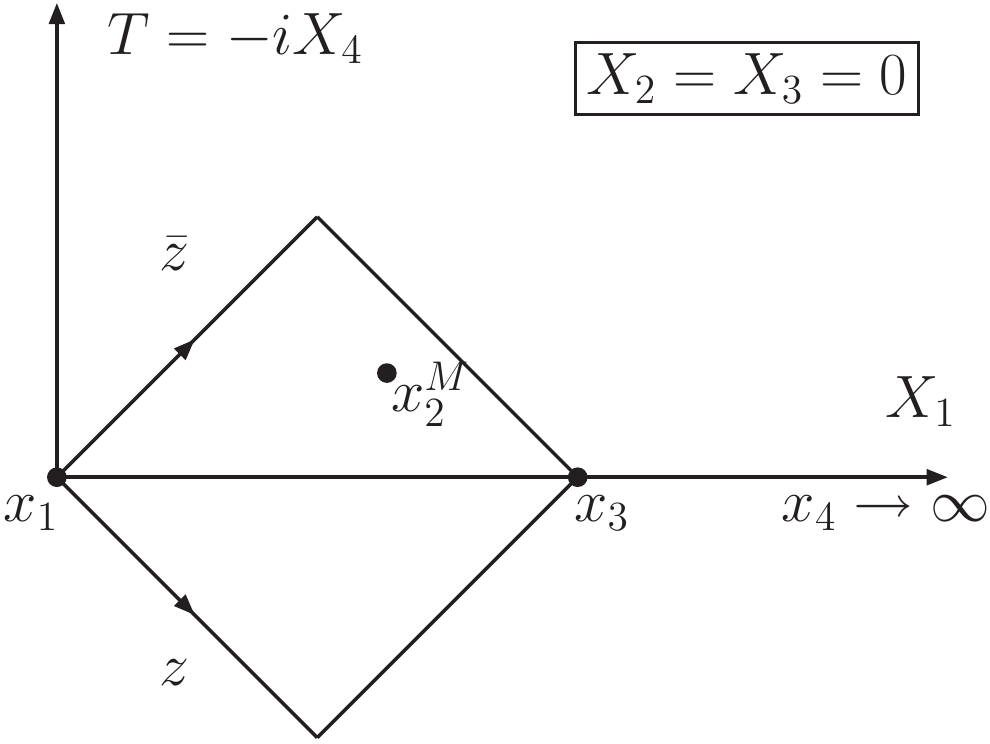}%
\caption{The \textquotedblleft spacelike diamond" (\ref{eq:diamond}) in which
the conformal blocks are real and regular, see the text.}%
\label{fig:Mink}%
\end{center}
\end{figure}

We will give a brief review of the derivation of Eq.\ (\ref{DO}) in
Appendix~\ref{lightening}. A short comment is here in order about the meaning
and range of $z$ and $\bar{z}$ (see Appendix~\ref{spacetime} for a more
detailed discussion). With points $x_{i}$ varying in the 4D Euclidean space,
these variables are complex conjugates of each other: $\bar{z}=z^{\ast}$.
Configurations corresponding to real $z=\bar{z}$ can be characterized as
having all 4 points lie on a planar circle. Below we will find it convenient
to analytically continue to the Minkowski signature, where $z$ and $\bar{z}$
can be treated as independent real variables. One possible spacetime
configuration which realizes this situation (the others being related by a
conformal transformation) is shown in Fig.~\ref{fig:Mink}. Here we put 3
points along a line in the $T=0$ Euclidean section:%
\begin{equation}
x_{1}=(0,0,0,0),\quad x_{3}=(1,0,0,0),\quad x_{4}=\infty\,, \label{eq:3out4}%
\end{equation}
while the 4-th point has been analitically continued to the Minkowski space:%
\begin{equation}
x_{2}\rightarrow x_{2}^{M}=(X_{1},0,0,T),\quad T=-iX_{4}. \label{eq:x2M}%
\end{equation}
One shows (see Appendix~\ref{spacetime}) that in this situation%
\begin{equation}
z=X_{1}-T,\quad\bar{z}=X_{1}+T\,. \label{eq:zzbarM}%
\end{equation}
We do not expect any singularities if $x_{2}^{M}$ stays inside the
\textquotedblleft spacelike diamond" region
\begin{equation}
0<z,\bar{z}<1\,, \label{eq:diamond}%
\end{equation}
formed by the boundaries of the past and future lightcones of $x_{1}$ and
$x_{3}$. Indeed, one can check that the conformal blocks are real smooth
functions in the spacelike diamond.%
%TCIMACRO{\FRAME{fhFU}{2.7968in}{2.1024in}{0pt}{\Qcb{The \textquotedblleft
%spacelike diamond" (\ref{eq:diamond}) in which the conformal blocks are real
%and regular, see the text.}}{\Qlb{fig:Mink}}{ct-fig1.pdf}%
%width 2.7968in;  height 2.1024in;  depth 0pt;  original-width 3.8303in;
%original-height 2.9525in;  cropleft "0";  croptop "1";  cropright "1";
%cropbottom "0";  filename 'ct-fig1.pdf';file-properties "XNPEU";}}}%
%BeginExpansion
%EndExpansion

\section{Crossing symmetry and the sum rule}

\label{sum-rule}

Let us consider the 4-point function (\ref{4pt}) with all 4 operators
identical: $\phi_{i}\equiv\phi$. We have:%
\begin{equation}
\langle\phi(x_{1})\phi(x_{2})\phi(x_{3})\phi(x_{4})\rangle=\frac
{g(u,v)}{|x_{12}|^{2d}|x_{34}|^{2d}},\quad d=[\phi].
\end{equation}
The LHS of this equation is invariant under the interchange of any two $x_{i}%
$, and so the RHS should also be invariant, which gives a set of
\textit{crossing symmetry }constraints for the function\textit{ }$g(u,v)$.
Invariance under $x_{1}\leftrightarrow x_{2}$ and $x_{1}\leftrightarrow x_{3}$
(other permutations do not give additional information) implies:
\begin{align}
&  g(u,v)=g(u/v,1/v)\quad\quad~~(x_{1}\leftrightarrow x_{2}),\label{cross1}\\
&  v^{d}g(u,v)=u^{d}g(v,u)\,\qquad~~(x_{1}\leftrightarrow x_{3}).
\label{cross2}%
\end{align}

At the same time, $g(u,v)$ can be expressed via the conformal block
decomposition (\ref{eq:g=sum}), which in the considered case takes the form:%
\begin{equation}
g(u,v)=1+\sum_{\mathcal{O}\in\phi\times\phi}\lambda_{\mathcal{O}}%
^{2}\,g_{\mathcal{O}}(u,v)\text{\thinspace.} \label{eq:dec}%
\end{equation}
Here in the first term we explicitly separated the contribution of the unit
operator, present in the $\phi\times\phi$ OPE. Since $\lambda_{\mathcal{O}}$
are real (see Section \ref{sec:corr}), all conformal blocks appear in
(\ref{eq:dec}) with positive coefficients.

Let us now see under which conditions Eq.~(\ref{eq:dec}) is consistent with
the crossing symmetry. The $x_{1}\leftrightarrow x_{2}$ invariance turns out
to be rather trivial. Transformation properties of any conformal block under
this crossing depend only on its spin \cite{do1}:%
\[
g_{\Delta,l}(u,v)=(-)^{l}g_{\Delta,l}(u/v,1/v).
\]
All the operators appearing in the OPE $\phi\times\phi$ have even
spin\footnote{A formal proof of this fact can be given by considering the
3-point function $\left\langle \phi(x)\phi(-x)\mathcal{O}_{(\mu)}%
(0)\right\rangle .$ By $x\rightarrow-x$ invariance, nonzero value of this
correlator is consistent with Eq.\ (\ref{3ptl2}) only if $l$ is even.}. Thus
the first crossing constraint (\ref{cross1}) will be automatically satisfied
for arbitrary coefficients $\lambda_{\mathcal{O}}^{2}$.

On the other hand, we do get a nontrivial condition by imposing that
(\ref{eq:dec}) satisfy the second crossing symmetry (\ref{cross2}). This
condition can be conveniently written in the form of the following \textit{sum
rule}:%

\begin{equation}
\boxed{ \begin{array}[l]{l} \displaystyle 1=\sum_{\Delta,l}p_{\Delta,l}F_{d,\Delta,l}(z,\bar{z}),\quad p_{\Delta ,l}>0\,,\\ \displaystyle F_{d,\Delta,l}(z,\bar{z})\equiv\frac{v^{d}g_{\Delta,l}(u,v)-u^{d}g_{\Delta,l}(v,u)\,}{u^{d}-v^{d}}, \end{array}}
\label{sumrule}%
\end{equation}
where the sum is taken over all $\Delta,l$ corresponding to the operators
$\mathcal{O}\in\phi\times\phi$, $p_{\Delta,l}=\lambda_{\mathcal{O}}^{2}$, and
$u,v$ are expressed via $z,\bar{z}$ via (\ref{uvzzbar}). As we will see below,
this sum rule contains a great deal of information. It will play a crucial
role in the derivation of our bound on the scalar operator dimensions.

Below we will always apply Eq.~(\ref{sumrule}) in the spacelike diamond
$0<z,\bar{z}<1$, see Section \ref{sec:blocks}$.$ We will find it convenient to
use the coordinates $a,b$ vanishing at the center of the diamond:%
\[
z=\frac{1}{2}+a+b,\quad\bar{z}=\frac{1}{2}+a-b.
\]
The sum rule functions $F_{d,\Delta,l}$ in this diamond:

\begin{enumerate}
\item are smooth;

\item are even in both $a$ and $b$, independently:%
\begin{equation}
F_{d,\Delta,l}(\pm a,\pm b)=F_{d,\Delta,l}(a,b)\,\text{;} \label{inv}%
\end{equation}

\item vanish on its boundary:%
\begin{equation}
F_{d,\Delta,l}(\pm1/2,b)=F_{d,\Delta,l}(a,\pm1/2)=0. \label{eq:vanish}%
\end{equation}

\end{enumerate}

Properties 1,2 are shown in Appendix~\ref{as}. Property 3 trivially follows
from the definition of $F_{d,\Delta,l}$, since both terms in the numerator
contain factors $z\bar{z}(1-z)(1-\bar{z})$.

A consequence of Property 3 is that the sum rule can never be satisfied with
finitely many terms in the RHS.

\subsection{The sum rule in the free scalar theory}

\label{sec:free}

To get an idea about what one can expect from the sum rule, we will
demonstrate how it is satisfied in the free scalar theory. In this case $d=1$,
and only operators of twist $\Delta-l=2$ are present in the OPE $\phi
\times\phi$ \cite{schroer},\cite{do1}. These are the operators%
\begin{equation}
\mathcal{O}_{\Delta,l}\propto\,\phi\,\partial_{\mu_{1}}\ldots\partial_{\mu
_{l}}\phi+\ldots\qquad(\Delta=l+2,\ l=0,2,4,\ldots). \label{primary-free}%
\end{equation}
The first term shown in (\ref{primary-free}) is traceless by $\phi$'s equation
of motion, but it is not conserved. The extra bilinear in $\phi$ terms denoted
by \ldots\ make the operator conserved for $l>0$ (in accord with the unitarity
bounds (\ref{unitarity})), without disturbing the tracelessness. Their exact
form can be found e.g. in \cite{Mikhailov}.

In particular, there is of course the dimension 2 scalar%
\[
\mathcal{O}_{2,0}=\frac{1}{\sqrt{2}}\phi^{2}\,,\,
\]
where the constant factor is needed for the proper normalization. At spin 2 we
have the energy-momentum tensor:%
\[
\mathcal{O}_{4,2}\propto\phi\partial_{\mu}\partial_{\nu}\phi-2\left[
\partial_{\mu}\phi\partial_{\nu}\phi-\frac{1}{4}\delta_{\mu\nu}(\partial
\phi)^{2}\right]  \,.\,
\]
The operators with $l>2$ are the conserved higher spin currents of the free
scalar theory.

The OPE coefficients of all these operators (or rather their squares) can be
found by decomposing the free scalar 4-point function into the corresponding
conformal blocks, Eq.\ (\ref{eq:dec}). We have \cite{do1},\cite{schroer}:%
\begin{equation}
p_{l+2,l}=2^{l+1}\frac{(l!)^{2}}{(2l)!}\qquad(l=2n)\,. \label{coeff}%
\end{equation}

\begin{figure}
[h]
\begin{center}
\includegraphics[
height=2.0055in,
width=2.9421in
]%
{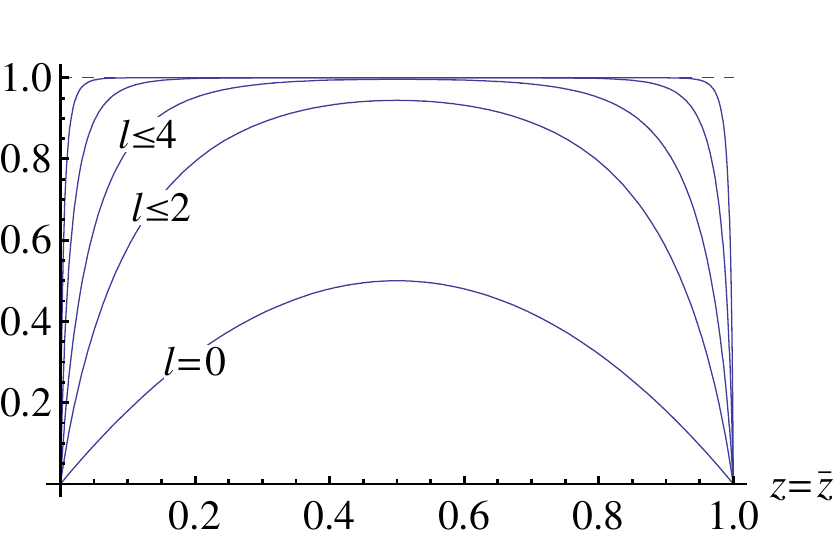}%
\caption{The RHS of the sum rule in the free scalar theory, summed over
$l\leq0,2,4,8,16$ (from below up) and\ plotted for $0\leq z=\bar{z}\leq1.$ The
asymptotic approach to $1$ (dashed line) is evident. Notice the symmetry with
respect to $z=1/2$, a consequence of (\ref{inv}).}%
\label{free}%
\end{center}
\end{figure}

Using these coefficients, we show in Fig.~\ref{free} how the sum rule
(\ref{sumrule}), summed over the first few terms, converges on the diagonal
$z=\bar{z}$ of the spacelike diamond. Several facts are worth noticing. First,
notice that the convergence is monotonic, i.e.~all $F_{d,\Delta,l}$ entering
the infinite series are positive. This feature is not limited to the free
scalar case and remains true for a wide range of $d$, $\Delta,$ $l$; it could
be used to limit the maximal size of allowed OPE coefficients (see footnote
\ref{noteinhom}).

Second, the convergence is uniform on any subinterval $z\in\lbrack
\varepsilon,1-\varepsilon]$, $\varepsilon>0$, but not on the full interval
$[0,1]$, because all the sum rule functions vanish at its ends, see
Eq.~(\ref{eq:vanish}). Finally, the convergence is fastest near the middle
point $z=1/2$, corresponding to the center $a=b=0$ of the spacelike diamond.
Below, when we apply the sum rule to the general case $d>1$, we will focus our
attention on a neighborhood of this point.
%TCIMACRO{\FRAME{fhFU}{2.9421in}{2.0055in}{0pt}{\Qcb{The RHS of the sum rule in
%the free scalar theory, summed over $l\leq0,2,4,8,16$ (from below up)
%and\ plotted for $0\leq z=\bar{z}\leq1.$ The asymptotic approach to $1$
%(dashed line) is evident. Notice the symmetry with respect to $z=1/2$, a
%consequence of (\ref{inv}).}}{\Qlb{free}}{free.pdf}%
%width 2.9421in;  height 2.0055in;  depth 0pt;  original-width 4.1632in;
%original-height 2.8245in;  cropleft "0";  croptop "1";  cropright "1";
%cropbottom "0";  filename 'free.pdf';file-properties "XNPEU";}} }%
%BeginExpansion
%EndExpansion

\section{Main results}

\label{results}

In this section we will present a derivation of the bound (\ref{main}), based
on the sum rule (\ref{sumrule}). We assume that we are given a unitary CFT
with a primary scalar operator $\phi$ of dimension $d>1.$ We consider the
4-point function $\left\langle \phi\phi\phi\phi\right\rangle $ and derive the
sum rule (\ref{sumrule}), where the sum is over all primary operators
appearing in the OPE $\phi\times\phi$. We will use only the most general
information about these operators, such as\footnote{The energy-momentum tensor
$T_{\mu\nu}$, which is a spin-2 primary of dimension 4, has to appear in the
OPE, with a known coefficient \cite{do1} $p_{4,2}=4d/(3\sqrt{c_{T}})$
depending on the central charge $c_{T}$ of the theory. However, we are not
making any assumptions about the central charge and will not take this
constraint into account. It may be worth incorporating such a constraint in
the future, since it could make the bound stronger. From the point of view of
phenomenology, estimates of the electroweak S-parameter prefer models with
small number of degree of freedom, hence small $c_{T}$.}:

\begin{enumerate}
\item only the operators satisfying the unitarity bounds (\ref{unitarity}) may appear;

\item their spins $l$ are even;

\item all the coefficients $p_{\Delta,l}$ are non-negative.
\end{enumerate}

We will prove the bound (\ref{main}) by contradiction. Namely, we will show
that if only scalar operators of dimension $\Delta>f(d)$ are allowed to appear
in the OPE, the sum rule cannot be satisfied no matter what are the
dimensions, spins, and OPE coefficients of all the other operators (as long as
they satisfy the above assumptions 1,2,3). Thus such a CFT cannot exist! In
the process of proving this, we will also derive the value of $f(d)$.

\subsection{Why is the bound at all possible?}

\label{sec:why}

Let us begin with a very simple example which should convince the reader that
some sort of bound should be possible, at least for $d$ sufficiently close to
$1$.

\begin{figure}[ptb]
\begin{center}
\includegraphics[
height=2.2459in, width=3.3416in ]{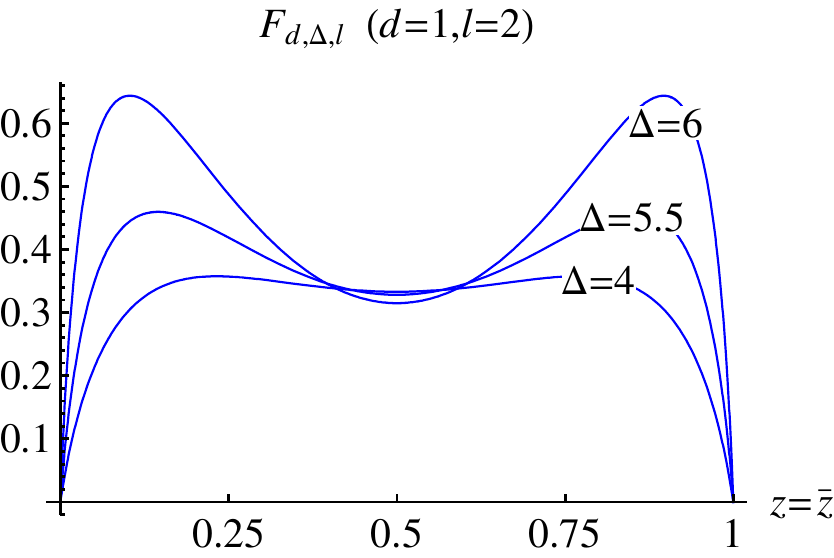}\quad\includegraphics[
height=2.2459in, width=3.3416in ]{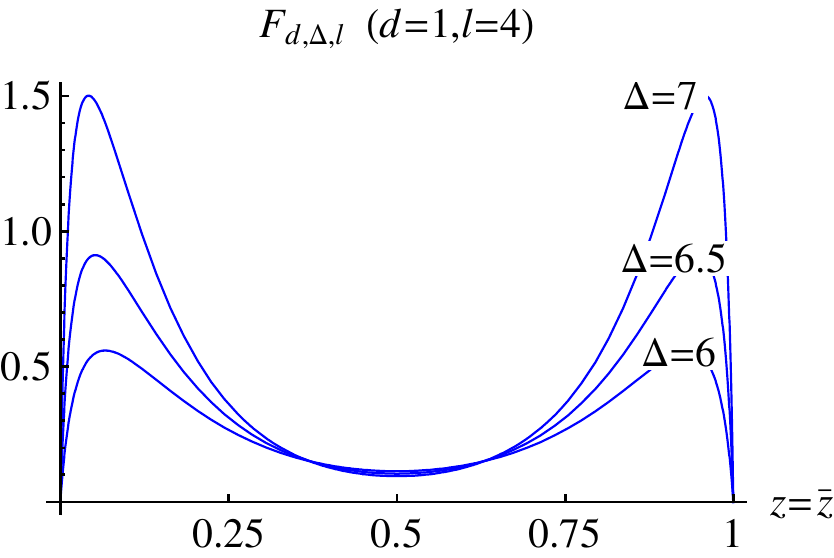}
\end{center}
\caption{The shape of $F_{d,\Delta,l}$ for $d=1$, $l=2,4$ and several values
of $\Delta$ satisfying the unitarity bound.}%
\label{fig:expl}%
\end{figure}

The argument involves some numerical exploration of functions $F_{d,\Delta,l}$
entering the sum rule (\ref{sumrule}), easily done e.g. with
\textsc{Mathematica}. These functions depend on two variables $z,\bar{z}$, but
for now it will be enough to explore the case $0<z=\bar{z}<1$, which
corresponds to the point $x_{2}$ lying on the diagonal $x_{1}-x_{3}$ of the
spacelike diamond in Fig.~\ref{fig:Mink}. We begin by making a series of plots
of $F_{d,\Delta,l}$ for $l=2,4$ and for $\Delta$ satisfying the unitarity
bound $\Delta\geq l+2$ appropriate for these spins (Fig. \ref{fig:expl}). The
scalar case $l=0$ will be considered below. We take $d=1$ in these plots.

What we see is that all these functions have a rather similar shape: they
start off growing monotonically as $z$ deviates from the symmetric point
$z=1/2,$ and after a while decrease sharply as $z\rightarrow0,1$. These
charecteristics become more pronounced as we increase $l$ and/or $\Delta$. We
invite the reader to check that, for $d=1,$ these properties are in fact true
for all $F_{d,\Delta,l}$ for even $l\geq2$ and $\Delta\geq l+2.$ By
continuity, they are also true for $d=1+\varepsilon$ as long as $\varepsilon
>0$ is sufficiently small.\footnote{One can check that they are true up to
$d\simeq1.12$. For larger $d$, $F_{d,4,2}^{\prime\prime}(z=1/2)$ becomes
negative.} Mathematically, we can express the fact that $F_{d,\Delta,l}$ is
downward convex near $z=1/2$ as:%
\begin{align}
&  F_{d,\Delta,l}^{\prime\prime}>0\text{\quad at }z=\bar{z}=1/2,\nonumber\\
&  l=2,4,6\ldots,\quad\Delta\geq l+2,\label{eq:l>=2}\\
&  1\leq d\leq1+\varepsilon\,.\nonumber
\end{align}

Even before addressing the existence of the bound, let us now ask and answer
the following more elementary question: could a CFT without any scalars in the
OPE\ $\phi\times\phi$ exist? Eq.~(\ref{eq:l>=2}) immediately implies that the
answer is \textit{NO}, at least if $d$ is sufficiently close to $1$.

The proof is by contradiction: in such a CFT, the sum rule (\ref{sumrule})
would have to be satisfied with only $l\geq2$ terms present in the RHS.
Applying the second derivative to the both sides of (\ref{sumrule}) and
evaluating at $z=\bar{z}=1/2$, the LHS is identically zero, while in the RHS,
by (\ref{eq:l>=2}), we have a sum of positive terms with positive
coefficients. This is a clear contradiction, and thus such a CFT does not exist.

\begin{figure}
[ptb]
\begin{center}
\includegraphics[
height=2.2459in,
width=3.3416in
]%
{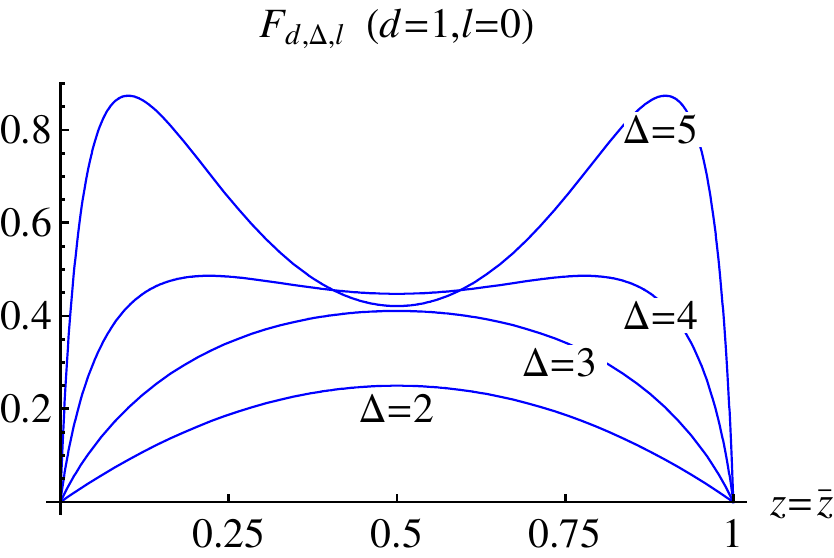}%
\caption{Same as Fig.~\ref{fig:expl}, for $l=0.$}%
\label{fig:expl0}%
\end{center}
\end{figure}

To rephrase what we have just seen, the sum rule must contain some terms with
negative $F^{\prime\prime}(z=1/2)$ to have a chance to be satisfied, and by
(\ref{eq:l>=2}) such terms can come only from $l=0.$ Thus, the next natural
step is to check the shape of $F_{d,\Delta,l}$ for $l=0$, which we plot for
several $\Delta\geq2$ in Fig.~\ref{fig:expl0}. We see that the second
derivative in question is negative at $\Delta=2$ (it better be since this
corresponds to the free scalar theory which surely exists!). By continuity, it
is also negative for $\Delta$ near $2.$ However, and this is crucial, it turns
\textit{positive} for $\Delta$ above certain critical dimension $\Delta_{c}$
between $3$ and $4$. It is not difficult to check that in fact $\Delta
_{c}\simeq3.61$ for $d$ near $1$.%

%TCIMACRO{\FRAME{ftbpFU}{3.3416in}{2.2459in}{0pt}{\Qcb{Same as
%Fig.~\ref{fig:expl}, for $l=0.$}}{\Qlb{fig:expl0}}{expl0.pdf}%
%width 3.3416in;  height 2.2459in;  depth 0pt;  original-width 3.333in;
%original-height 2.2312in;  cropleft "0";  croptop "1";  cropright "1";
%cropbottom "0";  filename 'expl0.pdf';file-properties "XNPEU";}} }%
%BeginExpansion
%EndExpansion

We arrive at our main conclusion: not only do some scalars have to be present
in the OPE, but at least one of them should have $\Delta\leq\Delta_{c}$!
Otherwise such a CFT will be ruled out by the same argument as a CFT without
any scalars in the OPE. In other words, we have just established the bound
$\Delta_{\min}\leq\Delta_{c}$ for $d$ near $1$.

Admittedly, this first result is extremely crude: for instance, the obtained
bound does not approach $2$ as $d\rightarrow1.$ However, what is important is
that it already contains the main idea of the method which will be developed
and used with increasing refinement below. This idea is that we have to look
for a differential operator which gives zero acting on the unit function in
the LHS of the sum rule, but stays positive when applied to the functions
$F_{d,\Delta,l}$ in the RHS.

Now, some readers may find it unappealing that the method as we presented it
above seems to be heavily dependent on the numerical evaluation of functions
$F_{d,\Delta,l}$ and their derivatives. Do we have an \textit{analytical}
proof establishing e.g.~the properties (\ref{eq:l>=2})? -- a purist of
mathematical rigor may ask.

Partly, the answer is yes, since the asymptotic behavior of $F_{d,\Delta,l}$
for large $\Delta$ and/or $l$ is easily accessible to analytical means (see
Appendix \ref{as}). These asymptotics establish Eq.~(\ref{eq:l>=2}) in the
corresponding limit. On the other hand, we do not have an analytic proof of
Eq.~(\ref{eq:l>=2}) for finite values of $\Delta$ and $l$. Notice that such a
proof must involve controlling hypergeometric functions near $z=1/2$. No
simple general expansions of hypergeometrics exist near this point (apart from
the one equivalent to summing up the full series around $z=0$). Thus we doubt
that a simple proof exists.

Nevertheless, and we would like to stress this, the fact that we can establish
Eq.~(\ref{eq:l>=2}) only numerically (with analytic control of the asymptotic
limits) does not make it less mathematically true! The situation can be
compared to proving the inequality $e<\pi$. An aesthete may look for a fully
analytical proof, but a practically minded person will just evaluate both
constants by computer. As long as the numerical precision of the evaluation is
high enough, the practical proof is no worse than the aesthete's (and much faster).

To summarize, we should be content that numerical methods allow us to extract
from general CFT properties (crossing, unitarity bounds, conformal block
decomposition, \ldots) precious information about operator dimensions which
would otherwise simply not be available.

\subsection{Geometry of the sum rule}

\label{geometry}

To proceed, it is helpful to develop geometric understanding of the sum rule.
Given $d$ and a spectrum $\left\{  \Delta,l\right\}  $ of $\mathcal{O}\in
\phi\times\phi$, and allowing for arbitrary positive coefficients
$p_{\Delta,l}$, the linear combinations in the RHS of (\ref{sumrule}) form, in
the language of functional analysis, a \textit{convex cone }$\mathcal{C}$
in\textit{ }the function space $\left\{  F(a,b)\right\}  $. For a fixed
spectrum, the sum rule can be satisfied for \textit{some} choice of the
coefficients if and only if the unit function $F(a,b)\equiv1$ belongs to this cone.

Obviously, when we expand the spectrum by allowing more operators to appear in
the OPE, the cone gets wider. Let us consider a one-parameter family of
spectra:%
\begin{equation}
\Sigma(\Delta_{\text{min}})=\left\{  \Delta,l~~|~~\Delta\geq\Delta
_{\text{min}}\,(l=0),\quad\Delta\geq l+2~(l=2,4,6\ldots)\right\}  \,.
\label{eq:family}%
\end{equation}
Thus we include all scalars of dimension $\Delta\geq\Delta_{\text{min}}$, and
all higher even spin primaries allowed by the unitarity bounds.

The crucial fact which makes the bound (\ref{main}) possible is that in the
limit $\Delta_{\text{min}}\rightarrow\infty$ the convex cone generated by the
above spectrum does \textit{not} contain the function $F\equiv1$! In other
words, CFTs without any scalars in the OPE $\phi\times\phi$ cannot exist, as
we already demonstrated in Section \ref{sec:why} for $d$ sufficiently close to
$1$.

\begin{figure}[ptb]
\centering\begin{minipage}[c]{0.3\textwidth}
\centering\includegraphics[height=1in]{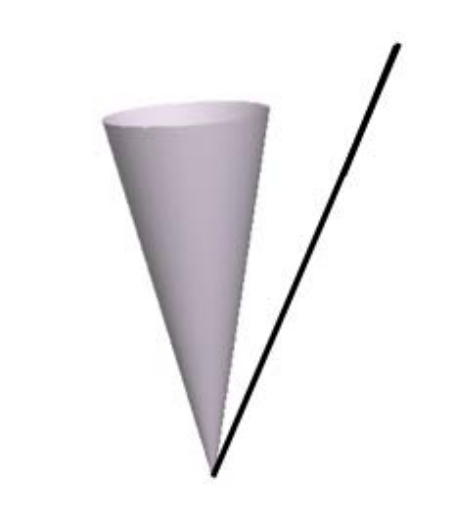}\\
$\Delta_{\rm min}>\Delta_c$\\
sum rule violated
\end{minipage}
\begin{minipage}[c]{0.3\textwidth}
\centering\includegraphics[height=1in]{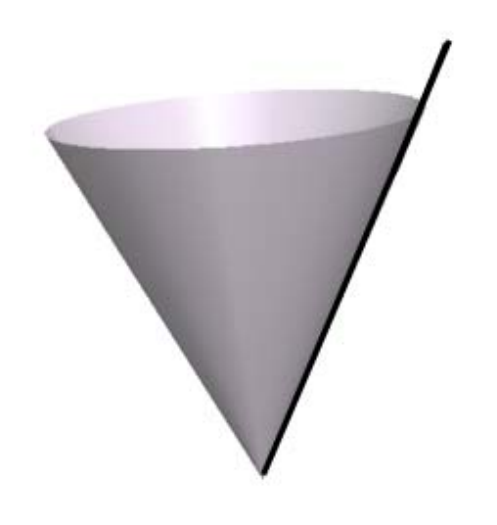} \\
$\Delta_{\rm min}=\Delta_c$
\\ critical case
\end{minipage}
\begin{minipage}[c]{0.3\textwidth}
\centering\includegraphics[height=1in]{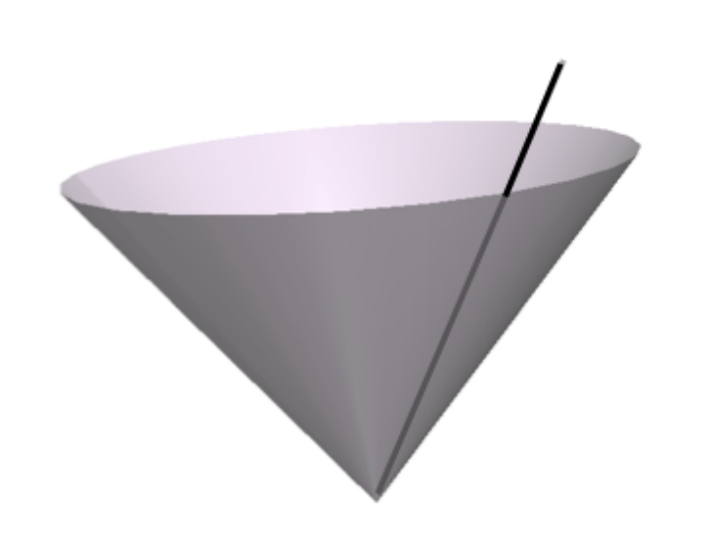}\\
$\Delta_{\rm min}<\Delta_c$\\
sum rule satisfied
\end{minipage}
\caption{The three geometric situations described in the text. The thick black
line denotes the vector corresponding to the function $F\equiv1$.}%
\label{fig:3cones}%
\end{figure}

As we lower $\Delta_{\text{min}}$, the spectrum expands, and the cone gets
wider. There exists a critical value $\Delta_{c}$ such that for $\Delta
_{\text{min}}>\Delta_{c}$ the cone is not yet wide enough and the function
$F\equiv1$ is still outside, while for $\Delta_{\text{min}}<\Delta_{c}$ the
$F\equiv1$ function is inside the cone. For $\Delta_{\text{min}}=\Delta_{c}$
the function belongs to the cone boundary. This geometric picture is
illustrated in Fig.\ \ref{fig:3cones}.

For $\Delta_{\text{min}}>\Delta_{c}$, the sum rule cannot be satisfied, and a
CFT corresponding to the spectrum $\Sigma(\Delta_{\text{min}})$ (or any
smaller spectrum) cannot exist. By contradiction, the bound (\ref{main}) with
$f(d)=\Delta_{c}\,$must be true in any CFT. The problem thus reduces to
determining $\Delta_{c}$.

Any concrete calculation must introduce a coordinate parametrization of the
above function space. We will parametrize the functions by an infinite vector
$\left\{  F^{(2m,2n)}\right\}  $ of even-order mixed derivatives at $a=b=0$:%
\begin{equation}
F^{(2m,2n)}\equiv\partial_{a}^{2m}\partial_{b}^{2n}F(a,b)\Bigl|_{a=b=0}\,.
\label{eq:dercor}%
\end{equation}
Notice that all the odd-order derivatives of the functions entering the sum
rule vanish at this point due to the symmetry expressed by Eq.~(\ref{inv}):%
\[
F^{(2m+1,2n)}=F^{(2m,2n+1)}=F^{(2m+1,2n+1)}=0\,.
\]
The choice of the $a=b=0$ point is suggested by this symmetry, and by the fact
that it is near this point that the sum rule seems to converge the fastest, at
least in the free scalar case, see Fig.~\ref{free}.

The derivatives (\ref{eq:dercor}) are relatively fast to evaluate numerically.
Presumably, there is also no loss of generality in choosing these coordinates
on the function space, since the functions entering the sum rule are analytic
in the spacelike diamond.

In terms of the introduced coordinates, the sum rule becomes a sequence of
linear equations for the coefficients $p_{\Delta,l}\geq0$. We have one
inhomogeneous equation:%
\begin{equation}
1=\sum p_{\Delta,l}F_{d,\Delta,l}^{(0,0)}\,, \label{eq:inhom}%
\end{equation}
and an infinite number of homogeneous ones:%
\begin{align}
0=  &  \sum p_{\Delta,l}F_{d,\Delta,l}^{(2,0)}\,,\nonumber\\
0=  &  \sum p_{\Delta,l}F_{d,\Delta,l}^{(0,2)}\,,\label{eq:hom}\\
&  \cdots\nonumber
\end{align}

We have to determine if, for a given $\Delta_{\text{min}}$, the above system
has a solution or not. It turns out that in the range $d\geq1$ and
$\Delta_{\text{min}}\geq2$ which is of interest for us, all $F_{d,\Delta
,l}^{(0,0)}>0$ in the RHS of the inhomogeneous equation (\ref{eq:inhom}). In
such a situation, if a nontrivial solution of the homogeneous system
(\ref{eq:hom}) is found, a solution of the full system (\ref{eq:inhom}),
(\ref{eq:hom}) can be obtained by a simple rescaling.\footnote{That is, unless
the series in the RHS of (\ref{eq:inhom}) diverge. However, this situation
does not occur in practice.}

\begin{figure}[ptb]
\centering\begin{minipage}[c]{0.3\textwidth}
\centering\includegraphics[height=1.7in]{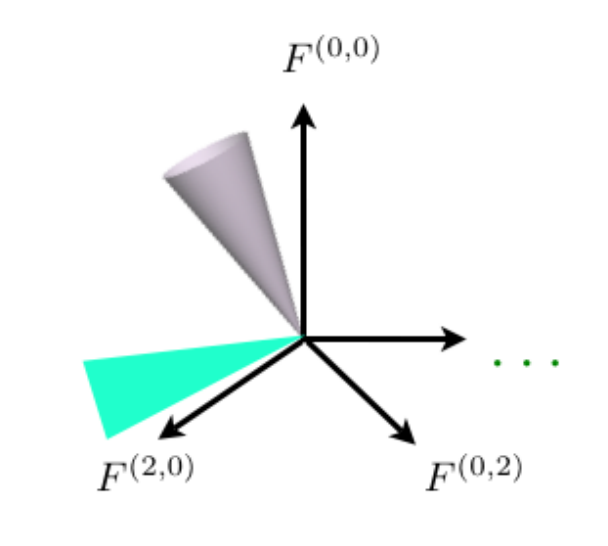}\\
$\Delta_{\rm min}>\Delta_c$
\end{minipage}
\begin{minipage}[c]{0.3\textwidth}
\centering\includegraphics[height=1.5in]{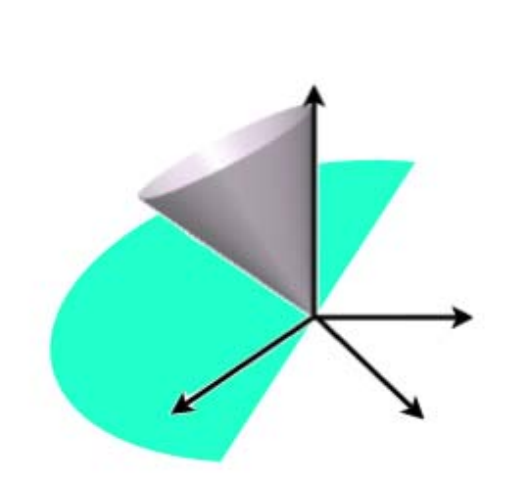} \\
$\Delta_{\rm min}=\Delta_c$
\end{minipage}
\begin{minipage}[c]{0.3\textwidth}
\centering\includegraphics[height=1.5in]{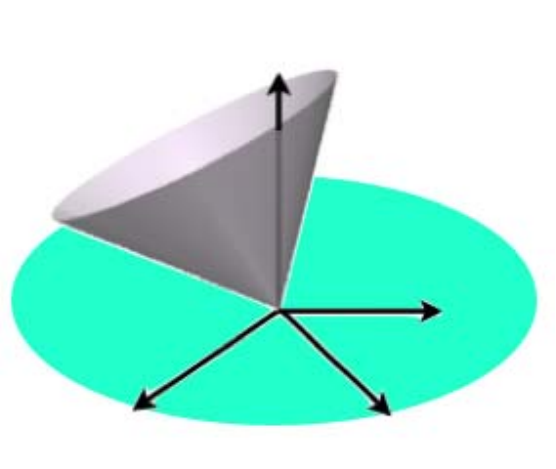}\\
$\Delta_{\rm min}<\Delta_c$
\end{minipage}
\caption{The shape of the projected cones in each of the 3 alternative cases
described in the text.}%
\label{fig:3coneproj}%
\end{figure}

Thus for our purposes it is enough to focus on the existence of nontrivial
solutions of the homogeneous system (\ref{eq:hom}).\footnote{\label{noteinhom}%
Eq.\ (\ref{eq:inhom}) can become useful when studying other questions. E.g.,
it suggests that arbitrarily large OPE coefficients may not be consistent with
the sum rule. It would be interesting to establish a model-independent
theoretical upper bound on the OPE coefficients, which could be a rigorous CFT
version of the NDA bounds in generic strongly coupled theories. For CFTs
weakly coupled to the SM (`unparticle physics' \cite{me-too-physics})
\textit{experimental} bounds on these OPE coefficients (also known as `cubic
unparticle couplings') have been recently considered in \cite{Feng}.}
Geometrically, this means that we are studying the projection of the convex
cone $\mathcal{C}$ on the $F^{(0,0)}=0$ plane. This \textit{projected cone,
}which is by itself a convex cone, may occupy a bigger or smaller portion of
the $F^{(0,0)}=0$ plane, or perhaps all of it. Each of the 3 cases shown in
Fig.\ \ref{fig:3cones} can be characterized in terms of the shape of the
projected cone, see Fig.\ \ref{fig:3coneproj}:

\begin{itemize}
\item $\Delta_{\mathrm{min}}>\Delta_{c}$: This case is realized if the opening
angle of the projected cone is `less than $\pi$', so that the homogeneous
equations have only the trivial solution $p_{\Delta,l}\equiv0$. In a more
technical language, the `opening angle less than $\pi$' condition means that
there exists a half-plane strictly containing the projected cone. If we write
the boundary of this half-plane as $\Lambda=0,$ the linear functional
$\Lambda$ will be strictly positive on the projected cone (except at its tip).

\item $\Delta_{\mathrm{min}}=\Delta_{c}$: This case is realized if the opening
angle of the projected cone is equal to $\pi$ in at least one direction. In
other words, the boundary of the projected cone must contains a linear
subspace passing through the origin. This subspace will be spanned by the
projections of the vectors saturating the sum rule; the vectors from the
\textquotedblleft bulk" cannot appear in the sum rule with nonzero coefficients.

\item $\Delta_{\mathrm{min}}<\Delta_{c}$: In this case the projected cone
covers the whole plane.
\end{itemize}

Using this classification, we are reduced to studying the boundary of the
projected cone.

For practical reasons we will have to work with finitely many derivatives,
i.e.~with a finite-dimensional subspace of the function space or,
equivalently, with a finite subset of the homogeneous system (\ref{eq:hom}).
The above geometric picture applies also within such a subspace. Satisfaction
of the sum rule on a subspace gives (in general) weaker but necessary
condition, so that we still get a valid bound (\ref{main}) with $f(d)=\Delta
_{c}$. As we expand the subspace by including more and more derivatives, the
critical scalar dimension $\Delta_{c}$ will go down, monotonically converging
to the optimal value corresponding to the full system.

\subsection{Warmup example: $d=1$}

\label{d=1}Let us use this philosophy to examine what the sum rule says about
the spectrum of operators appearing in the $\phi\times\phi$ OPE when $\phi$
has dimension $d=1$. Of course we know that $d=1$ corresponds to the free
scalar, see (\ref{limit}), and thus we know everything about this theory. In
particular, we know that only twist $2$ operators appear in the OPE, see
Section~\ref{sec:free}. Our interest here is to derive this result directly
from the sum rule. We expect the sum rule based approach to be robust: if we
make it work for $d=1$, chances are it will also give us a nontrivial result
for $d>1$. In contrast, the standard proof of (\ref{limit}) is not robust at
all: it is based on the fact that the 2-point function of a $d=1$ scalar is
harmonic, and can hardly be generalized to extract any information at $d>1$.

The attentive reader will notice that our considerations from Section
\ref{sec:why} were equivalent to retaining only the first equation out of the
infinite system (\ref{eq:hom}). This truncation did not control well the $d=1$
limit, since the obtained value of $\Delta_{c}\simeq3.61$ was well above the
free theory value $\Delta=2.$ The next natural try is to truncate to the first
\textit{two} equations in (\ref{eq:hom}). As we will see now, this truncation
already contains enough information to recover the free theory operator
dimensions from $d=1$.%

%TCIMACRO{\FRAME{ftbpFU}{4.2272in}{2.373in}{0pt}{\Qcb{The sum rule terms
%$F_{1,\Delta,l}$ in the $(F^{(2,0)},F^{(0,2)})$ plane. The shown curves
%correspond to $l=0,2,4,6$. The arrows are in the direction of increasing
%$\Delta$. The $l=0$ curve starts at $\Delta=1$ ($\Delta\lesssim1.01$ part is
%outside the plotted range); the $l=2,4,6$ curves---at $\Delta=l+2$. For large
%$\Delta$ the curves asymptote to the positive $F^{(2,0)}$ axis, see
%Appendix~\ref{as}. The shaded half-plane is the projected cone for a spectrum
%which includes the $\Delta=2$ scalar. }}{\Qlb{fig:d1proj}}{d=1proj.pdf}%
%width 4.2272in;  height 2.373in;  depth 0pt;  original-width 4.1632in;
%original-height 1.6812in;  cropleft "0";  croptop "1";  cropright "1";
%cropbottom "0";  filename 'd=1proj.pdf';file-properties "XNPEU";}} }%
%BeginExpansion
\begin{figure}
[ptb]
\begin{center}
\includegraphics[
height=2.373in,
width=4.2272in
]%
{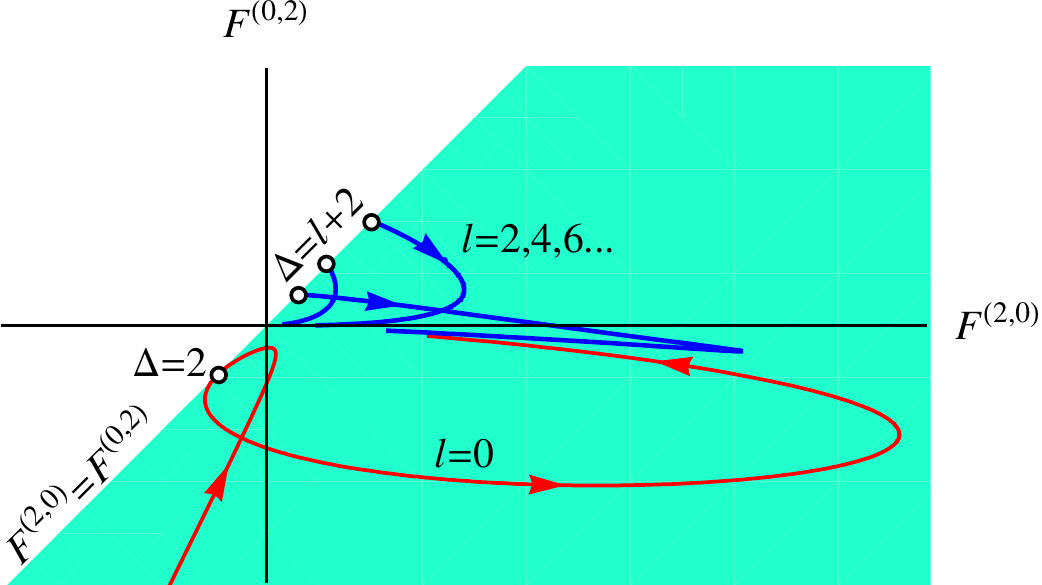}%
\caption{The sum rule terms $F_{1,\Delta,l}$ in the $(F^{(2,0)},F^{(0,2)})$
plane. The shown curves correspond to $l=0,2,4,6$. The arrows are in the
direction of increasing $\Delta$. The $l=0$ curve starts at $\Delta=1$
($\Delta\lesssim1.01$ part is outside the plotted range); the $l=2,4,6$
curves---at $\Delta=l+2$. For large $\Delta$ the curves asymptote to the
positive $F^{(2,0)}$ axis, see Appendix~\ref{as}. The shaded half-plane is the
projected cone for a spectrum which includes the $\Delta=2$ scalar. }%
\label{fig:d1proj}%
\end{center}
\end{figure}
%EndExpansion

Following the discussion in Section~\ref{geometry}, we consider the projected
cone---the cone generated by the vectors $F=F_{1,\Delta,l}$ projected into the
two-dimensional plane $(F^{(2,0)},F^{(0,2)})$. For each $l=0,2,4,\ldots$ we
get a curve in this plane, starting at the point corresponding to the lowest
value of $\Delta$ allowed by the unitarity bound (\ref{unitarity}), see
Fig.~\ref{fig:d1proj}. It can be seen from this figure that:

\begin{enumerate}
\item the vectors corresponding to the twist 2 operators $\Delta=l+2$ lie on
the line $F^{(2,0)}=F^{(0,2)}$,$\footnote{\label{note:easy}This fact is easy
to check analytically using the definition of $F_{d,\Delta,l}$ at $d=1$.}$
while all the other vectors lie to the right of this line;

\item the $l=0$, $\Delta=2$ vector points in the direction opposite to the
higher-spin twist $2$ operators.
\end{enumerate}

The boundary of the projected cone is thus given by the line $F^{(2,0)}%
=F^{(0,2)}$ \textit{if} the spectrum includes the $\Delta=2$ scalar and at
least one higher-spin twist $2$ operator (e.g.\ the energy-momentum tensor).
Otherwise the boundary will be formed by two rays forming an angle less than
$\pi$.\footnote{We ignore such subtleties as the possibility of a continuous
scalar spectrum ending at $\Delta=2$.} By the classification of
Section~\ref{geometry}, it is only in the former case that the sum rule can
have a solution. This case corresponds to $\Delta_{\text{min}}=\Delta_{c}$:
the boundary of the projected cone contains a linear subspace passing through
the origin. Thus we also have additional information: only the vectors from
the boundary, i.e.~those of the twist 2 operators, may be present in the sum
rule with nonzero coefficients.

The above argument appealed to the geometric intuition. For illustrative
purposes we will also give a more formal proof. Taking the difference of the
two first equations in (\ref{eq:hom}), we get:%
\[
0=\sum p_{\Delta,l}\left(  F_{d,\Delta,l}^{(2,0)}-F_{d,\Delta,l}%
^{(0,2)}\right)  ,\qquad p_{\Delta,l}\geq0\text{.}%
\]
By property 1 above, for $d=1$ all the terms in the RHS of this equation are
strictly positive unless $\Delta=l+2$. Thus, only twist $2$ operators may
appear with nonzero coefficients. End of proof.

It is interesting to note that in Fig.~\ref{fig:d1proj} the $l=0$ curve is
tangent to the line $F^{(2,0)}=F^{(0,2)}$ at $\Delta=2.^{\ref{note:easy}}$
Were it not so, we would not be able to exclude the existence of solutions to
the sum rule involving scalar operators of $\Delta<2$.

To conclude, we have shown that the spectrum of operators appearing in the sum
rule, and hence in the OPE, of a $d=1$ scalar consists solely of twist $2$
fields and that, moreover, a $\Delta=2$ scalar must be necessarily present in
this spectrum.

Isn't it amazing that we managed to find the whole spectrum using only the
first two out of the infinitely many equations (\ref{eq:hom})? One may ask if
by adding the complete information contained in the sum rule, a stronger
result can be proved, namely that the full 4-point function of an arbitrary
$d=1$ scalar is given by the free scalar theory expression. This would
constitute an independent proof of the fact that a $d=1$ scalar is necessarily
free. Such a proof can indeed be given \cite{Harcos}, but we do not present it
here since it is rather unrelated to our main line of reasoning.

\subsection{Simplest bound satisfying $f(1)=2$}

\label{simplest}

We will now present the simplest bound of the form (\ref{main}) which, unlike
the bound discussed in Section \ref{sec:why}, approaches $2$ as $d\rightarrow
1$. The argument uses the projection on the $(F^{(2,0)},F^{(0,2)})$ plane
similarly to the $d=1$ case from the previous section. Since that method gave
$\Delta_{\min}=2$ for $d=1$, by continuity we expect that it should give
$\Delta_{\min}\simeq2$ for $d$ sufficiently close to $1$.

\begin{figure}
[ptb]
\begin{center}
\includegraphics[
height=2.041in,
width=4.2134in
]%
{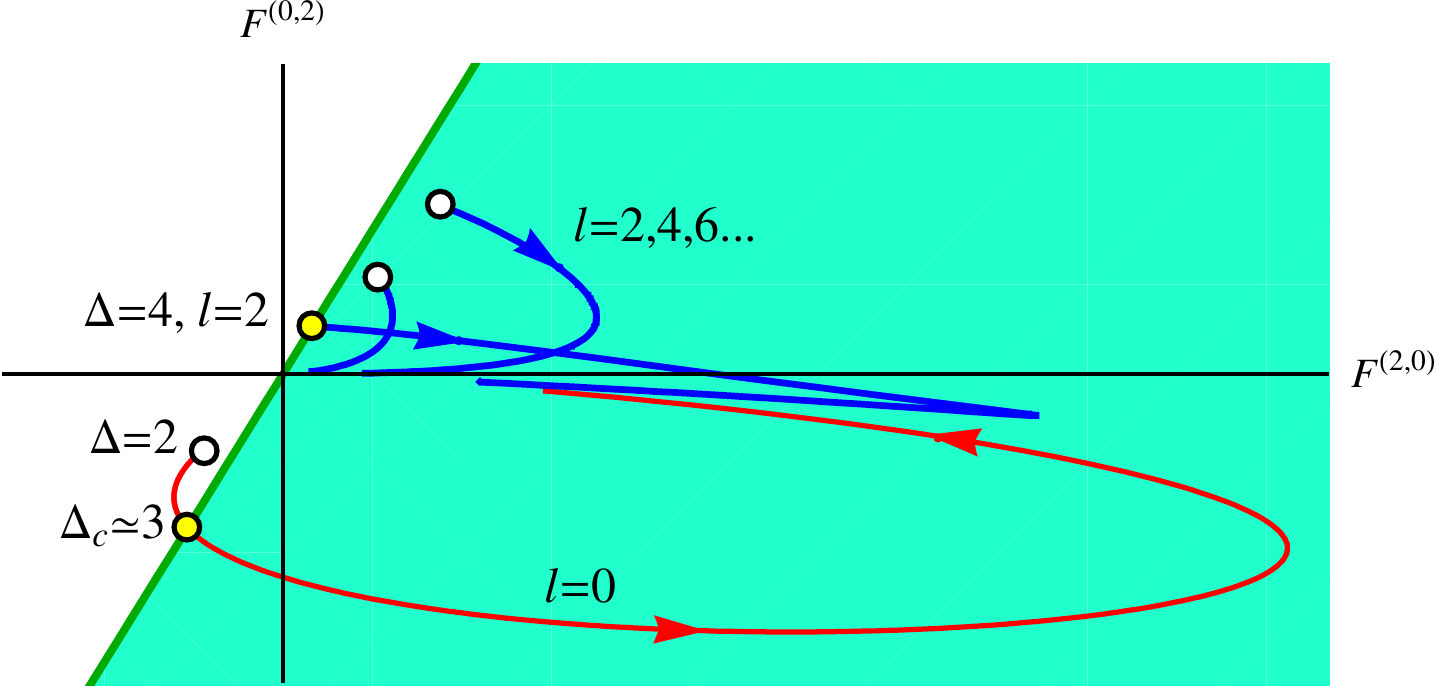}%
\caption{The analogue of Fig.~\ref{fig:d1proj} for $d=1.05$. In this plot we
started the $l=0$ curve at $\Delta=2$. The green line is the boundary of the
projected cone for $\Delta_{\text{min}}=\Delta_{c}\simeq3$, see
Fig.~\ref{fig:ill}. The slope of this line is determined by the
energy-momentum tensor vector.}%
\label{fig:d1.05proj}%
\end{center}
\end{figure}

\begin{figure}
[ptb]
\begin{center}
\includegraphics[
height=1.7417in,
width=1.6276in
]%
{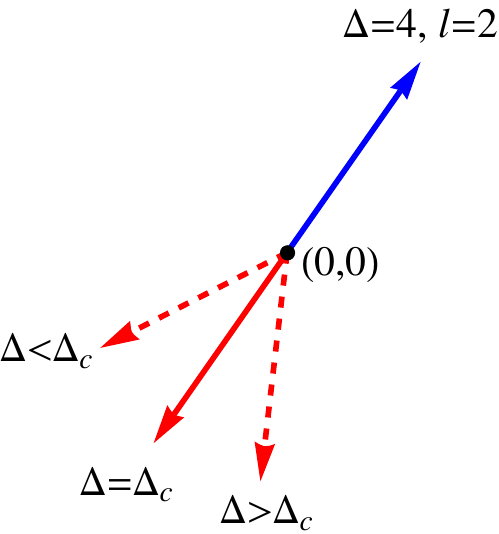}%
\caption{The relative position of the $l=0$ vectors (red) with respect to the
energy-momentum tensor vector (blue, pointing to upper right) determines the
shape of the projected cone, see the text. If the cone contains the blue
vector and both dashed red vectors, it covers the whole plane by their convex
linear combinations. }%
\label{fig:ill}%
\end{center}
\end{figure}

To demonstrate how the procedure works, we pick a $d$ close to $1$, say
$d=1.05$, and produce the analogue of the plot in Fig.~\ref{fig:d1proj}, see
Fig.~\ref{fig:d1.05proj}.%
%TCIMACRO{\FRAME{ftbpFU}{4.2134in}{2.041in}{0pt}{\Qcb{The analogue of
%Fig.~\ref{fig:d1proj} for $d=1.05$. In this plot we started the $l=0$ curve at
%$\Delta=2$. The green line is the boundary of the projected cone for
%$\Delta_{\text{min}}=\Delta_{c}\simeq3$, see Fig.~\ref{fig:ill}. The slope of
%this line is determined by the energy-momentum tensor vector.}}%
%{\Qlb{fig:d1.05proj}}{d=105proj.pdf}{\special{ language "Scientific Word";
%type "GRAPHIC";  maintain-aspect-ratio TRUE;  display "USEDEF";
%valid_file "F";  width 4.2134in;  height 2.041in;  depth 0pt;
%original-width 4.1632in;  original-height 2.0029in;  cropleft "0";
%filename 'd=105proj.pdf';file-properties "XNPEU";}} }%
%BeginExpansion
%EndExpansion
We see several changes with respect to Fig.~\ref{fig:d1proj}. The
energy-momentum tensor determines one part of the projected cone boundary (the
green line), while the spins $l=4,6,\ldots$ lie in the bulk of the cone.
Continuation of the green line to the other side of the origin intersects the
$l=0$ curve at the point corresponding to $\Delta=\Delta_{c}\simeq3.$ In the
terminology of Section~\ref{geometry}, this gives the critical value of
$\Delta_{\text{min}}$. Namely, if $\Delta_{\text{min}}>\Delta_{c}$ in the
spectrum (\ref{eq:family}), the projected cone will have an angle less than
$\pi$ and the sum rule will have no solutions. On the other hand, for
$\Delta_{\text{min}}<\Delta_{c}$ the projected cone covers the full plane, see
Fig.~\ref{fig:ill}, and a nontrivial solution to the first two equations of
the system (\ref{eq:hom}) will exist. For $\Delta_{\text{min}}=\Delta_{c}$ the
projected cone covers the half-plane shaded in Fig.~\ref{fig:d1.05proj}.%

%TCIMACRO{\FRAME{ftbpFU}{1.6276in}{1.7417in}{0pt}{\Qcb{The relative position of
%the $l=0$ vectors (red) with respect to the energy-momentum tensor vector
%(blue, pointing to upper right) determines the shape of the projected cone,
%see the text. If the cone contains the blue vector and both dashed red
%vectors, it covers the whole plane by their convex linear combinations. }%
%}{\Qlb{fig:ill}}{ill.pdf}{\special{ language "Scientific Word";
%type "GRAPHIC";  maintain-aspect-ratio TRUE;  display "USEDEF";
%valid_file "F";  width 1.6276in;  height 1.7417in;  depth 0pt;
%original-width 2.0029in;  original-height 1.7184in;  cropleft "0";
%filename 'ill.pdf';file-properties "XNPEU";}} }%
%BeginExpansion
%EndExpansion

One can check that the same situation is realized for any $d>1.$ In
particular, the slope of the critical cone boundary, described by the linear
equation
\begin{equation}
F^{(2,0)}-\lambda(d)F^{(0,2)}=0\,, \label{bdry}%
\end{equation}
is always determined by the energy-momentum tensor:%
\begin{equation}
\lambda(d)=\frac{F^{(2,0)}}{F^{(0,2)}}\,,\quad F=F_{d,4,2}. \label{kd1}%
\end{equation}
Once $\lambda(d)$ is fixed, the critical value of $\Delta_{\text{min}}$ is
determined from the intersection of the line (\ref{bdry}) with the $l=0$
curve:
\begin{equation}
F^{(2,0)}-\lambda(d)F^{(0,2)}=0,\quad F=F_{d,\Delta_{c},0}\text{\thinspace.}
\label{eq:inter}%
\end{equation}
The $l=0,$ $\Delta>\Delta_{c}$ points then lie strictly inside the half-plane
$F^{(2,0)}-\lambda(d)F^{(0,2)}\geq0.$ For $\Delta_{\text{min}}>\Delta_{c}$ the
cone angle is less than $\pi$, and the sum rule has no solution. Thus the
bound (\ref{main}) must be valid with $f(d)=\Delta_{c}$.

\begin{figure}
[ptb]
\begin{center}
\includegraphics[
trim=0.000000in 0.000000in -0.012688in 0.000000in,
height=1.919in,
width=2.917in
]%
{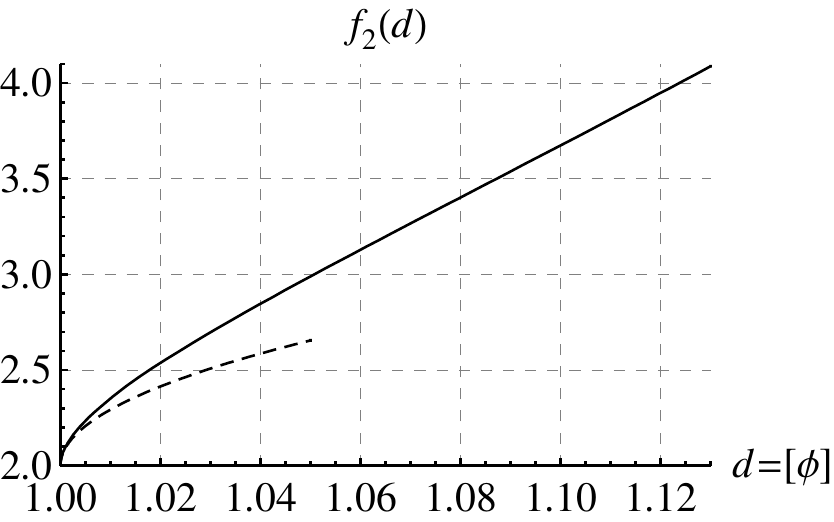}%
\caption{$f_{2}(d)=\Delta_{c}$ as determined by solving Eq.~(\ref{eq:inter}).
We plot it only for $d$ rather close to $1$ because in any case this bound
will be significantly improved below. The dashed line shows the asymptotic
behavior (\ref{deltaas}), which becomes a good approximation for
$d\lesssim1.001$.}%
\label{method1}%
\end{center}
\end{figure}

In Fig.~\ref{method1} we plot the corresponding value of $f(d)$ found
numerically from Eq.~(\ref{kd1}), (\ref{eq:inter}), denoted $f_{2}(d)$ to
reflect the order of derivatives used to derive this bound. As promised, the
free field theory value $\Delta=2$ is approached continuously as
$d\rightarrow1.$
%TCIMACRO{\FRAME{ftbpFU}{2.917in}{1.919in}{0pt}{\Qcb{$f_{2}(d)=\Delta_{c}$ as
%determined by solving Eq.~(\ref{eq:inter}). We plot it only for $d$ rather
%close to $1$ because in any case this bound will be significantly improved
%below. The dashed line shows the asymptotic behavior (\ref{deltaas}), which
%becomes a good approximation for $d\lesssim1.001$.}}{\Qlb{method1}%
%}{method1.pdf}{\special{ language "Scientific Word";  type "GRAPHIC";
%display "USEDEF";  valid_file "F";  width 2.917in;  height 1.919in;
%depth 0pt;  original-width 3.965in;  original-height 2.7015in;  cropleft "0";
%croptop "1";  cropright "1.0032";  cropbottom "0";
%filename 'method1.pdf';file-properties "XNPEU";}} }%
%BeginExpansion
%EndExpansion

The asymptotic behavior of $f_{2}(d)$ for $d\rightarrow1$ can be determined by
expanding the equations defining $\Delta_{c}$ in power series in $d-1$ and
$\Delta_{c}-2.$ We find:
\begin{align}
f_{2}(d)  &  =2+\gamma\sqrt{d-1}+{O}(d-1)\,,\nonumber\\
\gamma &  \equiv\lbrack2(K+1)/3]^{1/2}\simeq2.929,\label{deltaas}\\
K  &  \equiv(192\ln2-133)^{-1}.\nonumber
\end{align}
This asymptotics provides a good approximation for $d-1\lesssim10^{-3}$, see
Fig.~\ref{method1}. The square root dependence in (\ref{deltaas}) can be
traced to the fact that for $d=1$ the $l=0$ curve was \textit{tangent} to the
projected cone boundary at $\Delta=2$. The bound of Fig.~\ref{method1} will be
improved below by taking more derivatives into account, however the square
root behavior will persist (albeit with a different coefficient).

\subsection{Improved bounds: general method}

\label{general}

As we already mentioned in Section \ref{geometry}, the bound will improve
monotonically as we include more and more equations from the infinite system
(\ref{eq:hom}) in the analysis, i.e.~increase the number of derivatives
$F^{(2m,2n)}$ that we are controlling. We thus consider a finite basis
$\mathcal{B}$, adding several higher-order derivatives to the $F^{(2,0)}$ and
$F^{(0,2)}$ included in the previous section:%
\begin{equation}
\mathcal{B}=\{F^{(2m,2m)}\}=\{F^{(2,0)},F^{(0,2)},\ldots\}. \label{derset}%
\end{equation}
According to the discussion in Section \ref{geometry}, we have to study the
boundary of the projected cone in the finite-dimensional space with
coordinates (\ref{derset}). The logic in principle is the same as in the
previous section. We will have a family of curves corresponding to
$l=0,2,4\ldots$ generating the projected cone. As we lower $\Delta
_{\text{min}}$ in the spectrum (\ref{eq:family}), the projected cone grows.
For $\Delta_{\text{min}}<\Delta_{c}$ it will cover the whole space. However,
in this many-dimensional situation it is not feasible to look for $\Delta_{c}$
by making plots similar to Fig. \ref{fig:d1.05proj}. We need a more formal approach.

Such an approach uses the language of linear functionals, already encountered
in Section \ref{geometry}. A linear functional $\Lambda$ on the
finite-dimensional subspace with basis $\mathcal{B}$ is given by
\begin{equation}
\Lambda(F)=\sum_{\mathcal{B}}\lambda_{2m,2n}F^{(2m,2n)}, \label{fun0}%
\end{equation}
where $\lambda_{2m,2n}$ are some fixed numbers characterizing the functional.
They generalize the single parameter $\lambda(d)$ from Section \ref{simplest}
to the present situation.

Using linear functionals, the two non-critical cases of Fig.
\ref{fig:3coneproj} can be distinguished as follows:%
\begin{align}
\Delta_{\text{min}}>\Delta_{c}  &  \Longleftrightarrow\text{there \textit{IS}
a functional }\Lambda\text{ such that}\nonumber\\
&  \qquad\qquad\Lambda(F_{d,\Delta,l})>0\text{ for all }\Delta,l\in
\Sigma(\Delta_{\text{min}})\,\text{;}\label{sep2b}\\
\Delta_{\text{min}}<\Delta_{c}  &  \Longleftrightarrow\text{there is
\textit{NO }functional }\Lambda\text{ such that}\nonumber\\
&  \qquad\qquad\Lambda(F_{d,\Delta,l})\geq0\text{ for all }\Delta,l\in
\Sigma(\Delta_{\text{min}})\,\text{.} \label{sep2}%
\end{align}
A numerical procedure which for any given $\Delta_{\text{min}}$ finds such a
positive $\Lambda$ or shows that a non-negative $\Lambda$ does not exists will
be explained below. Assuming that we know how to do this, determination of
$\Delta_{c}$ becomes an easy task. First, we bracket $\Delta_{c}$ from above
and below by trying out a few values of $\Delta_{\text{min}}$ and checking to
which of the two above sets, $\Delta_{\text{min}}>\Delta_{c}$ or
$\Delta_{\text{min}}<\Delta_{c}$, they belong. Second, we apply the
division-in-two algorithm, \textit{i.e.~}reduce the length of the bracketing
interval by checking its middle point, etc. This achieves exponential
precision after a finite number of steps.

We will now explain the numerical procedure. Let us begin with the
non-negative functional defined by Eq.~(\ref{sep2}), and comment later about
the strictly positive case of Eq.~(\ref{sep2b}). Eq.~(\ref{sep2}) can be
viewed as a system of \textit{infinitely many} linear inequalities for the
coefficients $\lambda_{2m,2n}$. The infinitude is due to three reasons:

\begin{itemize}
\item there are infinitely many spins $l$;

\item for each spin $l$ the dimension $\Delta$ can be arbitrary large;

\item the dimension $\Delta$ varies continuously.
\end{itemize}

To be numerically tractable, this system needs to be truncated to a finite
system, removing each of the three infinities. We do it by imposing
inequalities in (\ref{sep2}) not for all $\Delta,l\in\Sigma(\Delta
_{\text{min}})$ but only for a `trial set' such that

\begin{itemize}
\item only finite number of spins $l\leq l_{\text{max}}$ are included$;$

\item only dimensions up to a finite $\Delta=\Delta_{\max}$ are included;

\item $\Delta$ is discretized.
\end{itemize}

To ensure that we are not losing important information by truncating at
$l_{\max}$ and $\Delta_{\max}$, we include into the trial set the vectors
corresponding to the large $l$ and large $\Delta$ asymptotics of the
derivatives. The relevant asymptotics have the form (see Eq.\ (\ref{as1}) in
Appendix \ref{as}):%
\begin{equation}
F_{d,\Delta,l}^{(2m,2n)}\sim\frac{const}{(2m+1)(2n+1)}(2\sqrt{2}%
l\,(1+x))^{2m+1}(2\sqrt{2}l)^{2n+1},\quad x\equiv\frac{\Delta-l-2}{l}\geq0\,,
\label{eq:asym}%
\end{equation}
where a constant $const>0$ is independent of $m$ and $n$. This asymptotics is
valid for $l\rightarrow\infty$, $x\ll l$ fixed.

Upon truncation to the trial set, Eq.~(\ref{sep2}) becomes a finite system of
linear inequalities, a particular case of the \textit{linear programming
problem}\footnote{A general linear programming problem consists in minimizing
a linear function of several variables subject to a set of linear constraints
(equalities and inequalities). Our problem is a particular case when all
constraints are inequalities and the function to be minimized is absent (or,
equivalently, it is constant).}. It is thus possible to determine if a
solution exists (and find it if it does) using one of several existing
efficient numerical algorithms (see \cite{lp}). In our work we used the
classic Simplex Method as realized by the \texttt{LinearProgramming} function
of \textsc{Mathematica}.

If, using the linear programming, we find that (\ref{sep2}) truncated to the
trial set has no solution, then \textit{a fortiori} the full non-truncated
system has no solution, i.e.\ non-negative $\Lambda$ does not exist. Thus we
can safely claim that the considered $\Delta_{\text{min}}$ brackets
$\Delta_{c}$ from below: $\Delta_{\text{min}}<\Delta_{c}$. The accuracy of
this bracketing will increase as we include more dimensions in the trial set
(e.g.~decreasing the discretization step).

Let us now consider bracketing from above, which requires finding a functional
satisfying (\ref{sep2b}). First, just as for (\ref{sep2}), we truncate to a
trial set. We'd like to use linear programming methods, however these methods
do not work for strict inequalities. Thus we strengthen $>0$ in (\ref{sep2b})
to $\geq\varepsilon$, where $\varepsilon$ is a fixed small positive
number\footnote{In principle, Eq.~(\ref{sep2c}) may become too constraining if
all components of the vector $F_{d,\Delta,l}$ are $O(\varepsilon)$, which may
happen in the large $\Delta$ limit. In this case one simply needs to rescale
$F_{d,\Delta,l}$ by a constant factor. Each function $F_{d,\Delta,l}$
determines a ray in the finite-dimensional space, and such a rescaling does
not change the content of the original Eq.~(\ref{sep2b}). In practice,
however, we never had to do such a rescaling.}:%
\begin{equation}
\Lambda(F_{d,\Delta,l})\geq\varepsilon\,\text{.} \label{sep2c}%
\end{equation}
Then we can use the linear programming to find a solution of the truncated system.

\begin{figure}
[ptb]
\begin{center}
\includegraphics[
height=1.6613in,
width=2.2648in
]%
{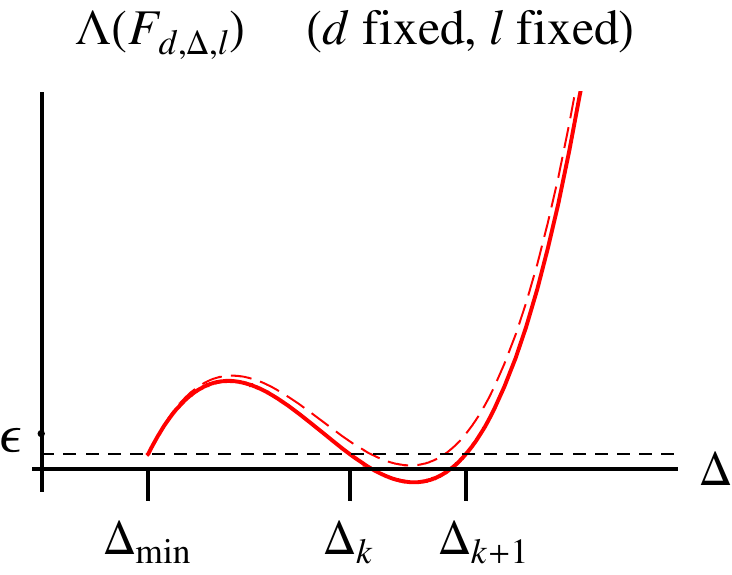}%
\caption{\textit{Solid curve}: schematic typical dependence of the functional
$\Lambda(F_{d,\Delta,l})$ on $\Delta$. The functional $\Lambda$ satisfies
(\ref{sep2c}) for all $\Delta$ from a discrete trial set, which includes the
dimensions $\Delta_{k}$ and $\Delta_{k+1}=\Delta_{k}+\delta\Delta$. Yet the
functional may become slightly negative for $\Delta_{k}<\Delta<\Delta_{k+1}$.
\textit{Dashed curve}: same for a functional corresponding to a smaller
$\delta\Delta$. The violation of (\ref{sep2}) at intermediate values has
disappered. }%
\label{fig:discr}%
\end{center}
\end{figure}

Unlike in the case of bracketing from below, to claim that indeed
$\Delta_{\text{min}}>\Delta_{c}$, we have to check that the found $\Lambda$
does not violate (\ref{sep2b}) for $\Delta,l$ not included in the trial set.
This will not happen for $l\geq l_{\text{max}}$ and $\Delta\geq\Delta
_{\text{max}}$ if we take these parameters sufficiently large and include the
asymptotics. The functional $\Lambda$ may however become slightly negative at
\textit{intermediate} $\Delta$ and $l$, as a consequence of discretizing
$\Delta$. This is not unexpected, since the rays $(\Delta,l)$ that determine
the boundary of the cone and thus the functional are determined with a
fuzziness proportional to the discretization step $\delta\Delta$. However,
this violation will disappear as $\delta\Delta\rightarrow0$ (for a fixed
$\varepsilon$), see Fig.~\ref{fig:discr}. By choosing the discretization step
smaller and smaller, we will be able to bracket $\Delta_{c}$ from above with
an arbitrary desired precision.%

%TCIMACRO{\FRAME{ftbpFU}{2.2648in}{1.6613in}{0pt}{\Qcb{\textit{Solid curve}:
%schematic typical dependence of the functional $\Lambda(F_{d,\Delta,l})$ on
%$\Delta$. The functional $\Lambda$ satisfies (\ref{sep2c}) for all $\Delta$
%from a discrete trial set, which includes the dimensions $\Delta_{k}$ and
%$\Delta_{k+1}=\Delta_{k}+\delta\Delta$. Yet the functional may become slightly
%negative for $\Delta_{k}<\Delta<\Delta_{k+1}$. \textit{Dashed curve}: same for
%a functional corresponding to a smaller $\delta\Delta$. The violation of
%(\ref{sep2}) at intermediate values has disappered. }}{\Qlb{fig:discr}%
%}{discr.pdf}{\special{ language "Scientific Word";  type "GRAPHIC";
%width 2.2648in;  height 1.6613in;  depth 0pt;  original-width 3.232in;
%original-height 2.3649in;  cropleft "0";  croptop "1";  cropright "1";
%cropbottom "0";  filename 'discr.pdf';file-properties "XNPEU";}} }%
%BeginExpansion
%EndExpansion

\subsection{Best results to date}

\label{sec:best}

\begin{figure}
[ptb]
\begin{center}
\includegraphics[
height=2.1629in,
width=3.3416in
]%
{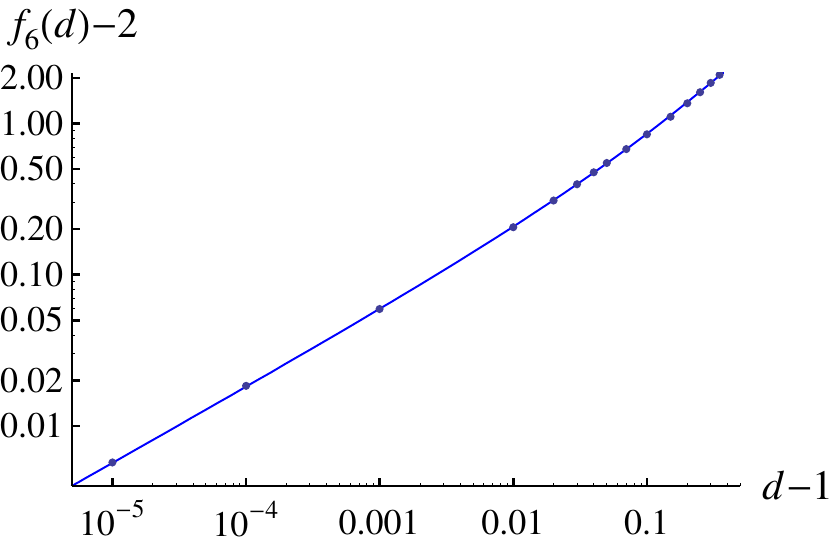}%
\caption{Log-log plot of the maximal allowed anomalous dimension of $\phi^{2}%
$, $f_{6}(d)-2,$ versus the anomalous dimension of $\phi$, $d-1$. The dots
correspond to the entries of Table \ref{tab:numerics}, while the curve is the
approximation (\ref{eq:appr-an}).}%
\label{fig:logplot}%
\end{center}
\end{figure}

In our numerical work we explored functionals (\ref{fun0}) with the leading
$a$-derivative up to $F^{(6,0)}$ and with various choices of subleading
derivatives. We now present our best results, which were obtained using the
full list of derivatives with $2m+2n\leq6$:%
\begin{equation}
\left\{  F^{(2m,2n)}%
~|~(2m,2n)=(6,0),(4,2),(2,4),(0,6),(4,0),(2,2),(0,4),(2,0),(0,2)\right\}  .
\label{best}%
\end{equation}
The bound $f(d)\equiv f_{6}(d)$ corresponding to this choice is plotted in
Figs.~\ref{fig:bound-intro} and \ref{fig:logplot}. Numerical values for
several values of $d$ are given in Table \ref{tab:numerics}. For each $d$ we
give the bound $f_{6}(d)$ and the coefficients $\lambda_{2m,2n}$ of the
functional used to obtain this bound. These functionals were found using the
linear programming method as described in the previous section. However, to
check that our bound is true, one does not need to know how we found these
functionals; it is enough to check that they indeed satisfy Eq.\ (\ref{sep2b})
with $\Delta_{\text{min}}=f_{6}(d)$.

We have also computed the bound $f_{6}(d)$ for other values of $d$ and found
that it changes smoothly, interpolating between the points given in Table
\ref{tab:numerics}. In the considered interval of $d$, the anomalous dimension
$f_{6}(d)-2$ turns out to be well approximated (within $\sim2\%$) by the
formula (see Fig.\thinspace\ref{fig:logplot}):%
\begin{equation}
f_{6}(d)-2\simeq1.79\sqrt{d-1}+2.9(d-1)\,\qquad(1<d<1.35). \label{eq:appr-an}%
\end{equation}
While our method would give a bound also for $d>1.35$, we did not explore this
range. The reason is that for $d\gtrsim1.33$ our bound exceeds $\Delta=4$ and
starts getting not very interesting, taking into account the phenomenological
motivations from Section \ref{pheno}.%

%TCIMACRO{\FRAME{ftbpFU}{3.3416in}{2.1629in}{0pt}{\Qcb{Log-log plot of the
%maximal allowed anomalous dimension of $\phi^{2}$, $f_{6}(d)-2,$ versus the
%anomalous dimension of $\phi$, $d-1$. The dots correspond to the entries of
%Table \ref{tab:numerics}, while the curve is the approximation
%(\ref{eq:appr-an}).}}{\Qlb{fig:logplot}}{logplot.pdf}%
%width 3.3416in;  height 2.1629in;  depth 0pt;  original-width 3.333in;
%original-height 2.1473in;  cropleft "0";  croptop "1";  cropright "1";
%cropbottom "0";  filename 'logplot.pdf';file-properties "XNPEU";}} }%
%BeginExpansion
%EndExpansion

The coefficients $\lambda_{2m,2n}$ in Table \ref{tab:numerics} have been
rounded with 6 significant digits; we have checked that the resulting slight
violation of (\ref{sep2b}) is very small\footnote{The rounded functionals
violate (\ref{sep2b}) by about $10^{-4}$ for a few isolated values of
$\Delta,l$. This should be compared to the typical ${O}(1\div1000)$ range of
$\Lambda(F)$ away from these points.}. The anomalous dimensions $f_{6}(d)-2$
in Table \ref{tab:numerics} approximate the optimal values, attainable using
the subspace (\ref{best}), \textit{from above} to within $1\%$ (due to the
finite accuracy of the division-in-two algorithm used to bracket $\Delta_{c}%
$); they have been rounded \textit{up} with 3 significant digits.
%A non-rounded version of Table \ref{tab:numerics} and a sample
%\textsc{Mathematica} code which can be used to reproduce this table are
%available at \cite{download}.

In obtaining these results, we used the trial set in the sense of the previous
subsection with $l_{\max}=10$, $\Delta_{\max}=20$ $(l=0)$, $\Delta_{\max
}=l+10$ $(l\geq2)$. We discretized $\Delta$ with step $\delta\Delta=0.01$,
decreased to $\delta\Delta=0.0025$ around a few critical dimensions where the
functional approaches zero, as in Fig.~\ref{fig:discr} for $\Delta_{k}%
<\Delta<\Delta_{k+1}$. To take into account the asymptotic behavior
(\ref{eq:asym}), we have included into the trial set vectors with
\begin{equation}
F^{(2m,2n)}=\left\{
\begin{array}
[c]{l}%
\left[  (2m+1)(2n+1)(1+x)^{2n}\right]  ^{-1},\quad~2m+2n=6\,,\\
0,\qquad\qquad\qquad\qquad\qquad\qquad\qquad2m+2n<6\,,
\end{array}
\right.  \label{eq:incl-asym}%
\end{equation}
obtained from (\ref{eq:asym}) by rescaling and taking the $l\rightarrow\infty$
limit. The parameter $x$ in (\ref{eq:incl-asym}) was varying from $x=0$ to
$10$ with $\delta x=0.01$. We set $\varepsilon=10^{-4}$ in (\ref{sep2c}).

\begin{figure}
[ptb]
\begin{center}
\includegraphics[
height=2.2869in,
width=3.341in
]%
{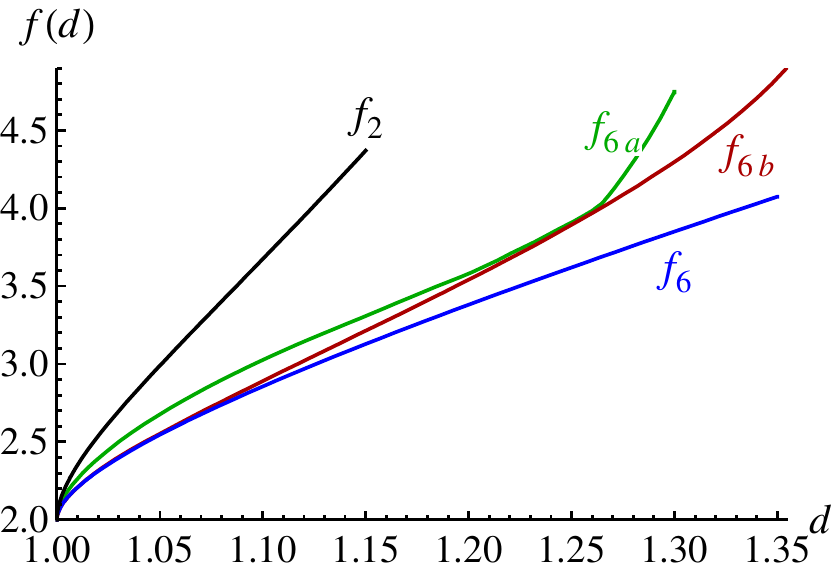}%
\caption{From this plot one can get an idea how the bound monotonically
improves as more and more derivatives are taken into acccount in the infinite
system (\ref{eq:hom}). The black (upper) curve is the simplest bound of
Section \ref{simplest}, obtained by using only the second derivatives. The
next two curves (green and red) correspond to the two sets (\ref{eq:twosets}).
The lowest-lying blue curve is our best current bound obtained using the set
(\ref{best}).}%
\label{fig:comp}%
\end{center}
\end{figure}

We expect that including more derivatives in the list (\ref{best}) should
somewhat improve the bound, especially for $d$ close to the upper end of the
considered interval$.$ We have observed a similar improvement trying out the
functionals with the same leading $a$-derivative $(6,0)$ as in (\ref{best}),
but with smaller sets of subleading derivatives:%
\begin{align}
\text{Set 6a}\text{: }  &  (2m,2n)=(6,0),(4,0),(2,0),(0,2); \label{eq:twosets}%
\\
\text{Set 6b}\text{: }  &
(2m,2n)=(6,0),(4,0),(2,0),(0,2),(0,4),(0,6)\,.\nonumber
\end{align}
For illustration, we plot the corresponding bounds in Fig.~\ref{fig:comp},
including also the simplest second-derivative bound from Section
\ref{simplest}.

\begin{table}[th]%
\[%
\begin{array}
[c]{cc|ccccccccc}%
d-1 & f_{6}-2 & \lambda_{6,0} & \lambda_{4,2} & \lambda_{2,4} & \lambda_{0,6}
& \lambda_{4,0} & \lambda_{2,2} & \lambda_{0,4} & \lambda_{2,0} &
\lambda_{0,2}\\[5pt]\hline
10^{-5} & 0.00573 & {\scriptstyle1.} & {\scriptstyle-0.97747} &
{\scriptstyle-1.06327} & {\scriptstyle-0.047622} & {\scriptstyle-116.282} &
{\scriptstyle277.34} & {\scriptstyle49.2726} & {\scriptstyle3344.18} &
{\scriptstyle-7170.94}\\
10^{-4} & 0.0185 & {\scriptstyle1.} & {\scriptstyle0.} &
{\scriptstyle-0.00052291} & {\scriptstyle-0.999251} & {\scriptstyle-153.677} &
{\scriptstyle48.2058} & {\scriptstyle126.869} & {\scriptstyle7344.28} &
{\scriptstyle-8370.35}\\
10^{-3} & 0.0593 & {\scriptstyle1.} & {\scriptstyle-1.24135} &
{\scriptstyle-0.845225} & {\scriptstyle-0.0234949} & {\scriptstyle-101.866} &
{\scriptstyle276.743} & {\scriptstyle36.8807} & {\scriptstyle2626.56} &
{\scriptstyle-6384.32}\\
0.01 & 0.207 & {\scriptstyle1.} & {\scriptstyle-0.738985} &
{\scriptstyle0.839453} & {\scriptstyle-1.00868} & {\scriptstyle-104.348} &
{\scriptstyle14.8941} & {\scriptstyle88.7308} & {\scriptstyle4989.57} &
{\scriptstyle-5288.51}\\
0.02 & 0.31 & {\scriptstyle1.} & {\scriptstyle-1.04266} &
{\scriptstyle1.22926} & {\scriptstyle-1.01672} & {\scriptstyle-87.2012} &
{\scriptstyle0.} & {\scriptstyle73.9343} & {\scriptstyle4384.97} &
{\scriptstyle-4356.11}\\
0.03 & 0.397 & {\scriptstyle1.} & {\scriptstyle-0.669013} &
{\scriptstyle0.980721} & {\scriptstyle-1.02839} & {\scriptstyle-100.265} &
{\scriptstyle-8.59119} & {\scriptstyle82.6461} & {\scriptstyle5012.32} &
{\scriptstyle-4772.52}\\
0.04 & 0.476 & {\scriptstyle1.} & {\scriptstyle-1.14492} &
{\scriptstyle1.51543} & {\scriptstyle-1.03388} & {\scriptstyle-77.4978} &
{\scriptstyle-20.9209} & {\scriptstyle63.1812} & {\scriptstyle4200.66} &
{\scriptstyle-3696.39}\\
0.05 & 0.548 & {\scriptstyle1.} & {\scriptstyle6.31335} &
{\scriptstyle-5.62554} & {\scriptstyle-1.0635} & {\scriptstyle-221.053} &
{\scriptstyle-276.119} & {\scriptstyle338.147} & {\scriptstyle11062.1} &
{\scriptstyle-4888.67}\\
0.07 & 0.678 & {\scriptstyle1.} & {\scriptstyle8.20962} &
{\scriptstyle-7.39456} & {\scriptstyle-1.07537} & {\scriptstyle-236.106} &
{\scriptstyle-425.566} & {\scriptstyle439.223} & {\scriptstyle13130.} &
{\scriptstyle-3371.37}\\
0.1 & 0.849 & {\scriptstyle1.} & {\scriptstyle10.4578} &
{\scriptstyle-9.55059} & {\scriptstyle-1.08398} & {\scriptstyle-255.95} &
{\scriptstyle-620.004} & {\scriptstyle579.352} & {\scriptstyle16504.5} &
{\scriptstyle-1662.47}\\
0.15 & 1.11 & {\scriptstyle1.} & {\scriptstyle11.981} & {\scriptstyle-10.4519}
& {\scriptstyle-1.14201} & {\scriptstyle-261.146} & {\scriptstyle-768.116} &
{\scriptstyle649.227} & {\scriptstyle19160.9} & {\scriptstyle-46.5554}\\
0.2 & 1.36 & {\scriptstyle1.} & {\scriptstyle12.7909} & {\scriptstyle-10.5811}
& {\scriptstyle-1.20555} & {\scriptstyle-259.763} & {\scriptstyle-863.149} &
{\scriptstyle676.812} & {\scriptstyle20924.6} & {\scriptstyle1544.21}\\
0.25 & 1.61 & {\scriptstyle1.} & {\scriptstyle14.283} & {\scriptstyle-11.2729}
& {\scriptstyle-1.28025} & {\scriptstyle-262.93} & {\scriptstyle-1027.33} &
{\scriptstyle746.012} & {\scriptstyle24108.2} & {\scriptstyle3944.39}\\
0.3 & 1.86 & {\scriptstyle1.} & {\scriptstyle18.0218} & {\scriptstyle-14.0589}
& {\scriptstyle-1.36917} & {\scriptstyle-281.038} & {\scriptstyle-1467.79} &
{\scriptstyle996.762} & {\scriptstyle32918.8} & {\scriptstyle9435.72}\\
0.35 & 2.1 & {\scriptstyle1.} & {\scriptstyle24.6535} & {\scriptstyle-19.3331}
& {\scriptstyle-1.49588} & {\scriptstyle-292.263} & {\scriptstyle-2367.92} &
{\scriptstyle1493.39} & {\scriptstyle51357.2} & {\scriptstyle20484.}%
\end{array}
\]
\caption{For several $d$ we give numerical values of $f_{6}\equiv f_{6}(d)$
appearing in the bound (\ref{main}) and the coefficients $\lambda_{2m,2n}$ of
a functional satisfying Eq.~(\ref{sep2}) with $\Delta_{\text{min}}=f_{6}$, see
the text. The functionals are normalized via $\lambda_{6,0}=1$.}%
\label{tab:numerics}%
\end{table}%

%TCIMACRO{\FRAME{ftbpFU}{3.341in}{2.2869in}{0pt}{\Qcb{From this plot one can
%get an idea how the bound monotonically improves as more and more derivatives
%are taken into acccount in the infinite system (\ref{eq:hom}). The black
%(upper) curve is the simplest bound of Section \ref{simplest}, obtained by
%using only the second derivatives. The next two curves (green and red)
%correspond to the two sets (\ref{eq:twosets}). The lowest-lying blue curve is
%our best current bound obtained using the set (\ref{best}).}}{\Qlb{fig:comp}%
%}{comparison.pdf}{\special{ language "Scientific Word";  type "GRAPHIC";
%width 3.341in;  height 2.2869in;  depth 0pt;  original-width 3.232in;
%original-height 2.2047in;  cropleft "0";  croptop "1";  cropright "1";
%cropbottom "0";  filename 'comparison.pdf';file-properties "XNPEU";}} }%
%BeginExpansion
%EndExpansion

\section{Comparison to known results}

\label{sec:comparison}

To further test our method, in this section we shall compare our bound to the
operator dimensions in calculable CFTs. Several nontrivial tests are offered
by exactly solvable CFTs in 2D. We will discuss these examples in subsection
1. On the other hand there are fewer calculable examples in 4D, and they all
turn out to satisfy our bound in a somewhat trivial way. The point is that our
bound on $\Delta$ lies abundantly above the line $\Delta=2d$, and none of the
calculable models is significantly above this line. For instance in
supersymmetric gauge theories the operators whose dimension is exactly
calculable are chiral. In that case the relation $d=2r/3$ holds, with $r$
representing the $R$-charge. Then, given a chiral scalar operator $\phi$ of
dimension $d$, additivity of $R$-charge implies that $\phi^{2}$ has dimension
$2d$, precisely on the $\Delta=2d$ line, and thus our bound is trivially satisfied.

In the case of large $N$ theories Green's functions factorize at leading $1/N$
order, implying $\Delta=2d+{O}(1/N)$. This relation does not provide a
stringent test of our result unless $d-1<{O}(1/N)$, which corresponds to an
elementary, free, field at leading order in $1/N$. This situation can
potentially be realized in variants of the Belavin-Migdal-Banks-Zaks (BZ)
fixed point \cite{BM},\cite{BZ}. The simplest case of a non-Abelian gauge
theory with matter consisting of charged fermions obviously does not provide a
nontrivial check of our bound. This is because the gauge invariant operator of
lowest dimension is $\bar{\psi}\psi$ with dimension already close to $3$. In
order to have a chance to find a model that nearly saturates our bound we must
necessarily add a scalar gauge singlet field $\phi$ to the theory and look for
a new BZ fixed point\footnote{See Section 5.1 of \cite{intriligator} for a
related discussion.}. Consider then a BZ model based on gauge group $SU(N)$
with $N_{\mathrm{F}}$ fermionic flavors in the fundamental coupled to $\phi$.
The Lagrangian is
\begin{equation}
\mathcal{L}=-\frac{1}{4g^{2}}G_{\mu\nu}G^{\mu\nu}+{\bar{q}}\!\!\not \!
\!Dq+y\phi\bar{q}q+\lambda\phi^{4}\,.
\end{equation}
By dialing $N_{\mathrm{F}}/N=11/2-\epsilon$ with $1/N\lesssim\epsilon\ll1$ the
1-loop $\beta$-function is small and is compensated by the 2-loop contribution
which comes with opposite sign ($b>0$)
\begin{equation}
8\pi^{2}\mu\frac{d}{d\mu}\frac{1}{g^{2}}=\frac{2}{3}N\epsilon-b\frac
{g^{2}N^{2}}{8\pi^{2}}=0\,,\qquad\quad\frac{g^{2}N}{8\pi^{2}}\sim\epsilon\,.
\end{equation}
One can then easily check that $\beta(y)$ and $\beta(\lambda)$ possess,
already at 1-loop order, non-trivial zeroes satisfying $y^{2}\sim\lambda$ and
${y^{2}}/{8\pi^{2}}\sim\epsilon/(NN_{\mathrm{F}})\sim\epsilon/N^{2}$. Notice
also that, for such value of $y^{2}$, its contribution to the 2-loop gauge
$\beta$-function is subleading in $1/N$, and therefore does not significantly
affect the location of the zero of $\beta(g)$. The anomalous dimensions are
given by
\begin{align}
\gamma_{\phi}  &  \equiv d-1=c_{1}\frac{y^{2}N^{2}}{8\pi^{2}}=a_{1}%
\epsilon\,,\\
\gamma_{\phi^{2}}  &  \equiv\Delta-2=2\gamma_{\phi}+c_{2}\frac{\lambda}%
{8\pi^{2}}=2a_{1}\epsilon+a_{2}\frac{\epsilon}{N^{2}}\,.
\end{align}
Again our bound is largely satisfied, just because $\gamma_{\phi}$ arises at
leading nontrivial order, at 1-loop. Notice that our bound in the small
$\gamma_{\phi}$ region is roughly $\gamma_{\phi^{2}}\leq c\sqrt{\gamma_{\phi}%
}$,$~c\simeq1.79$: in order to saturate it, $\gamma_{\phi^{2}}$ and
$\gamma_{\phi}$ should respectively arise at 1- and 2-loop order. This is
never going to be the case if $\phi$ has Yukawa couplings to fermions, but it
could in principle be so if $\phi$ is only coupled via quartic scalar
couplings, as these do not lead to wave function renormalization at 1-loop. It
is however easy to see that also this option does not help us to produce a
nontrivial saturation of our bound. Indeed the fixed point condition
necessarily implies that $\phi$ should enter at most linearly in scalar
quartic couplings with charged scalars, otherwise the beta function
$\beta(\lambda)$ for the self-coupling would be strictly positive. Then even
if a fixed point existed, with such limited, just linear, interaction
$\gamma_{\phi}$ and $\gamma_{\phi^{2}}$ would vanish at 1-loop. Saturation of
our bound would then require $\gamma_{\phi^{2}}=2$-loops and $\gamma_{\phi}%
=4$-loops which seems unlikely to happen. Notice also that $\phi^{2}$ will mix
with other invariant bilinears constructed with the charged scalars, so the
\textquotedblleft dimension of $\phi^{2}$\textquotedblright\ here means the
lowest eigenvalue of an in principle complicated matrix of anomalous dimensions.

One possible conclusion from the above discussion is that in order to saturate
our bound, even at $\gamma_{\phi}$ near $0$, we necessarily need a theory at
small $N$. In 4D we unfortunately have no other examples to play with. One
obvious next try is to consider fixed points in $4-\epsilon$ dimensions. Even
if our bound strictly applies only to 4D theories, the comparison to fixed
points in $4-\epsilon$ is almost compulsory. The result, as we now show, is
partly encouraging and partly frustrating. Consider the $O(N)$ theory in
$4-\epsilon$ with Lagrangian
\begin{equation}
\mathcal{L}=\frac{1}{2}\partial_{\mu}\phi_{a}\partial^{\mu}\phi_{a}%
-\frac{\lambda}{4!}(\phi_{a}\phi_{a})^{2}\,.
\end{equation}
As was first studied in {\cite{Wilson}}, this model has a fixed point at
$\lambda(N+8)/48\pi^{2}=\epsilon$. There are two operators playing the role of
$\phi^{2}$, the singlet $\mathcal{O}_{S}=\phi_{a}\phi_{a}$ and the symmetric
traceless tensor $\mathcal{O}_{T}=\phi_{a}\phi_{b}-(1/N)\delta_{ab}(\phi
_{c}\phi_{c})$. The computation of anomalous dimension gives {\cite{Wilson}}
\begin{align}
d_{\phi}  &  =d_{\text{free}}+\gamma_{\phi}=\left(  1-\frac{\epsilon}%
{2}\right)  +\frac{N+2}{4(N+8)^{2}}\epsilon^{2}\,,\\
\Delta_{S}  &  =\Delta_{\text{free}}+\gamma_{S}=\left(  2-{\epsilon}\right)
+\frac{N+2}{N+8}\epsilon\,,\\
\Delta_{T}  &  =\Delta_{\text{free}}+\gamma_{T}=\left(  2-{\epsilon}\right)
+\frac{2}{N+8}\epsilon\,,
\end{align}
where in brackets we have indicated the free field scaling dimensions in
$D=4-\epsilon$ dimensions, $d_{\text{free}}$ and $\Delta_{\text{free}}$, for
$\phi$ and $\phi^{2}$ respectively. In analogy, and \textquotedblleft naive
continuity\textquotedblright, with our study in 4D we should compare the
anomalous dimension of the composite and elementary fields. Indeed the
anomalous dimension of $\phi$ arises only at two-loops so that we have
$\gamma_{T,S}\propto\sqrt{\gamma_{\phi}}$ like in our bound! One always has
$\gamma_{T}<\gamma_{S}$ and the most interesting relation is that between
$\gamma_{T}$ and $\gamma_{\phi}$
\begin{equation}
\gamma_{T}=\frac{4}{\sqrt{N+2}}\sqrt{\gamma_{\phi}}\equiv c_{N}\sqrt
{\gamma_{\phi}}\,.
\end{equation}
For $N\geq3$, $c_{N}<1.79$ consistent with our 4D bound. On the other hand
$c_{1},c_{2}>1.79$, above our bound. It is not clear what to make of this
apparent contradiction, given that our bound surely applies only to 4D while
here we are discussing a theory in $4-\epsilon$. On one side one would be
tempted to argue that our bound smoothly extends to $4-\epsilon$, namely
\begin{equation}
\gamma_{\phi^{2}}\leq c(\epsilon)\sqrt{\gamma_{\phi}}%
\end{equation}
with $c(\epsilon)$ well behaved near $\epsilon=0$. If our 4D result is correct
then this cannot be the case. Instead it is possible that the relation between
$\gamma_{\phi^{2}}$ and $\gamma_{\phi}$ away from $\epsilon=0$ is more
complicated than our result. Indeed we can view our 4D result as a bound on
$\gamma_{\phi^{2}}/\sqrt{\gamma_{\phi}}$ at $\sqrt{\gamma_{\phi}}\gg\epsilon$.
The full result could be
\begin{equation}
\gamma_{\phi^{2}}\leq\sqrt{\gamma_{\phi}}A(\sqrt{\gamma_{\phi}}/\varepsilon),
\end{equation}
where $A(x)$ is a function which interpolates between our coefficient
$c\simeq1.79$ at $x=\infty$, and a larger coefficient at $x=0$, with a
crossover around $x\sim1.$ For instance:
\begin{equation}
A(x)=c+\frac{\delta c}{x^{2}+1},\qquad\delta c>0,
\end{equation}
could do the job.

\subsection{Bounds in 2D CFT and comparison with exact results}

A wealth of information accumulated about exactly solvable CFTs in 2D
\cite{df} allows for a nontrivial check of our method. Much of our discussion
in Sections 3,4,5 carries over to 2D with minimal, simplifying, changes. In 2D
CFTs, we must make distinction between the global conformal group $SL(2,C),$
and the infinite-dimensional Virasoro algebra of local conformal
trasnformations, of which $SL(2,C)$ is a finite-dimensional subgroup
\ Virasoro algebra plays crucial role in solving these theories exactly, but
it has no analogue in 4D, and the results have to be expressed in $SL(2,C)$
terms to allow for comparison.

When we speak about primaries, descendants, conformal blocks in 2D theories,
we must specify with respect to which group we define these concepts, Virasoro
or $SL(2,C)$. The former is standard in the 2D CFT literature, while it is the
latter that is directly analogous to 4D situation.\footnote{$SL(2,C)$
primaries are sometimes called \textit{quasi-primaries} in the 2D CFT
literature.}

Every Virasoro primary is a $SL(2,C)$ primary, but the converse is not true.
E.g.~the stress tensor in any 2D\ CFT is a Virasoro descendant of the unit
operator. To find $SL(2,C)$ primaries, we need to decompose the sequence of
all Virasoro descendants of each Virasoro primary (the so called Verma module)
into irreducible $SL(2,C)$ representations. While this is possible in
principle, it may not be straightforward in practice. Nevertheless we know
that $SL(2,C)$ primaries have dimensions of the form%
\[
\Delta_{SL(2,C)}=\Delta_{\text{Vir}}+n,\quad n=0\text{ or }n\geq2,
\]
where $\Delta_{\text{Vir }}$ is a Virasoro primary dimension, and $n$ is an
integer. This is because the Virasoro operators which are not in $SL(2,C)$
raise the dimension by at least 2 units.

The unitarity bound for bosonic fields in 2D is%
\[
\Delta\geq l\text{,}%
\]
where $l=0,1,2,\ldots$ is the Lorentz spin. The $SL(2,C)$ conformal blocks in
2D were found in \cite{do1},\cite{do2}\footnote{In contrast, explicit
expressions for Virasoro conformal blocks are not known in general.}; in the
same coordinates as before we have%
\begin{equation}
g_{\Delta,l}(u,v)=\frac{(-)^{l}}{2^{l}}\left[  \,f_{\Delta+l}(z)f_{\Delta
-l}(\bar{z})+(z\leftrightarrow\bar{z})\right]  . \label{DO2D}%
\end{equation}
Using the unitarity bound, the known conformal blocks, and the sum rule
(\ref{sumrule}), which is valid in any dimension, we can try to answer the
same question as in 4D. Namely, for a $SL(2,C)$ scalar primary $\phi$ of
dimension $d$, what is an upper bound on the dimension $\Delta_{\min}$ of the
first scalar operator appearing in the OPE $\phi\times\phi$? I.e. we want a 2D
analogue of Eq.~(\ref{main}). Since the free scalar is dimensionless in 2D,
the region of interest is $d\rightarrow d_{\text{free}}=0$.

\begin{figure}
[ptb]
\begin{center}
\includegraphics[
height=2.3428in,
width=3.3416in
]%
{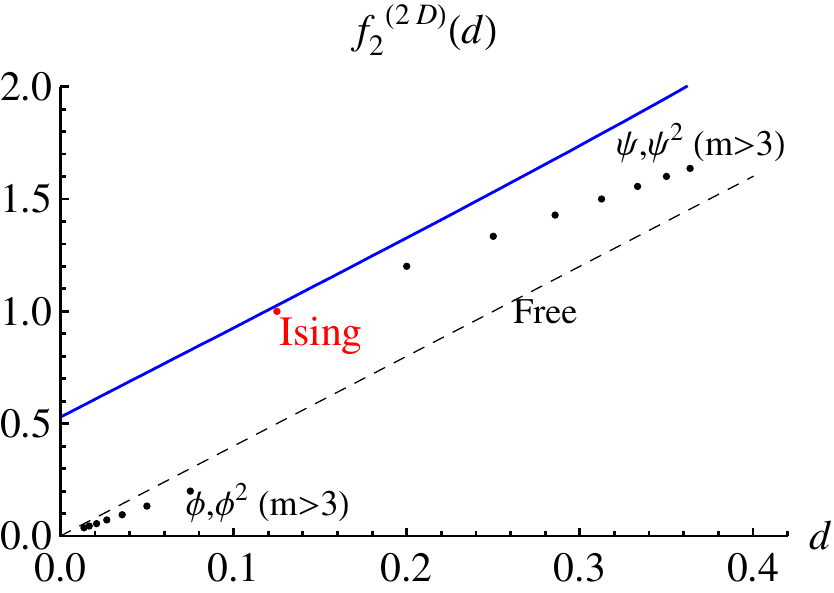}%
\caption{The solid (blue) line represents the simplest upper bound, in an
arbitrary 2D CFT, on the dimension $\Delta_{\min}$ of the first scalar in the
OPE $\mathcal{O}\times\mathcal{O}$ of a dimension $d$ scalar with itself. The
dots show the position of the minimal model OPEs $\phi\times\phi$ and
$\psi\times\psi$ (see the text) in this plane. The dashed line corresponds to
the free theory OPE (\ref{vop}). The bound is respected in all cases.}%
\label{2D}%
\end{center}
\end{figure}

In Fig.~\ref{2D} we show such a bound on $\Delta_{\min}$ as a function of $d$
obtained using the second derivatives $F^{(2,0)}$, $F^{(0,2)}$ (thus this
bound is analogous to the simple 4D bound from Section \ref{simplest}). The
dependence looks approximately linear:%
\[
f(d)=f_{2}^{(2D)}(d)\simeq0.53+4d\quad\quad\text{(2D, 2nd
derivatives)\thinspace.}%
\]
Unlike in 4D, this simplest bound does not approach the canonical value zero
as $d\rightarrow d_{\text{free}}=0$; we do not know if this has any deep
meaning. Improvements of this bound using the method discussed in Section
\ref{general} are possible; see below.%

%TCIMACRO{\FRAME{ftbpFU}{3.3416in}{2.3428in}{0pt}{\Qcb{The solid (blue) line
%represents the simplest upper bound, in an arbitrary 2D CFT, on the dimension
%$\Delta_{\min}$ of the first scalar in the OPE $\mathcal{O}\times\mathcal{O}$
%of a dimension $d$ scalar with itself. The dots show the position of the
%minimal model OPEs $\phi\times\phi$ and $\psi\times\psi$ (see the text) in
%this plane. The dashed line corresponds to the free theory OPE (\ref{vop}).
%The bound is respected in all cases.}}{\Qlb{2D}}{2d_cft.pdf}%
%width 3.3416in;  height 2.3428in;  depth 0pt;  original-width 3.333in;
%original-height 2.3281in;  cropleft "0";  croptop "1";  cropright "1";
%cropbottom "0";  filename '2D_CFT.pdf';file-properties "XNPEU";}} }%
%BeginExpansion
%EndExpansion

We will now see how the bound of Fig.~\ref{2D} checks with the known operator
dimensions and OPEs in solvable unitary CFTs in 2D. Our first example is the
free scalar in 2D. This CFT contains the so called \textit{vertex operator}
primaries given by an exponential of the fundamental field:%
\begin{equation}
V_{\alpha}=e^{i\alpha\phi},\quad\lbrack V_{\alpha}]=\alpha^{2}. \label{vop}%
\end{equation}
The basic OPE of $V_{\alpha}$ with itself has the form:%
\[
V_{\alpha}\times V_{\alpha}=V_{2\alpha}\,\,.
\]
Thus we have $d=\alpha^{2}$, $\Delta=4\alpha^{2}$, which gives the dashed line
in Fig. \ref{2D},\footnote{Strictly speaking the bound in Fig. \ref{2D} was
derived for real fields. However, we can apply it to the real parts which
satisfy the OPE $\operatorname{Re}V_{\alpha}\times\operatorname{Re}V_{\alpha
}\sim1+\operatorname{Re}V_{2\alpha}$.} below the bound.

A more interesting example involves the \textit{minimal model} family of
exactly solvable 2D CFT. The unitary minimal models (see \cite{df}%
,\cite{Polchinski}) are numbered by an integer $m=3,4,\ldots$, and describe
the universality class of the multicritical Ginzburg-Landau model:%
\begin{equation}
\mathcal{L}\sim\left(  \partial\phi\right)  ^{2}+\lambda\phi^{2m-2}\,.
\label{GL}%
\end{equation}
For $m=3$, the Ising model is in the same universality class. The central
charge of the model,%
\[
c=1-\frac{6}{m(m-1)},
\]
monotonically approaches the free scalar value $c_{\text{free}}=1$ as
$m\rightarrow\infty.$ Intuitively, as $m$ increases, the potential becomes
more and more flat, allows more states near the origin ($c$ grows), and
disappears as $m\rightarrow\infty$ (free theory).

Minimal models are called so because they have finitely many Virasoro primary
fields (the number of $SL(2,C)$ primaries is infinite). All Virasoro primaries
are scalar fields $O_{r,s}$ numbered by two integers $1\leq s\leq r\leq m-1$,
whose dimension is%
\begin{equation}
\Delta_{r,s}=\frac{(r+m(r-s))^{2}-1}{2m(m+1)}\,. \label{dims}%
\end{equation}
The $O_{1,1}$ is the unit operator ($\Delta_{1,1}=0)$, while the field
$\phi\equiv O_{2,2}$ has the smallest dimension among all nontrivial
operators:%
\begin{equation}
d_{\phi}=\Delta_{2,2}=\frac{3}{2m(m+1)}. \label{dimphi}%
\end{equation}
This field is identified with the Ginzburg-Landau field in (\ref{GL}). For
$m=3$ we have $\Delta_{2,2}=1/8$, which is the spin field dimension in the
Ising model.

It is convenient to extend the Virasoro primary fields to a larger range
$1\leq r\leq m-1$, $1\leq s\leq m$, subject to the identification%
\begin{equation}
(r,s)\leftrightarrow(m-r,m+1-s)\,. \label{id}%
\end{equation}
The \textit{fusion rules}, which say which operators appear in the OPE
$O_{r_{1}s_{1}}\times O_{r_{2}s_{2}}$ (but do not specify the coefficients)
can now be written in a relatively compact form:%
\begin{align}
&  O_{r_{1}s_{1}}\times O_{r_{2}s_{2}}\sim\sum O_{r,s}\label{fusion}\\
&  r=|r_{1}-r_{2}|+1,|r_{1}-r_{2}|+3,\ldots\min(r_{1}+r_{2}-1,2m-1-r_{1}%
-r_{2})\nonumber\\
&  s=|s_{1}-s_{2}|+1,|s_{1}-s_{2}|+3,\ldots\min(s_{1}+s_{2}-1,2m+1-s_{1}%
-s_{2})\nonumber
\end{align}
For any $m,$ the fusion rules respect a discrete $Z_{2}$ symmetry%
\begin{equation}
O_{r,s}\rightarrow\pm O_{r,s}, \label{Z2}%
\end{equation}
where $\pm=(-1)^{s-1}$ for $m$ odd, $(-1)^{r-1}$ for $m$ even (this choice is
dictated by consistency with (\ref{id})). This symmetry corresponds to the
$\phi\rightarrow-\phi$ symmetry of the Ginzburg-Landau model; in particular
$\phi=O_{2,2}$ is odd under (\ref{Z2}).

We are interested in OPEs of the form $O\times O\sim1+\tilde{O}+\ldots$ where
both $O$ and $\tilde{O}$ have small dimensions. Two such interesting OPEs are%
\begin{align}
\phi\times\phi &  \sim1+\phi^{2}+\ldots\label{phiphi}\\
\psi\times\psi &  \sim1+\psi^{2}+\ldots,\quad\psi\equiv O_{1,2},\quad d_{\psi
}=\frac{1}{2}-\frac{3}{2(m+1)}. \label{psipsi}%
\end{align}
Here $\phi^{2}$ and $\psi^{2}$ are just notation for the lowest dimension
operators appearing in the RHS. Note that for $m=3$ we have $\psi\equiv\phi$
via (\ref{id}). Using the fusion rule (\ref{fusion}) and the operator
dimensions (\ref{dims}) it is not difficult to make identification:%
\begin{align}
m=3\text{:\quad}  &  \phi^{2}\equiv O_{1,3},~\Delta_{\phi^{2}}=1,\quad
\text{(Ising)}\label{dimphi2}\\
m>3\text{:\quad}  &  \phi^{2}\equiv O_{3,3},~\Delta_{\phi^{2}}=\frac
{4}{m(m+1)}\,,\nonumber\\
&  \psi^{2}\equiv O_{1,3},~\Delta_{\psi^{2}}=2-\frac{4}{m+1}.
\end{align}
In particular, we see that the $\psi\times\psi$ OPE does not contain $\phi
^{2}$, which is precisely the reason why we are considering it\footnote{In
general, $\phi^{2}$ does not appear in the OPE $O_{r,s}\times O_{r,s}$ for
$r=1$ or $s=1$. The operators $\psi$ has the lowest dimension among all these
fields.}.

We are now ready for the check. Operator dimensions in both OPEs
(\ref{phiphi}) and (\ref{psipsi}) are subject to the bound of Fig.~\ref{2D},
where we marked the corresponding points up to $m=10$. We see that the bound
is respected in all the cases, although the Ising model point lies remarkably
close to the boundary. We have tried to improve the bound of Fig.~\ref{2D} by
including more derivatives in the functional, according to the general method
described in Section \ref{general}. Although we have seen some improvement for
lower valued of $d$, there was practically no improvement around the Ising
spin dimension $d=1/8$, so that also the improved bound was respected. One
could wonder if the fact that the Ising model (almost) saturates the bound has
any special significance.

We have searched for other exactly solvable 2D CFTs which could provide checks
of our 2D bound. E.g.~some interesting OPEs can be extracted from the WZNW
models. However, as far as we could see, none of them come close to saturating
the bound.

In conclusion, we would like to mention that some bounds for
dimensions of operators appearing in the OPE of two primaries in 2D
CFTs were derived in the past by Lewellen \cite{lewellen} and
Christe and Ravanini \cite{ravanini}. Those bounds were however of a
different nature than our bound (\ref{main}). Roughly, the
Lewellen-Christe-Ravanini (LCR) bounds say that \textit{IF} a
primary appears in the OPE, its dimension is not bigger than a
certain bound. This is of course not the same as our result, which
says that a certain primary \textit{MUST} be present in the OPE,
with the dimension not bigger than a certain bound.

The methods of LCR are based on studying the monodromy of the
conformal blocks near their singularities in the complex plane, as
opposed to the more detailed information about the shape and size of
conformal blocks at intermediate \textit{regular} points used by us.
They have to make a crucial assumption that only a finite number of
singularity types exist, which means that the total number $N$ of
primaries appearing in the OPE, or at least the total number of such
primaries having different $\Delta\operatorname{mod}1$, is finite.
This assumption is realized in Rational CFTs, but not in general.
The LCR bounds become increasingly weak for large $N$ and disappear
in the limit $N\rightarrow\infty.$ Thus it is doubtful that such
methods could be useful in our problem, since we would like to be
free of any assumptions about the spectrum of higher primaries.

\section{Comparison to phenomenology}

\label{sec:connection}

In this section we will comment on the precise relation of our main result
(\ref{main}) with the phenomenological discussion of Section \ref{pheno}. That
discussion led to the constraints (\ref{dbound}),~(\ref{deltadbound}) on the
dimension $d$ of the Higgs field operator $H$ and on the dimension $\Delta
_{S}$ of the first \textit{singlet} in the $H\times H^{\dagger}$ OPE, denoted
$H^{\dagger}H$. Are there any low-dimension non-singlets in this OPE?

The standard considerations related to the $\rho$-parameter lead us to assume
the \textquotedblleft custodial" $SO(4)=SU(2)_{L}\times SU(2)_{R}$ as the
global symmetry of the CFT \footnote{Our conclusions would however not be
affected by assuming just $SU(2)\times U(1)$.}. The real components $h_{a}$ of
the Higgs field:
\[
H=\left(
\begin{array}
[c]{c}%
h_{1}+ih_{2}\\
h_{3}+ih_{4}%
\end{array}
\right)  \,,
\]
form a multiplet of primary scalars in the fundamental of $SO(4)$. Their basic
OPE will in general have the structure%
\begin{equation}
h_{a}(x)h_{b}(0)\sim\frac{1}{|x|^{2d}}\left(  \delta_{ab}1+C_{S}%
|x|^{\Delta_{S}}\delta_{ab}\,(H^{\dagger}H)(0)+C_{T}|x|^{\Delta_{T}%
}\mathcal{T}_{(ab)}(0)+C_{J}|x|^{2}x^{\mu}J_{\mu}^{[ab]}(0)+\ldots\right)  \,.
\label{eq:OPEglobal}%
\end{equation}
Here we indicated the possibility for two symmetry structures in the even-spin
sector: along with the $SO(4)$ singlet $H^{\dagger}H$, a scalar $\mathcal{T}%
_{ab}$ transforming as a \textit{traceless} \textit{symmetric} tensor,
\textit{i.e.}~the $(1,1)$ representation, will in general be present. E.g.~in
free theory $\mathcal{T}_{ab}=h_{a}h_{b}\!-(1/4)\delta_{ab}\left(  h_{c}%
h_{c}\right)  .$ On the other hand, in the odd-spin sector the first
contribution will be associated with the conserved, dimension $3$, $SO(4)$
current $J_{\mu}^{[ab]},$ which is in the \textit{antisymmetric} tensor
representation $(1,0)+(0,1).$ Its coefficient $C_{J}$ will be related to the
normalization of the $SO(4)$ Ward identity.

\begin{figure}[ptb]
\centering\begin{minipage}[c]{0.4\textwidth}
\centering\includegraphics[height=3in]{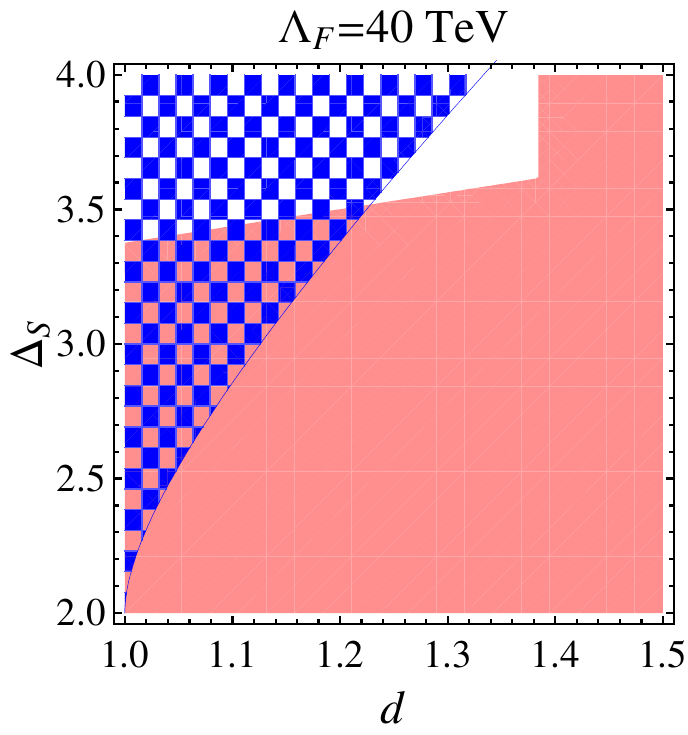}
\end{minipage}\begin{minipage}[c]{0.4\textwidth}
\centering\includegraphics[height=3in]{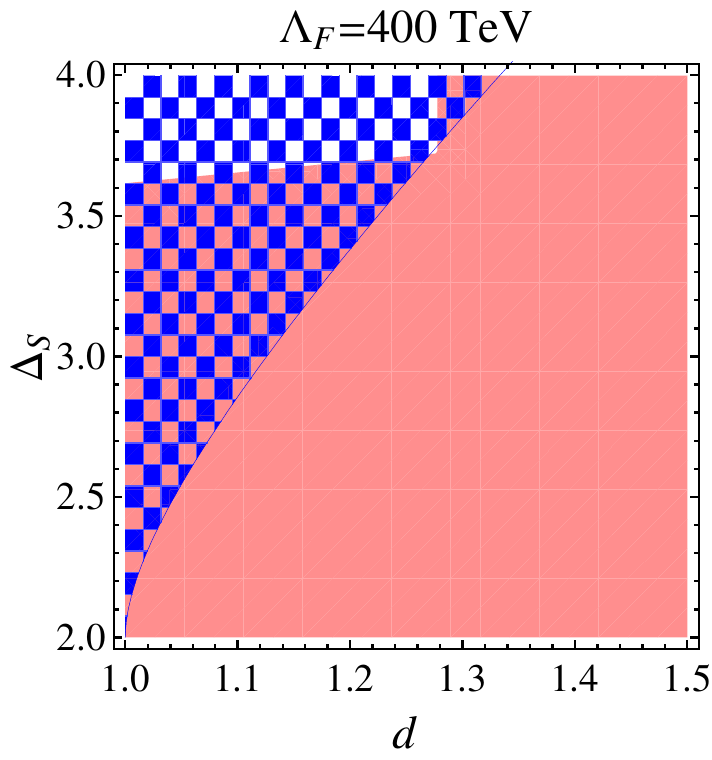}
\end{minipage}\caption{\textit{Solid red}: The region of $(d,\Delta_{S})$
plane disfavored by phenomenological considerations of Section \ref{pheno}. In
this region one or both constraints (\ref{dbound}),~(\ref{deltadbound}) are
not satisfied. We consider two cases: $\Lambda_{\text{F}}=40$ TeV $($left) and
$\Lambda_{\text{F}}=400$ TeV (right). \textit{Checkered blue}: the region of
the same plane which \emph{would be} excluded by our bound (\ref{main}) with
$f=f_{6}$ if we knew that $\Delta_{S}<\Delta_{T}$.}%
\label{LambdaF}%
\end{figure}

Clearly (\ref{eq:OPEglobal}) looks more complicated than (\ref{OPE0}).
However, if we choose $a=b=1$ (say) in (\ref{eq:OPEglobal}), the current and
other odd-spin fields drop out due to antisymmetry, and we get an OPE\ of the
form (\ref{OPE0}). We just need to identify $\phi\equiv h_{1}$, $\phi
^{2}\equiv$ $H^{\dagger}H$ or $\mathcal{T}_{11}$ depending on which of these
two scalars has smaller dimension. Eq.~(\ref{main}) then implies%
\[
\min(\Delta_{S},\Delta_{T})\leq f(d)\,.
\]
We get a bound on $\Delta_{S}$ only assuming $\Delta_{S}<\Delta_{T}$. This is
not satisfactory; in fact the reverse $\Delta_{S}>\Delta_{T}$ seems likely (as
it happens in the Wilson-Fisher fixed points in $4-\varepsilon$ dimensions,
see Section \ref{sec:comparison}). However, we have reasons to believe (or
hope) that our bound on $\min(\Delta_{S},\Delta_{T})$ is close to being
saturated by some existing 4D CFT. One reason is that the truncation discussed
in Section \ref{sec:best} shows clear signs of convergence. The other reason
is that in $4-\epsilon$ we do have $\gamma_{\phi^{2}}\sim\sqrt{\gamma_{\phi}}$
as dictated by our bound at small $\gamma_{\phi}$. Taking this into
consideration it makes sense to compare our bound $f(d)$ to the
phenomenological constraints on $\Delta_{S}$. We have done that in
Fig.~\ref{LambdaF} assuming two different constraints on the scale
$\Lambda_{\text{F}}$, a weaker one $\Lambda_{\text{F}}>40$ TeV and a robust
one $\Lambda_{\text{F}}>400$ TeV. According to the previous reasoning, even
for the less likely situation $\Delta_{S}<\Delta_{T}$, these plots indicate
that there exists space for relaxing the flavor problem. The optimal value for
the dimension of the Higgs field turns out to be between 1.2 and 1.3. Is it
possible that a situation like this will be realized in nature? Hints of the
answer to this question may soon come with the LHC.

In the meantime, we believe that it should be possible to disentangle the
contributions of $H^{\dagger}H$ and $\mathcal{T}$ in (\ref{eq:OPEglobal}) and
obtain a bound on $\Delta_{S}$ free of the assumption that $\Delta_{S}%
<\Delta_{T}.$ As the above discussion clearly shows, such a bound cannot be
found by considering only \textit{diagonal} $\phi\times\phi$ OPEs where $\phi$
is a fixed $h_{a}$ component. One should try to use additional information
contained in the \textit{nondiagonal }OPEs with $a\neq b$, something which we
did not do in this paper. As it is apparent from (\ref{eq:OPEglobal}), the
global symmetry current and higher odd-spin operators will generally
contribute to these nondiagonal OPEs. More work is needed to determine if the
contributions of the odd-spin operators can be controlled in a
model-independent way.

\section{Discussion and Outlook}

\label{sec:conclusions}

In this paper we have shown that prime principles of Conformal Field Theory,
such as unitarity, OPE, and conformal block decomposition, imply the existence
of an upper bound $f(d)$ on the dimension $\Delta_{\min}$ of the first scalar
operator $\phi^{2}$ in the OPE of a scalar $\phi$ of a given dimension $d$.

We developed a method which allows numerical determination of $f(d)$ with
arbitrary desired accuracy. The method is based on the \textit{sum rule}, a
function-space identity satisfied by the conformal block decomposition of the
4-point function $\left\langle \phi\phi\phi\phi\right\rangle $, which follows
from the crossing symmetry constraints. In practical application of the method
the sum rule is Taylor-expanded: replaced by finitely many equations for the
derivatives. The bound $f(d)$ improves monotonically as more and more
derivatives are included; see Figs.~\ref{fig:bound-intro}, \ref{fig:ill} for
the best current bound obtained using derivatives up to the 6th order, and a
sequence of weaker bounds obtained using fewer derivatives.

We have checked that our bound is satisfied, by a large margin, in all weakly
coupled 4D CFTs that we are able to construct. We have also derived an
analogous bound in 2D and checked it against exact 2D CFT results. Again, the
bound is satisfied, and in a less trivial way than in 4D, since the Ising
model almost saturates it.

Our results open up several interesting research directions:

\begin{enumerate}
\item It should be relatively straightforward to improve our bounds, both in
4D and in 2D, using our method but including more derivatives in the analysis.
These improved bounds should monotonically converge to the optimal bound,
corresponding to the infinite number of derivatives; we can already see signs
of such convergence in Fig.~\ref{fig:ill}.

\item One should search for more examples of CFTs which come close to
saturating the bound, especially in 4D.

\item The Dolan-Osborn closed-form expressions for conformal blocks are
available only in even dimensions (up to $D=6$). It is important to find
expressions in 3D, comparable in simplicity to (\ref{DO}) and (\ref{DO2D}).
Then one could derive an analogous bound in 3D and confront it with the
operator dimensions of known 3D CFTs, such as the $O(N)$ universality classes.
Although these theories are not exactly solvable, rather precise estimates for
critical exponents and operator dimensions have been obtained using
$\varepsilon$-expansion, high-temperature expansion, and Monte-Carlo
simulations \cite{ZJ}. For example, in 3D Ising model we have $\gamma_{\phi
}=0.0183(4)$ and $\gamma_{\phi^{2}}=0.412(1)$ \cite{3D}.

\item It would be interesting to understand what is the appropriate
extrapolation of our bound to $4-\varepsilon$ dimension. This should explain
why the comparison with Wilson-Fischer fixed points in Section
\ref{sec:comparison} was not perfect.

\item A very important but difficult problem is to find a genuine
generalization of our bound to the situation when the CFT has a global
symmetry. The case of $SO(4)$ symmetry, readily generalized to $SO(N)$, has
been discussed in Section \ref{sec:connection}.
\end{enumerate}

\section*{Acknowledgements}

We thank Pasquale Calabrese, Gergely Harcos, Marcus Luty, Hugh Osborn and
Yassen Stanev for useful discussions and communications. This work is
partially supported by the Swiss National Science Foundation under contract
No. 200021-116372. V.S.R. was also partially supported by the EU under RTN
contract MRTN-CT-2004-503369 and ToK contract MTKD-CT-2005-029466; he thanks
the Institute of Theoretical Physics of Warsaw University for hospitality
during final months of writing this paper.

\appendix

\section{Reality property of Euclidean 3-point functions}

\label{reality}

In this appendix we would like to briefly discuss the reality property of
3-point functions, which was used at some point in our discussion. First of
all, one can always choose a basis where each operator corresponds to an
hermitian operator in the Minkowski space description. We work in such a
basis. The reality properties of the Euclidean $n$-point functions for such
operators are quickly deduced by analitic continuation of the Minkowskian
correlators at space-like separation. Consider indeed the 3-point function
\begin{equation}
G^{\alpha_{1},\alpha_{2},\alpha_{3}}(x_{1},x_{2},x_{3})\equiv\langle
O_{1}^{\alpha_{1}}(x_{1})O_{2}^{\alpha_{2}}(x_{2})O_{3}^{\alpha_{3}}%
(x_{3})\rangle\,,
\end{equation}
where $\alpha_{i}$ collectively denote the spin indices (we only consider
bosons). When $x_{12}$, $x_{23}$ and $x_{31}$ are spacelike, the operators
commute by causality, thus implying that $G$ is real
\begin{equation}
G^{\alpha_{1},\alpha_{2},\alpha_{3}}(x_{1},x_{2},x_{3})^{\ast}=\langle
O_{3}^{\alpha_{3}}(x_{3})O_{2}^{\alpha_{2}}(x_{2})O_{1}^{\alpha_{1}}%
(x_{1})\rangle=\langle O_{1}^{\alpha_{1}}(x_{1})O_{2}^{\alpha_{2}}(x_{2}%
)O_{3}^{\alpha_{3}}(x_{3})\rangle=G^{\alpha_{1},\alpha_{2},\alpha_{3}}%
(x_{1},x_{2},x_{3})\,.
\end{equation}
Continuation to the Euclidean then amounts to
\begin{align}
x^{0}  &  \rightarrow-ix_{E}^{0}\qquad x^{k}\rightarrow x_{E}^{k}%
\quad(k=1,2,3)\,,\\
O^{\alpha}  &  \rightarrow(-i)^{n_{\alpha}}O_{E}^{\alpha}\,,
\end{align}
where $n_{\alpha}$ is the number of $0$ indices in $\{\alpha\}$. Using the
above rules, analytically continuing $G^{\alpha_{1},\alpha_{2},\alpha_{3}}$
from the spacelike patch we find the Euclidean functions
\begin{equation}
G_{E}^{\alpha_{1},\alpha_{2},\alpha_{3}}(x_{E1},x_{E2},x_{E3})=(i)^{n_{\alpha
_{1}}+n_{\alpha_{2}}+n_{\alpha_{3}}}G^{\alpha_{1},\alpha_{2},\alpha_{3}%
}(-ix_{E1}^{0},x_{E1}^{k};-ix_{E2}^{0},x_{E2}^{k};-ix_{E3}^{0},x_{E3}^{k})\,.
\label{euclidean3}%
\end{equation}
Now, by Lorentz and translation invariance $G^{\alpha_{1},\alpha_{2}%
,\alpha_{3}}(x_{1},x_{2},x_{3})$ depends on the invariants $x_{ij}^{2}$, while
the tensor indices are covariantly reproduced by combinations of $x_{ij}^{\mu
}$ with the invariant tensors $\eta^{\mu\nu}$ and $\epsilon^{\mu\nu\rho\sigma
}$. It is now evident that if $\epsilon^{\mu\nu\rho\sigma}$ does not appear in
$G^{\alpha_{1},\alpha_{2},\alpha_{3}}$ then the factor $(i)^{n_{\alpha_{1}%
}+n_{\alpha_{2}}+n_{\alpha_{3}}}$ in eq. (\ref{euclidean3}) is exactly
compensated by a factor $(-i)^{n_{\alpha_{1}}+n_{\alpha_{2}}+n_{\alpha_{3}}}$
from the coordinate dependence of the tensor structure. In such a situation
$G_{E}^{\alpha_{1},\alpha_{2},\alpha_{3}}$ is therefore real. On the other
hand, contributions to $G_{E}$ that are proportional to one (equivalently an
odd number) power of $\epsilon^{\mu\nu\rho\sigma}$ are pure imaginary. For the
case that interests us in which two operators, say $O_{1}$ and $O_{2}$ have
zero spin and $O_{3}\equiv O_{3}^{\mu_{1}\dots\mu_{j}}$, the tensor structure
of $G^{\mu_{1}\dots\mu_{j}}(x_{1},x_{2},x_{3})$ can only involve $x_{12}%
^{\mu_{k}}$, $x_{23}^{\mu_{k}}$ and $\eta^{\mu\nu}$ and thus the corresponding
euclidean function must be real. On the other hand the three point function
for vector fields admits in conformal field theory a contribution proportional
to $\epsilon^{\mu\nu\rho\sigma}$ which is precisely proportional to the
triangle anomaly diagram \cite{Schreier:1971um}. Therefore the euclidean
3-point function for vector fields in CFT is in general complex. To conclude
we notice that one can simply reproduce the result just discussed by formally
assigning the following transformation property to the invariant tensors
\begin{equation}
\eta_{\mu\nu}\rightarrow\delta_{\mu\nu}\qquad\qquad\epsilon_{\mu\nu\rho\sigma
}\rightarrow-i\epsilon_{\mu\nu\rho\sigma}\,.
\end{equation}

\section{Closed-form expressions for conformal blocks}

\label{lightening}

The Dolan-Osborn result (\ref{DO}) is crucial for us, and we would like to say
a few words about how it is derived, following \cite{do2}\footnote{The first
derivation \cite{do1} is a brute force resummation of contributions of all
conformal descendants of $\mathcal{O}$ and is not particularly enlightening.}.
The main idea is that the conformal block is, in a certain sense, a spherical
harmonic of the conformal group. In particular, it satisfies an eigenvalue
equation%
\begin{equation}
\mathcal{D}_{x_{1},x_{2}}\text{\texttt{CB}}_{\mathcal{O}}\,=-c_{\Delta
,l}\text{\texttt{CB}}_{\mathcal{O}}\,,\quad c_{\Delta,l}=l(l+2)+\Delta
(\Delta-4), \label{eigen}%
\end{equation}
where $\mathcal{D}$ is a second-order partial differential operator acting on
the coordinates $x_{1,2}$. This operator encodes the action of the quadratic
Casimir operator of the conformal group\footnote{Here and in (\ref{diff}) we
use the sign conventions of \cite{fradkin} for the conformal generators.},%
\[
C=\frac{1}{2}M_{\mu\nu}M_{\mu\nu}-D^{2}-\frac{1}{2}(P_{\mu}K_{\mu}+K_{\mu
}P_{\mu})\equiv L_{A}L_{A},
\]
on the state $\phi_{1}(x_{1})\phi_{2}(x_{2})|0\rangle.$ The defining equation
is%
\[
C\cdot\phi_{1}(x_{1})\phi_{2}(x_{2})\equiv\lbrack L_{A},[L_{A},\phi_{1}%
(x_{1})\phi_{2}(x_{2})]]=\mathcal{D}_{x_{1},x_{2}}\phi_{1}(x_{1})\phi
_{2}(x_{2}).
\]
On the other hand, $c_{\Delta,l}$ in (\ref{eigen}) are nothing but the
eigenvalues of the Casimir acting on conformal primaries \cite{ferrara}:
\[
C\cdot\mathcal{O}_{(\mu)}(0)=-c_{\Delta,l}\mathcal{O}_{(\mu)}(0).
\]
Since $C$ is a Casimir, the same eigenvalue equation is simultaneously
satisfied for all descendants of $\mathcal{O}$. Thus Eq.\ (\ref{eigen}) is a
consequence of the OPE (\ref{OPE}) and of the definition of conformal blocks.

The explicit form of $\mathcal{D}_{x_{1},x_{2}}$ can be found using the known
expressions for the action of conformal generators on scalar primaries:%
\begin{align}
\lbrack P_{\mu},\phi(x)]  &  =i\partial_{\mu}\phi\,,\label{diff}\\
\lbrack D,\phi(x)]  &  =i(\Delta_{\phi}+x^{\mu}\partial_{\mu})\phi
\,,\nonumber\\
\lbrack M_{\mu\nu},\phi(x)]  &  =i(x_{\mu}\partial_{\nu}-x_{\nu}\partial_{\mu
})\phi\,,\nonumber\\
\lbrack K_{\mu},\phi(x)]  &  =i(x^{2}\partial_{\mu}-2x_{\mu}x\cdot
\partial-2x_{\mu}\Delta_{\phi})\phi\,.\nonumber
\end{align}
Eq.\ (\ref{eigen}) can then be rewritten as a differential equation for
$g_{\Delta,l}(u,v).$ The clever change of variables $u,v\rightarrow z,\bar{z}$
performed in \cite{do2} allows to find explicit solutions. In these variables
the differential equation takes the form%
\begin{align*}
&  \qquad\qquad\qquad\qquad\qquad\mathcal{D}_{z,\bar{z}}g_{\Delta,l}=\frac
{1}{2}c_{\Delta,l}g_{\Delta,l},\\
&  \mathcal{D}_{z.\bar{z}}=z^{2}(1-z)\partial_{z}^{2}+\bar{z}^{2}(1-\bar
{z})\partial_{\bar{z}}^{2}-\left(  z^{2}\partial_{z}+\bar{z}^{2}\partial
_{\bar{z}}\right)  +\frac{2z\bar{z}}{\bar{z}-z}\left[  (1-z)\partial
_{z}-(1-\bar{z})\partial_{\bar{z}}\right]  \,.
\end{align*}
The OPE fixes the asymptotic behavior:%
\[
g_{\Delta,l}\sim\frac{(-)^{l}}{2^{l}}\frac{(z\bar{z})^{\frac{\Delta-l}{2}}%
}{z-\bar{z}}(z^{l+1}-\bar{z}^{l+1}),\quad(z,\bar{z}\rightarrow0).
\]
With these boundary conditions the solution is unique and is given by
(\ref{DO}).

\section{$z$ and $\bar{z}$}

\label{spacetime}

\begin{figure}
[h]
\begin{center}
\includegraphics[
height=3.8331cm,
width=7.1039cm
]%
{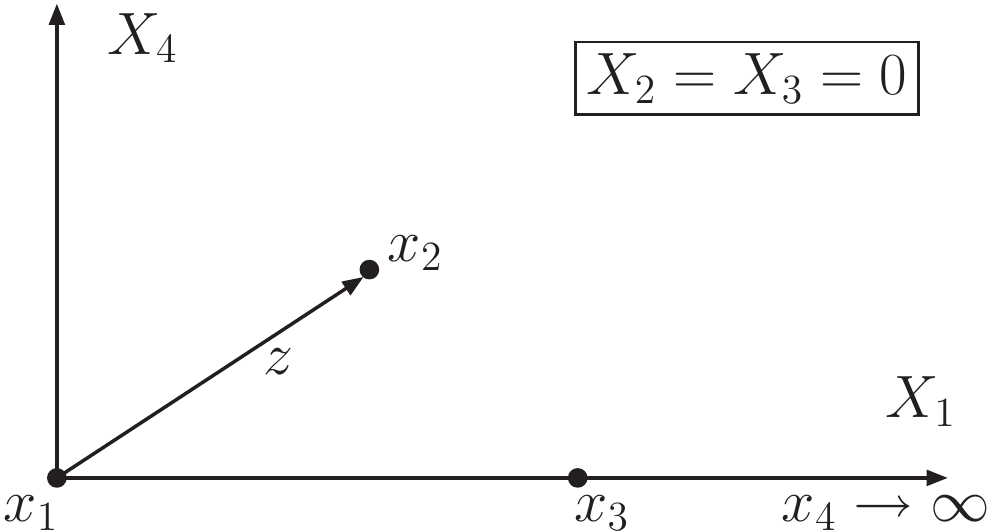}%
\caption{In Euclidean, $\bar{z}=z^{\ast}.$ For the configuration of points
chosen in this figure, $z=X_{1}+iX_{4}$ is the complex coordinate of $x_{2}$
in the $X_{1}-X_{4}$ plane. }%
\label{fig:E}%
\end{center}
\end{figure}

Here we comment upon the ranges of the variables $z$ and $\bar{z}$ introduced
in Section~\ref{sec:blocks}. Since $z$ and $\bar{z}$ are functions of the
conformally-invariant cross-ratios, we can use conformal symmetry to fix some
of the coordinate freedom. For instance, we can put three out of four points
along a straight line, and moreover send one of them to infinity, as in
Eq.\ (\ref{eq:3out4}). After that there still remains freedom to perform
rotations leaving this line invariant, so that we can fix%
\begin{equation}
x_{2}=(X_{1},0,0,X_{4})\,. \label{eq:x2E}%
\end{equation}
Now it becomes trivial to compute the cross-ratios:%
\begin{align*}
u  &  =x_{12}^{2}=X_{1}^{2}+X_{4}^{2}\,,\\
v  &  =x_{23}^{2}=(X_{1}-1)^{2}+X_{4}^{2}\,.
\end{align*}
Moreover, we can easily solve Eq.\ (\ref{uvzzbar}) for $z$ and $\bar{z}$:%
\begin{equation}
z=X_{1}+iX_{4}\text{,\quad}\bar{z}=z^{\ast}\,. \label{eq:zzbarE}%
\end{equation}
Thus we conclude that in the Euclidean, $\bar{z}$ is always the complex
conjugate of $z$. Moreover, $z$ is real if and only if all four points lie on
a circle (which is a conformal image of the straight line in the above
parametrization, see Fig.\ \ref{fig:E}).

Finally, Eq.\ (\ref{eq:x2M}),(\ref{eq:zzbarM}) are obtained from
(\ref{eq:x2E}),(\ref{eq:zzbarE}) by Wick-rotating to the Minkowski
time.\vspace{2cm}%

%TCIMACRO{\FRAME{fhFU}{7.1039cm}{3.8331cm}{0pt}{\Qcb{In Euclidean, $\bar
%{z}=z^{\ast}.$ For the configuration of points chosen in this figure,
%$z=X_{1}+iX_{4}$ is the complex coordinate of $x_{2}$ in the $X_{1}-X_{4}$
%plane. }}{\Qlb{fig:E}}{ct-fig2.pdf}{\special{ language "Scientific Word";
%type "GRAPHIC";  maintain-aspect-ratio TRUE;  display "USEDEF";
%valid_file "F";  width 7.1039cm;  height 3.8331cm;  depth 0pt;
%original-width 3.9548in;  original-height 2.1153in;  cropleft "0";
%filename 'ct-fig2.pdf';file-properties "XNPEU";}} }%
%BeginExpansion
%EndExpansion

\section{Asymptotic behavior}

\label{as}In this Appendix we find large $l$ and $\Delta$ asymptotics of
derivatives of $F_{d,\Delta,l}$ at $a=b=0$. It is useful to rewrite the
definition of $F_{d,\Delta,l}$ as follows:%
\begin{equation}
F_{d,\Delta,l}(a,b)=h_{d}(a,b)\frac{\tilde{g}_{d,\Delta,l}(a,b)-\tilde
{g}_{d,\Delta,l}(-a,b)}{a\,}\,, \label{rewrite}%
\end{equation}
where we introduced the functions%
\begin{align*}
\tilde{g}_{d,\Delta,l}  &  \equiv\lbrack(1-z)(1-\bar{z})]^{d}g_{\Delta,l}\,,\\
h_{d}(a,b)  &  \equiv\frac{a}{(z\bar{z})^{d}-[(1-z)(1-\bar{z})]^{d}}\,.
\end{align*}
These functions are smooth in the spacelike diamond. Moreover, it is not
difficult to see that
\[
\tilde{g}(a,-b)=\tilde{g}(a,b),\;\;h_{d}(\pm a,\pm b)=h_{d}(a,b)\,.
\]
In particular, from (\ref{rewrite}) we see the property (\ref{inv}) from
Section \ref{sum-rule}.

Let us introduce the parameter%
\[
\delta\equiv\Delta-l-2.
\]
As we will see below, there are three relevant asymptotic limits to consider:

\begin{itemize}
\item $l$ large, $\delta={O}(1)\,;$

\item $l$ large, $\delta$ large, $\delta\ll l^{2}\,;$

\item $\delta$ large, $\delta\gg l^{2}\,.$
\end{itemize}

In all these cases the large asymptotic behavior of derivatives will come from
differentiating $g_{\Delta,l}$, which we write in the form%
\begin{equation}
g_{\Delta,l}=const(-)^{l}\frac{z\bar{z}}{b}\left[  \,k_{2l+\delta
+2}(z)k_{\delta}(\bar{z})-(b\rightarrow-b)\right]  \,. \label{gdl}%
\end{equation}
In this Appendix by $const$ we denote various \textit{positive} constants
which may depend on $d,$ $\delta$ or $l$ but are independent of the derivative
order $\partial_{a}^{2m}\partial_{b}^{2n}$. These constant factors are
irrelevant for controlling the positivity of the linear functionals defined on
the cones.

Starting from the following integral representation for the hypergeometric
function (see \cite{bateman})%
\[
_{2}F_{1}\left(  a,b,c;x\right)  =\frac{\Gamma(c)}{\Gamma(b)\Gamma(c-b)}%
\int_{0}^{\infty}e^{-bt}(1-e^{-t})^{c-b-1}(1-x\,e^{-t})^{-a}\,dt\quad(\text{Re
}c>\text{Re }b>0)
\]
and using the steepest descent method, we derive the large $\beta$
asymptotics:%
\begin{align*}
&  k_{\beta}(x)\,=\,e^{\,(\beta/2)h(x)}\,[q(x)+{O}(1/\beta)]\,,\\
h(x)\,  &  =\,\ln\left(  \frac{4(1-\sqrt{1-x}\,)^{2}}{x}\right)  ,\quad
q(x)\,=\,\frac{x}{2\left(  1-\sqrt{1-x}\right)  \,\sqrt[4]{1-x}}\,.
\end{align*}
The leading asymptotic behavior appears when all the derivatives fall on the
exponential factors in $f_{\beta}$ containing large exponents. Various
prefactors appearing in (\ref{rewrite}) and (\ref{gdl}) are not differentiated
in the leading asymptotics. However, the $a^{-1}$ and $b^{-1}$ factors are
responsible for changing the order of the needed derivative, as follows:%
\begin{align*}
&  F_{d,\Delta,l}^{(2m,2n)}\sim\frac{const}{2m+1}(g_{\Delta,l})^{(2m+1,2n)}\\
&  \qquad\quad~\sim\frac{const(-)^{l}}{(2m+1)(2n+1)}(\exp A)^{(2m+1,2n+1)}%
\,,\\[5pt]
&  A=(l+\delta/2)h(1/2+a+b)+(\delta/2)h(1/2+a-b)\,.
\end{align*}
To find the leading asymptotics, we expand $A$ near $a=b=0$:%
\begin{align}
A  &  =(l+\delta)[h(1/2)+a\,h^{\prime}(1/2)]+lb\,h^{\prime}(1/2)+(\delta
/2)b^{2}h^{\prime\prime}(1/2)+\ldots,\label{expexp}\\
&  h^{\prime}(1/2)=2\sqrt{2},\quad h^{\prime\prime}(1/2)=-2\sqrt{2}.\nonumber
\end{align}
In the case $\delta\ll l^{2}$ the last term in (\ref{expexp}) plays no role,
and we get:%
\begin{equation}
F_{d,\Delta,l}^{(2m,2n)}\sim\frac{const(-)^{l}}{(2m+1)(2n+1)}[h^{\prime
}(1/2)(l+\delta)]^{2m+1}[h^{\prime}(1/2)l]^{2n+1},\quad\delta\ll l^{2}\,.
\label{as1}%
\end{equation}
This asymptotic is applicable for $l$ large, while $\delta$ can be small or
large, as long as the condition $\delta\ll l^{2}$ is satisfied; i.e.~it covers
the first two cases mentioned above. If on the other hand $\delta\gg l^{2}$,
it is the last term in (\ref{expexp}) which determined the asymptotics of
$b$-derivatives, and we get%
\[
F_{d,\Delta,l}^{(2m,2n)}\sim const(-)^{l}\frac{(2n-1)!!}{2m+1}[h^{\prime
}(1/2)\delta]^{2m+1}[h^{\prime\prime}(1/2)\delta]^{n},\quad\delta\gg l^{2}\,.
\]
Because $h^{\prime\prime}(1/2)<0$, the last asymptotics changes sign depending
on the parity of $n$.

\end{document}